\RequirePackage{lmodern}
\documentclass[12pt, oneside, extrafontsizes]{memoir}  

\setstocksize{11in}{8.5in}
\settrimmedsize{11in}{8.5in}{*}
\settrims{0in}{0in}
\setlrmarginsandblock{38mm}{25mm}{*}
\setulmarginsandblock{25mm}{25mm}{*}
\setheadfoot{26pt}{26pt}
\setheaderspaces{*}{13pt}{*}
\checkandfixthelayout
\DoubleSpacing
\setsecnumdepth{subsubsection}
\headstyles{default}
\chapterstyle{ell}
\setsecheadstyle{\scshape\LARGE\raggedright}

\usepackage[colorlinks,bookmarksnumbered,bookmarksdepth=subsubsection,unicode=true]{hyperref}
\newsubfloat{figure}  
\hypersetup{
pdfauthor = {Charles C Onu},
pdftitle = {Making deep neural networks work for medical audio: representation, compression and domain adaptation},
pdfsubject = {Subject},
pdfkeywords = {machine learning, },
pdfcreator = {LaTeX with the hyperref package},
pdfproducer = {},
linkcolor = [HTML]{000000},
citecolor = [HTML]{0000FF},
}
\usepackage{amsmath}
\usepackage{amsfonts}
\usepackage{amssymb}
\usepackage{amsthm}
\usepackage{dsfont}
\usepackage{xspace}
\usepackage{tikz}
\usetikzlibrary{shapes,arrows}
\usepackage{standalone}
\usepackage{algorithm}
\usepackage{algpseudocode}
\usepackage{appendix}
\usepackage{array}

\AtBeginEnvironment{subappendices}{%
\chapter*{Appendix}
\addcontentsline{toc}{chapter}{Appendices}
\counterwithin{figure}{section}
\counterwithin{table}{section}
}

\newcommand{\va}[0]{\vect{a}}
\newcommand{\vb}[0]{\vect{b}}
\newcommand{\vc}[0]{\vect{c}}

\newcommand{\ve}[0]{\vect{e}}
\newcommand{\vf}[0]{\vect{f}}

\newcommand{\vh}[0]{\vect{h}}

\newcommand{\vo}[0]{\vect{o}}

\newcommand{\vr}[0]{\vect{r}}

\newcommand{\vv}[0]{\vect{v}}
\newcommand{\vx}[0]{\vect{x}}

\newcommand{\vy}[0]{\vect{y}}

\newcommand{\mM}[0]{\matr{M}}

\newcommand{\mU}[0]{\matr{U}}
\newcommand{\mV}[0]{\matr{V}}
\newcommand{\mW}[0]{\matr{W}}

\DeclareMathAlphabet\mathbfcal{OMS}{cmsy}{b}{n}

\DeclareMathOperator{\TTL}{TTL}
\newcommand{\tGn}{\tG_{[n]}}
\newcommand{\tGno}{\tG_{[n+1]}}

\def\1{\bm{1}}
\newcommand{\bigO}{\mathcal{O}}

\def\va{\textbf{\textit{a}}}
\def\vb{\textbf{\textit{b}}}
\def\vc{\textbf{\textit{c}}}

\def\ve{\textbf{\textit{e}}}
\def\vf{\textbf{\textit{f}}}

\def\vh{\textbf{\textit{h}}}

\def\vo{\textbf{\textit{o}}}

\def\vr{\textbf{\textit{r}}}

\def\vu{\textbf{\textit{u}}}
\def\vv{\textbf{\textit{v}}}

\def\vx{\textbf{\textit{x}}}
\def\vy{\textbf{\textit{y}}}

\def\mM{\textbf{\textit{M}}}

\def\mS{\textbf{\textit{S}}}

\def\mU{\textbf{\textit{U}}}
\def\mV{\textbf{\textit{V}}}
\def\mW{\textbf{\textit{W}}}
\def\mX{\textbf{\textit{X}}}

\DeclareMathAlphabet{\mathsfit}{\encodingdefault}{\sfdefault}{m}{sl}
\SetMathAlphabet{\mathsfit}{bold}{\encodingdefault}{\sfdefault}{bx}{n}
\newcommand{\tens}[1]{\mathbfcal{#1}}

\def\tB{{\tens{B}}}

\def\tG{{\tens{G}}}

\def\tT{{\tens{T}}}

\def\tV{{\tens{V}}}

\def\tX{{\tens{X}}}
\def\tY{{\tens{Y}}}

\newcommand{\E}{\mathbb{E}}

\newcommand{\R}{\mathbb{R}}

\newcommand{\cut}[1]{}

\newcommand{\xhdr}[1]{{\noindent\bfseries #1}.}

\newcolumntype{L}{>{\arraybackslash}m{9cm}}

\usepackage{multirow}
\newcommand\MyBox[2]{
  \fbox{\lower0.75cm
    \vbox to 1.7cm{\vfil
      \hbox to 1.7cm{\hfil\parbox{1.4cm}{#1\\#2}\hfil}
      \vfil}%
  }%
}

\newcommand{\cem}[1]{#1} 

\theoremstyle{plain}

\theoremstyle{definition}

\usepackage[utf8]{inputenc}
\usepackage{csquotes}
\usepackage{showidx}
\makeindex

\usepackage[backend=bibtex]{biblatex}
\begin{document}


\pretitle{\begin{center}\cftchapterfont\huge}
\posttitle{\end{center}}
\preauthor{\begin{center}\huge}
\postauthor{\end{center}}
\predate{\begin{center}\large}
\postdate{\end{center}}



\begin{titlingpage}
    \centering
    
    {\LARGE \textbf{Making deep neural networks work for medical audio: representation, compression and domain adaptation}}\\[2cm]
    
    {\large A DISSERTATION}\\[0.5cm]
    {\large SUBMITTED TO THE SCHOOL OF COMPUTER SCIENCE}\\
    {\large McGILL UNIVERSITY}\\[2cm]
    
    BY\\[0.5cm]
    {\Large \textbf{Charles C. Onu}}\\[2cm]
    
    {\large IN PARTIAL FULFILLMENT OF THE REQUIREMENTS}\\
    {\large FOR THE DEGREE OF}\\
    {\large \textbf{Doctor of Philosophy}}\\[2cm]
    
    {\large Supervised by Prof. Doina Precup}\\[2cm]
    
    {\large December, 2024}
\end{titlingpage}

\pagenumbering{roman}


\clearpage
\section*{Acknowledgements}
I owe a profound debt of gratitude to the many individuals who have supported me throughout this PhD journey.

First and foremost, I extend my heartfelt thanks to my advisor, Prof Doina Precup. Her guidance has been the cornerstone of my academic and personal growth. She provided unwavering support—technically and personally—and her encouragement carried me through every challenge. Her mentorship has truly been a privilege. I also wish to thank Prof Yoshua Bengio and Prof Joelle Pineau, whose mentorship, support and advice were invaluable at critical moments on this journey.

I am deeply grateful to my parents, Patrick and Roseline Onu, whose love, sacrifices, and steadfast guidance have been the foundation of my life. To my brother, Dr Ikenna Onu, an unwavering supporter, and my other siblings, Ijeoma, Chidinma, and Uchenna, I am thankful for the love.

My clinical collaborators, Dr Uchenna Ekwochi, Dr Peter Ubuane, Dr Datonye Briggs, Dr Guilherme Sant’Anna, and Dr Robert Kearney, provided invaluable expertise and insights that enriched this research. I thank all patients-parents and infants-whose participation was essential to the success of this work.

Lastly, I am deeply grateful to the Ubenwa team, especially Innocent Udeogu and Samantha Latremouille who offered tireless commitment and collaboration. I thank all my friends and colleagues: Ifesinachi Nwoko, Kevin Ekwenwa, Munachiso Ilokah, Eyenimi Ndiomu, Sumana Basu, Upasana Dasgupta, Jon Lebensold and many others. Each of them has brightened this journey in unique ways.


\clearpage
\section*{abstract}
This thesis addresses the technical challenges of applying machine learning to understand and interpret medical audio signals. The sounds of our lungs, heart, and voice convey vital information about our health. Yet, in contemporary medicine, these sounds are primarily analyzed through auditory interpretation by experts using devices like stethoscopes. Automated analysis offers the potential to standardize the processing of medical sounds, enable screening in low-resource settings where physicians are scarce, and detect subtle patterns that may elude human perception, thereby facilitating early diagnosis and treatment.

Focusing on the analysis of infant cry sounds to predict medical conditions, this thesis contributes on four key fronts. First, in low-data settings, we demonstrate that large databases of adult speech can be harnessed through neural transfer learning to develop more accurate and robust models for infant cry analysis. Second, in cost-effective modeling, we introduce an end-to-end model compression approach for recurrent networks using tensor decomposition. Our method requires no post-hoc processing, achieves compression rates of several hundred-fold, and delivers accurate, portable models suitable for resource-constrained devices. Third, we propose novel domain adaptation techniques tailored for audio models and adapt existing methods from computer vision. These approaches address dataset bias and enhance generalization across domains while maintaining strong performance on the original data. Finally, to advance research in this domain, we release a unique, open-source dataset of infant cry sounds, developed in collaboration with clinicians worldwide.

This work lays the foundation for recognizing the infant cry as a vital sign and highlights the transformative potential of AI-driven audio monitoring in shaping the future of accessible and affordable healthcare.
\clearpage
\section*{Résumé}
Cette thèse aborde les défis techniques liés à l’application de l’apprentissage automatique pour comprendre et interpréter les signaux audio médicaux. Les sons de nos poumons, de notre cœur et de notre voix transmettent des informations essentielles sur notre santé. Pourtant, en médecine contemporaine, ces sons sont principalement analysés par des experts, à l’aide d’appareils comme les stéthoscopes. L’analyse automatisée offre la possibilité de standardiser le traitement des sons médicaux, de permettre le dépistage dans les régions à faibles ressources où les médecins sont rares et de détecter des motifs subtils qui pourraient échapper à la perception humaine, facilitant ainsi un diagnostic et un traitement précoces.

En nous concentrant sur l’analyse des pleurs de nourrissons pour prédire des conditions médicales, cette thèse apporte des contributions sur quatre axes principaux : Premièrement, dans des contextes de données limitées, nous démontrons que de grandes bases de données de voix d’adultes peuvent être exploitées via l’apprentissage par transfert pour développer des modèles plus précis et robustes pour l’analyse des pleurs de nourrissons. Deuxièmement, en matière de modélisation rentable, nous introduisons une approche de compression de modèles récurrents de bout en bout basée sur la décomposition tensorielle. Notre méthode, sans traitement postérieur, atteint des taux de compression de plusieurs centaines de fois et produit des modèles précis et portables, adaptés aux dispositifs contraints en ressources. Troisièmement, nous proposons des techniques d’adaptation de domaine conçues pour les modèles audio, et nous adaptons des méthodes existantes issues de la vision par ordinateur. Ces approches résolvent le problème des biais des ensembles de données et améliorent la généralisation dans de nouveaux domaines, tout en conservant de solides performances sur les données d'origine. Enfin, pour favoriser la recherche future dans ce domaine, nous publions un ensemble de données unique et open-source de pleurs de nourrissons, développé en collaboration avec des cliniciens du monde entier.

Ce travail pose les bases pour reconnaître le cri du nourrisson comme un signe vital et met en lumière le potentiel transformateur de la surveillance audio par IA dans l’avenir des soins de santé accessibles et abordables.

\newpage
\section*{Contribution to original knowledge}
\addcontentsline{toc}{chapter}{Contribution to original knowledge}
The thesis makes original contributions to enabling scalable and robust AI-driven medical audio analysis by addressing data scarcity, computational efficiency, domain generalization, and providing a foundational dataset for infant cry recognition:

\begin{enumerate}
    \item \textbf{Neural Transfer Learning for Infant Cry Analysis}: This thesis pioneers the application of neural transfer learning from adult speech datasets to analyze infant cries for medical purposes, such as diagnosing perinatal asphyxia. By utilizing representations learned from large, publicly available adult speech datasets, the work proposes pre-training methodologies (both supervised and self-supervised) that bridge the gap created by the limited availability of annotated infant cry data. The result is models that outperform traditional counterparts in both accuracy and robustness. This contribution validates that adult speech is an effective source domain for transfer learning in infant cry analysis.

    \item \textbf{End-to-end Model Compression with Tensor Decomposition}: The thesis introduces a method for compressing recurrent neural networks (RNNs) using tensor decomposition techniques. This approach reformulates matrix operations in neural networks into low-rank tensor representations, achieving compression rates of 50-300x without significant loss in performance. Unlike traditional post-hoc compression methods, this end-to-end trainable approach reduces computational complexity while maintaining the model’s ability to generalize. The compressed models are lightweight and suitable for deployment on resource-constrained hardware, facilitating real-world applications of medical audio analysis on devices like smartphones and low-power edge devices.

    \item \textbf{Addressing Dataset Bias and Domain Shift in Medical Audio}: The research addresses a critical challenge in medical audio analysis—dataset bias and domain shift caused by variations in audio recordings collected across different hospitals and geographical locations. By customizing domain adaptation techniques from computer vision to audio, the work develops methods to learn domain-invariant features while retaining task-specific discriminative power. The results demonstrate significant improvements in cross-domain generalization, ensuring reliable performance across diverse datasets and clinical environments.

    \item \textbf{CryCeleb: A Dataset for Infant Cry Recognition}: To address the scarcity of well-annotated infant cry datasets, the thesis presents a first-of-its-kind dataset, CryCeleb, a large and diverse dataset comprising over six hours of recordings from 786 infants across multiple clinical settings. This dataset enables research in speaker verification and infant recognition, with applications in neonatal care. Accompanying the dataset, the CryCeleb 2023 competition encouraged global participation to advance cry-based biometric recognition techniques. The work highlights the challenges of cry recognition, such as high intra-class variability, and provides a foundation for future exploration of infant cry sounds.
    \end{enumerate}

\newpage
\section*{Contribution of authors}
\addcontentsline{toc}{chapter}{Contribution of authors}
This thesis includes work conducted in collaboration with co-authors, and published in different scientific conferences and journals. Below is a chapter-by-chapter breakdown of contributions:

\begin{itemize}
    \item \textbf{Chapter 1} was written by me specifically for this dissertation, and reviewed by Doina Precup. 

    \item \textbf{Chapter 2} was written by me specifically for this dissertation, and reviewed by Doina Precup.
    
    \item \textbf{Chapter 3} is based on the paper "Neural transfer learning from adult speech to infant cry", published in INTERSPEECH'2019 \cite{onu2019neural}. I conceived the core idea behind this work, designed the scientific methodology, led the experimentation, led the analysis and interpretation of results and wrote the first draft of the manuscript. Jonathan Lebensold contributed to the writing and running of experiments. All authors revised the manuscript.

    \item \textbf{Chapter 4} is based on the paper "A Fully Tensorized Recurrent Neural Network", published as a pre-print \cite{onu2020fullytensorized}. I proposed the core idea of this work, developed the theoretical framework, implemented the new tensorized neural network, wrote the training code, conducted the experiments, contributed to the analysis of results and wrote the first draft of the manuscript. Jacob Miller contributed to the design of the theoretical framework and analysis of the results. All authors revised the manuscript.

    \item \textbf{Chapter 5} is based on the paper "Self-supervised learning for infant cry analysis" published at the International Conference on Acoustics, Speech, and Signal Processing (ICASSP) 2024 \cite{gorin2023self} and "A cry for help: Early detection of brain injury in newborns" currently under journal review \cite{onu2023help}. I conceived the core idea of this work, led the design and implementation of the international clinical data acquisition study, led the design and development of the data acquisition software, contributed to the design of the machine learning methodology, and contributed to the experiments and analysis of results. I wrote the first draft of the manuscript. Samantha Latremouille contributed to the clinical study design and implementation. Innocent Udeogu contributed to the development of the software. Arsenii Gorin, Junhao Wang, Sajjad Abdoli and Cem Subakan contributed to the experiments and interpretation of results. All authors reviewed the manuscript.

    \item \textbf{Chapter 6} is based on the paper "Learning domain-invariant classifiers for infant cry sounds" published as a pre-print \cite{onu2023learningdomaininvariantclassifiersinfant}. I conceived the idea behind this work, led the design of the methodology, led the analysis and interpretation of results and wrote the first draft of the manuscript. Hemanth Sheetha conducted the experiments. All authors contributed to the analysis of the results and revision of the manuscript.

    \item \textbf{Chapter 7} is based on the paper "CryCeleb: A Speaker Verification Dataset Based on Infant Cry Sounds", published in the International Conference on Acoustics, Speech, and Signal Processing (ICASSP) 2024 \cite{budaghyan2024crycelebspeakerverificationdataset}. I conceived the main idea of this work, contributed to the design of the methodology and the analysis of results. David Budaghyan and Arsenii Gorin ran the experiments and conducted the competition. All authors contributed to the manuscript.

    \item \textbf{Chapter 8} was written by me specifically for this dissertation, and reviewed by Doina Precup.
\end{itemize}

\newpage
\section*{Contributions not included in the thesis}
\addcontentsline{toc}{chapter}{Contributions not included in the thesis}
The dissertation author made other research contributions not included in this thesis. Below are a list of these:

\begin{itemize}
    \item "Automated prediction of extubation success in extremely preterm infants: the APEX multicenter study" published in Nature Pediatric Research, 2023 \cite{kanbar2023automated}. 
    \item "Characteristics of newborns with hypoxic ischemic encephalopathy treated in NICUs at three different income-level countries" published at the Congress of joint European Neonatal Societies, 2023 \cite{latremouille2023characteristics}.
    \item "Mapping computer science research in Africa: using academic networking sites for assessing research activity" published Springer Scientometrics Journal, 2021 \cite{harsh2021mapping}.
    \item "Undersampling and Bagging of Decision Trees in the Analysis of Cardiorespiratory Behavior for the Prediction of Extubation Readiness in Extremely Preterm Infants" published 40th Annual International Conference of the IEEE Engineering in Medicine and Biology Society (EMBC), 2018 \cite{kanbar2018undersampling}.
\end{itemize}

\clearpage
\setcounter{tocdepth}{2}
\tableofcontents

\newpage
\chapter*{List of Acronyms}
\addcontentsline{toc}{chapter}{List of Acronyms}
\begin{itemize}
  \item \textbf{RNN}: Recurrent Neural Network
  \item \textbf{LSTM}: Long Short-Term Memory
  \item \textbf{GRU}: Gated Recurrent Unit
  \item \textbf{MFCC}: Mel-Frequency Cepstral Coefficients
  \item \textbf{SVM}: Support Vector Machine
  \item \textbf{CNN}: Convolutional Neural Network
  \item \textbf{DNN}: Deep Neural Network
  \item \textbf{TDNN}: Time Delay Neural Network
  \item \textbf{CRNN}: Convolutional Recurrent Neural Network
  \item \textbf{ECAPA-TDNN}: Emphasized Channel Attention, Propagation, and Aggregation Time Delay Neural Network
  \item \textbf{GAN}: Generative Adversarial Network
  \item \textbf{SSL}: Self-Supervised Learning
  \item \textbf{ANN}: Artificial Neural Network
  \item \textbf{AAM}: Additive Angular Margin
  \item \textbf{HAFN}: High Adaptation Feature Normalization
  \item \textbf{SAFN}: Standard Adaptation Feature Normalization
  \item \textbf{IoT}: Internet of Things
  \item \textbf{ASR}: Automatic Speech Recognition
  \item \textbf{SHAP}: SHapley Additive exPlanations
  \item \textbf{LIME}: Local Interpretable Model-Agnostic Explanations
\end{itemize}

\newpage
\listoffigures
\newpage
\listoftables

\clearpage
\pagenumbering{arabic}
\chapter{Introduction}
\section{Motivation}
The analysis and interpretation of sound is critical in medicine. One of the earliest medical devices to be invented was the stethoscope, a device for listening to cardiothoracic and abdominal sounds. To this day, it plays a central role in routine health care, an indication that the sound of our lungs, heart and voice tells a deep story about our health. 

In this work, we are interested in 3 main problems that arise when applying machine learning to analyze and understand medical audio data. (1) First challenge is that of the low data regime. Audio is not typically stored as part of a patient's medical records, even in cases where it played a factor in an eventual diagnosis. So having sufficient data to build ML models is a challenge. (2) The second challenge we're interested in is the question of efficient models. It is extremely useful to have models that are not only accurate but can also run fast on relatively inexpensive hardware. Thus, the question of model compression becomes important. (3) The final issue we study is dataset bias and domain shift. Medical audio data collected and annotated for the same task but in different "domains" may be distributed differently due to many factors. This discrepancy in the distributions results to models trained in one domain transferring poorly to another. We investigate approaches to mitigate the impact of dataset bias in medical audio models.

We study this in the context of models for infant cry analysis. Since the 1960s, neonatal clinicians have known that newborns suffering from certain neurological conditions exhibit altered crying patterns such as the high-pitched cry in birth asphyxia~\cite{Michelsson1971, partanen1967auditory}. Despite an annual burden of over 1.5 million infant deaths and disabilities~\cite{lawn20054}, early detection of neonatal brain injuries due to asphyxia remains a challenge, particularly in developing countries where the majority of births are not attended by a trained physician~\cite{montagu2011poor}.

Cry-based neurological monitoring opens the door for low-cost, easy-to-use, non-invasive and contact-free screening of at-risk babies, especially when integrated into simple devices like smartphones or neonatal ICU monitors. This would provide a reliable tool where there are no alternatives, but also curtail the need to regularly exert newborns to physically-exhausting or radiation-exposing assessments such as brain CT scans. Previous research, however, has been limited by the lack of a large, high-quality clinical database of cry recordings, constraining the application of state-of-the-art machine learning \cite{Reyes-Galaviz2004, Reyes-Galaviz2008, Manfredi2006AAnalysis}. 

In this thesis, we aim to tackle key technical challenges on the path to enabling AI-driven sound monitoring as one key step to a future of more affordable healthcare.
\\
\\
\section{Summary of research contributions}
\subsection{Neural transfer learning from adult speech to infant cry} We explore neural transfer learning in developing rich representations for the task of classifying perinatal asphyxia from infant cry sounds. As with most medical problems, annotated data is limited. We used the only known publicly available (but relatively small) dataset, the Chillanto Infant Cry Database \cite{Reyes-Galaviz2004} containing 1,389 1-sec recordings from 69 infants. By studying the usefulness of representations from several audio datasets and training schema, we successfully develop a model that surpasses the de facto support vector machine baseline, despite the minimal amount of data available for training the neural net.

This draws from breakthroughs in computer vision. Models pre-trained on large image corpora have been found to achieve representations in intermediate layers of the network that extract semantically salient aspects of the input image such as corners, edges, colors, as well as more complex elements such as mesh patterns \cite{zeiler2014visualizing}. When solving a new task, the use of a pre-trained model's parameters to initialize a network before finetuning has led to state-of-the-art performance in many (sometimes unrelated) tasks \cite{donahue2014decaf, sharif2014cnn}. In medical imaging, transfer learning models consistently outperform models trained on analytical (or "hand-engineered") features \cite{morid2021scoping}. 

Motivated by this, we investigate the usefulness of pre-trained neural representations of speech for classifying perinatal asphyxia. Though adult speech and infant crying are different kinds of audio signals, the hypothesis is that intermediate model representations of the former could serve as a good initialization for accurate cry analysis models. For pre-training we employ 5 speech databases: Voice Cloning Toolkit (VCTK) \cite{veaux2017}, Speakers in the Wild (SITW)  \cite{mitchell2016}, Speech Commands  \cite{warden2018}, AudioSet \cite{gemmeke2017}, and ESC-50 \cite{piczak2015}. We find that the classical baseline of a support vector machine (SVM) was difficult to surpass. One of the 3 speech-based neural transfer models outperformed its accuracy, suggesting that the small amount of data available is on the margins of utility of this approach. Nevertheless, during robustness evaluations, we observe that the neural transfer models exhibit more stable performance over ablations in both time and frequency domains, indicating one reason to choose neural models over SVM in a practical application. Furthermore, an analysis of the variance explained by the principal components of each model's features further shows that neural transfer models learn a richer and more non-linear relationship between input and prediction as compared to their counterparts learned from random initialization. This work validates the hypothesis that features extracted from adult speech are also useful for infant cry models, and presents a framework for doing this successfully. Some of the results in this work is presented in the paper:

\begin{itemize}
    \item Onu, C. C., Lebensold, J., Hamilton, W. L., \& Precup, D. (2019). Neural transfer learning for cry-based diagnosis of perinatal asphyxia. arXiv preprint arXiv:1906.10199. \cite{onu2020neural}.
\end{itemize}

\subsection{End-to-end model compression with tensor decomposition} Here we present an end-to-end trainable approach for model compression derived from methods of tensor decomposition. The compression of deep neural networks (DNNs) has been of interest for a long time. It has been shown that DNNs are typically parameterized in a redundant fashion, allowing the prediction of values of some parameters of a trained model given knowledge of the others~\cite{denil2013}. Model distillation~\cite{ba2014,hinton2015}, pruning \cite{li2016pruning, han2015deep} and quantization \cite{zhang2018lq, rastegari2016xnor} are common approaches. These methods are post-hoc or can be quite involving. And sometimes, the process of compressing the DNN is separate from training. The may require retraining to mitigate loss of already learned information.

In this work, we develop an approach for training high capacity models in an end-to-end fashion with only a small number of parameters. We achieve this by reformulating all matrix operations in a neural network into a low-rank tensorized format. This allowed significant compression to be achieved not only with high-dimensional inputs, but also with high-dimensional hidden states. We apply this to 2 variants of recurrent neural networks (RNNs), resulting to 50-300x reduction in number of trainable model parameters at only a minimal impact on model performance. Our work builds upon ~\cite{tjandra2017} to achieve further compression by jointly tensorizing the weights within each RNN cell. We show how this process leads to a novel form of weight sharing, which is verified experimentally to have tangible benefits for performance and compression. We further illustrate that the use of a tensor-train parameterization represents an implicit regularization capable of improving training and generalization. Our tensorized RNN implementation is available as open-source code, providing modules that can be used as a drop-in replacement for standard RNN models or as building blocks for other DNNs\footnote{https://github.com/onucharles/tensorized-rnn}. We apply this to the task of speaker verification an important first step in a cry analysis system. This work is presented in the paper:

\begin{itemize}
    \item Onu, C. C., Miller, J. E., \& Precup, D. (2020). A fully tensorized recurrent neural network. arXiv preprint arXiv:2010.04196. \cite{onu2020fullytensorized}.
\end{itemize}

\subsection{Addressing dataset bias and domain shift in medical audio} Here, we address the challenge of domain shift and dataset bias in a novel medical audio database of infant cry sounds. In real-world datasets, it is rarely the case that distributions of source and target data are the same \cite{candela2009dataset}, typically resulting in degradation of model performance in the target domain. Multiple prior work have developed neuro injury detection models from cry sounds using neural transfer learning\cite{onu2020neural} and self-supervised learning\cite{gorin2023selfsupervised}. Although these methods show effectiveness on in-domain test sets, they fail to generalize as well to new hospital data. 

In this work, we study the nature of domain shift in a clinical database of infant cry sounds acquired across different geographies and quantify its impact. Then, we develop and test effective methods for mitigating the underlying bias in this dataset. Precisely, we adapt unsupervised domain adaptation methods from computer vision to learn an audio representation that is domain-invariant to hospitals, yet task-discriminative. By experimenting with 5 different methods we illustrate that the best methods not only improve target accuracy but also accuracy in the source domain. Secondly, we validate previous clinical findings about the newborn cry as a universal language -- the pitch of baby cries is similarly distributed regardless of geography. We propose a relatively simple and promising approach for DA in infant cry data, target noise injection (TNI), for unsupervised domain adaptation which requires neither labels nor training data from the target domain. Our method requires no architectural changes nor complex, min-max optimization, employs a simple cross-entropy loss function, and requires neither labels nor cry recordings from the target domain -- only target noise samples \cite{NANNI2020101084}.

We apply this technique to the Ubenwa infant cry database, a relatively large database of infant cry sounds  acquired in collaboration with Ubenwa AI from geographically diverse settings – 5 hospitals across 3 continents. We develop a training methodology for pathology detection in this audio database. Our system extracts interpretable acoustic biomarkers that support clinical decisions and is able to accurately detect neurological injury from newborns’ cries with a high AUC of 92.5\% (88.7\% sensitivity at 80\% specificity). Furthermore, when developing per-hospital models, our best-performing domain-adapted model significantly improves target accuracy by 7.2\%, without negatively affecting the source domain.

This work is presented in the papers:

\begin{itemize}
    \item Onu, C. C., Sheetha, H. K., Gorin, A., Precup, D. (2023). Learning domain-invariant classifiers for infant cry sounds. \cite{onu2023learningdomaininvariantclassifiersinfant}
    \item Onu, C. C., Latremouille, S. Gorin, A., Wang, J., Ekwochi, U., Ubuane, P. O., Kehinde, O. A., Salisu, M. A., Briggs, D., Bengio, Y., Precup, D. (2023). A cry for help: Early detection of brain injury in newborns. arXiv preprint arXiv:2310.08338, \cite{onu2023help}
\end{itemize}

\subsection{A public dataset for cry-based infant recognition}
In our final contribution, we present CryCeleb \cite{ubenwa_2023}, a novel dataset comprising over 6 hours of infant cry recordings from 786 newborns, designed for speaker verification and infant identification. The dataset is a response to the scarcity of real-world, well-annotated infant cry data available for training robust models. The dataset is a result of nearly 3 years of clinical studies in collaboration with clinicians across 3 continents. The recordings are labeled with anonymized infant IDs, and the dataset includes multiple recordings for some infants at different time points -- immediately after birth and just before hospital discharge. This dataset was introduced alongside the CryCeleb 2023 competition, which challenged participants to develop models that can determine if two cry recordings belong to the same infant.

In the CryCeleb 2023 challenge, we used an ECAPA-TDNN model, pre-trained on adult speech, as a baseline for cry verification. Fine-tuning this model on the CryCeleb dataset led to a reduction in the Equal Error Rate (EER) from around 38\% to approximately 28\%. The competition attracted significant interest, with 435 submissions across 37 countries. Contributions improved upon the baseline using a variety of techniques including test-time data augmentation, parameter and label smoothing, employing triplet loss over softmax. The resulting models however still lagged behind state-of-the-art adult speaker verification systems. This highlighted the difficulty of recognizing infants from cries and the need for continued research to enhance the performance of infant cry verification and to better understand the characteristics of cry signals.
\begin{itemize}
    \item Budaghyan, D., Onu, C. C., Gorin, A., Subakan, C., Precup, D. (2023). CryCeleb: A Speaker Verification Dataset Based on Infant Cry Sounds. 
    \cite{budaghyan2024crycelebspeakerverificationdataset}
\end{itemize}

\section{Organization}
We organize our discussion according to the different contributions that constitute this thesis. Chapter \ref{chap:background}, "Background", provides an overview of key concepts underlying this research. Chapter \ref{chap:contribution1}: "Building Rich Audio Representations for Improved Downstream Accuracy and Robustness" presents the development of transfer learning techniques to create robust audio representations from adult speech datasets to improve the analysis of infant cry sounds. Chapter \ref{chap:contribution2}: "End-to-End Compression for Efficient Modeling" explores tensor decomposition methods for compressing recurrent neural networks, allowing efficient and portable model deployment. Chapter \ref{chap:ssl-infantcry}: "A Clinical Study for the Early Detection of Birth Asphyxia Using Infant Cry Sounds" details a multi-center clinical study and investigates the application of machine learning for the early detection of neurological conditions in newborns. Chapter \ref{chap:cryceleb}: "CryCeleb - A Dataset for Cry-Based Infant Recognition" introduces and describes the CryCeleb dataset for cry-based infant recognition tasks. Chapter \ref{chap:contribution3}: "Understanding and Addressing Dataset Bias and Domain Shift" develops domain adaptation methods to address bias and improve model generalization across diverse datasets. Finally, in Chapter \ref{chap:conclusion}, "Conclusion", we provide a synopsis of the findings of this research, and a discussion on limitations and avenues for future work.

\chapter{Background}
\label{chap:background}
This dissertation is concerned with the effective application of deep neural networks to analyze sound in medicine. This chapter introduces fundamental concepts in both audio signal processing and deep learning as they relate to the research presented in later chapters. Our goal is to provide a concise overview of these topics, supplemented with references to primary sources for further exploration.

\section{Sound, speech and cries}

The analysis of medical audio, relies on a foundational understanding of sound. This section provides essential background knowledge for analyzing sound as well as its extension to speech and infant cries -- as it applies to our research.

\subsection{Fundamentals of sound}
\subsubsection{What is sound?}
Sound is a mechanical wave that results from the vibration of particles in a medium, typically air \cite{feynman1963feynman}. These vibrations create pressure waves that propagate through space and can be perceived by the human auditory system. The fundamental mathematical representation of a sinusoidal sound wave is given by:
\begin{equation}
    s(t) = A \cos(2\pi f t + \phi)
\end{equation}

\subsubsection{Physical properties of sound}
Sound is characterized by several fundamental properties \cite{rossing2002springer}:

\begin{itemize}
    \item \textbf{Frequency} ($f$): The number of oscillations per second, measured in Hertz (Hz). It determines the perceived pitch of the sound.
    \item \textbf{Amplitude} ($A$): The magnitude of pressure variation in the sound wave. Higher amplitudes correspond to louder sounds.
    \item \textbf{Wavelength} ($\lambda$): The distance between successive pressure peaks in a wave, related to frequency by the equation:
    \begin{equation}
        c = f\lambda
    \end{equation}
    where $c$ is the speed of sound in the medium.
    \item \textbf{Timbre}: The quality of sound that differentiates sources with the same pitch and loudness, influenced by harmonics and spectral content.
    \item \textbf{Duration}: The time span of a sound event, which affects its perception and recognition.
\end{itemize}

\subsubsection{Representation of sound signals}
Sound signals are commonly represented in different formats depending on the task at hand.

\begin{itemize}
    \item \textbf{Time-domain waveforms}: A plot of amplitude versus time.
    \item \textbf{Frequency-domain representations}: A breakdown of a signal into its component frequencies using the Fourier Transform (FT):
    \begin{equation}
        X(f) = \int x(t) e^{-j2\pi ft} dt
    \end{equation}
    where $X(f)$ is the frequency representation of $x(t)$.
    \item \textbf{Spectrograms}: A time-frequency representation showing how different frequencies evolve over time, often used in speech and cry analysis.
\end{itemize}

We will expand on frequency and time-frequency representations in section \ref{sec:background-asp} below.

\subsubsection{Human perception and its impact on sound analysis}
The human auditory system does not process sound linearly. The field of psychoacoustics studies how humans perceive loudness, pitch, and timbre \cite{moore2012introduction} developing certain key principles. These principles not only inform our understanding of human perception, but also influence how we process sound signals digitally.

\begin{enumerate}
    \item \textbf{Frequency masking}: The human ear perceives some frequencies more strongly than others, and certain sounds can "mask" others if they are close in frequency\cite{moore2012introduction}. 
    
    \item \textbf{Nonlinear loudness perception}: The perceived loudness of a sound depends not just on its amplitude but also on its frequency. This is why the human ear is more sensitive to mid-range frequencies (around 1–5 kHz) than to very low or high frequencies \cite{fletcher1933loudness}.
    
    \item \textbf{Critical bands}: The auditory system processes sound in frequency bands, called critical bands, rather than analyzing individual frequencies separately. It is within these bands where masking and perception effects occur \cite{fletcher1940auditory}. 
\end{enumerate}

\subsection{Speech and its characteristics}

\subsubsection{What is speech?}
Speech is the vocalized form of human communication, produced by the coordinated movement of the lungs, vocal folds, and articulators (tongue, lips, and jaw) \cite{titze1994principles}. Speech production involves three main processes:

\begin{enumerate}
    \item \textbf{Respiration}: Airflow is generated in the lungs and travels up the vocal tract.
    \item \textbf{Phonation}: The vibration of the vocal cords generates fundamental frequencies.
    \item \textbf{Articulation}: The travelling sound waves are ultimately shaped by the tongue, lips, into speech.
\end{enumerate}


\subsubsection{Linguistic and paralinguistic features of speech}

Speech generated by the process above is a unique type of sound as it contains both verbal (linguistic) and non-verbal (paralinguistic) information \cite{schuller2013paralinguistics}. Linguistic features of speech refer to structured language elements like phonetics, phonology, syntax and semantics; while the paralinguistic features refer to more expressive information such as intonation, rhythm, emotion, speaker identity, etc.

In a sense, the broad aim of the field of automatic speech recognition (ASR) is to uncover these underlying features of given speech using automated algorithms.

\subsection{Cries and their unique features}

A cry is a vocalization produced by infants, often as a reflexive or emotional response to distress, discomfort, or physiological needs. Cry analysis has historically focused on certain acoustic and temporal features of cries including fundamental frequency ($F_0$) variations, duration and rhythmic structure, and spectral features \cite{truby1965cry, Golub1985ACry}. We extend these methods to build more robust models powered by deep neural networks in the context of our first contribution in chapter \ref{sec:cry-representation}.

\subsection{Medical sounds}
Sounds abound in medicine and provide clinical insights in most cases. Speech patterns could be an indicator of different disorders and neurodenerative conditions such as dysarthria \cite{darley1969differential}, alzheimer's disease \cite{appell1982alzheimer}, parkinson's disease \cite{illman1981parkinson}, and multiple sclerosis \cite{darley1972multiple}. Crying patterns in infants has been linked to pain and other congenital and acquired disorders such as perinatal asphyxia \cite{Michelsson1971CryInfants}, hyperbilirubinemia \cite{wasz1985twenty}, and meningitis \cite{Michelsson1977}. Other physiological sounds such as heart sounds (murmurs, arrhythmias), lung sounds (wheezing, crackles) are regularly used as diagnostic inputs in the day-to-day practice of medicine.

\section{Processing audio signals}
\label{sec:background-asp}
Audio signals contain complex information spread across different frequency components. Unlike static data, such as images or text, sound must be analyzed both in the time domain and the frequency domain to extract meaningful insights. For example, two musical notes may have similar amplitudes but different frequencies. In frequency analysis, we analyze how energy is distributed across frequencies. This is done using Fourier transform-based methods.

\subsection{Fourier transforms}

\textbf{Fourier Transform (FT)}. The Fourier Transform (FT) \cite{fourier1822theorie} decomposes a signal into its sinusoidal frequency components. For a continuous-time signal $x(t)$, the FT is defined as:

\begin{equation}
    X(f) = \int_{-\infty}^{\infty} x(t) e^{-j2\pi ft} dt
\end{equation}

where: $X(f)$ represents the frequency content at frequency $f$, $x(t)$ is the original time-domain signal and $j$ is the imaginary unit.
\\
\textbf{Discrete Fourier Transform (DFT)}. Since real-world signals are stored and processed digitally, in practice, we use the Discrete Fourier Transform (DFT) \cite{oppenheim1999discrete} for numerical computation:

\begin{equation}
    X[k] = \sum_{n=0}^{N-1} x[n] e^{-j 2\pi k n / N}
\end{equation}

where: $x[n]$ is the discrete signal of length $N$, $X[k]$ is the frequency component at index $k$.
\\
\textbf{Fast Fourier Transform (FFT)}. Computing the DFT directly is computationally expensive ($O(N^2)$ operations) \cite{oppenheim1999discrete}. The Fast Fourier Transform (FFT) \cite{cooley1965algorithm} is an efficient algorithm that reduces the complexity to $O(N \log N)$, making it practical for real-time use. Despite its efficiency, FFT alone is insufficient for analyzing signals with time-varying frequency content, which leads us to time-frequency representations such as spectrograms.

\subsection{Spectrograms and time-frequency representations}

A spectrogram is a representation of how frequency components evolve over time \cite{flanagan1972speech}. It is computed by applying the Short-Time Fourier Transform (STFT), which divides the signal into overlapping time segments and computes the Fourier Transform for each segment:

\begin{equation}
    STFT(x(t)) = \sum_{n} x(n) w(n - t) e^{-j2\pi fn}
\end{equation}

where $w(n)$ is a window function such as Hann or Hamming windows \cite{harris1978use} which ensures smooth transitions between segments.

Spectrograms capture both temporal and spectral characteristics and are useful for visualizing sound.

\subsection{Log-Mel spectrogram and Mel-frequency cepstral coefficients (MFCCs)}

\paragraph{Log-Mel Spectrogram} The Mel scale \cite{stevens1937scale} is a perceptually motivated frequency scale that mimics human auditory perception. Humans perceive differences in lower frequencies more sharply than in higher frequencies, so we apply a non-linear transformation to the frequency axis. To compute a log-Mel spectrogram:
\begin{enumerate}
    \item Compute the spectrogram using STFT.
    \item Apply a Mel filterbank, a series of overlapping triangular filters spaced according to the Mel scale:
    \begin{equation}
        m = 2595 \log_{10} \left(1 + \frac{f}{700} \right)
    \end{equation}
    \item Convert power values to log scale to approximate human loudness perception.
\end{enumerate}

\paragraph{Mel-Frequency Cepstral Coefficients (MFCCs)} While spectrograms and log-Mel spectrograms provide detailed frequency content, they still contain redundant information. MFCCs extract compact features by applying a Discrete Cosine Transform (DCT) to log-Mel features \cite{Davis1980ComparisonSentences}:
\begin{equation}
    c_n = \sum_{m=0}^{M-1} S_m \cos \left[ n(m - 0.5) \frac{\pi}{M} \right]
\end{equation}

where: $S_m$ is the log-Mel spectrogram value at filter $m$, $c_n$ is the $n$th MFCC coefficient and $M$ is the number of Mel filters.

\subsection{Support vector machines in audio signal processing}

Support vector machines (SVMs) \cite{Cortes1995Support-VectorNetworks} have historically been a popular choice for analyzing audio features, particularly before the rise of deep learning-based models. SVMs have been widely used in tasks such as speech recognition \cite{ganapathiraju1998support}, emotion detection\cite{dumas2001emotional}, music genre classification\cite{tzanetakis2002musical}, and others. Their effectiveness stems from their ability to separate high-dimensional feature representations extracted from audio, such as Mel-frequency cepstral coefficients (MFCCs).

\subsubsection{Support vector machines}

A support vector machine is a supervised learning algorithm used for classification and regression tasks. The key idea behind an SVM is to find a hyperplane that best separates data points belonging to different classes. The hyperplane is chosen to maximize the \textit{margin}, which is the distance between the nearest data points (support vectors) of each class \cite{Cortes1995Support-VectorNetworks}.

Mathematically, given a dataset with labeled instances $(x_i, y_i)$, where $x_i \in \mathbb{R}^n$ are feature vectors and $y_i \in \{-1, 1\}$ are class labels, an SVM seeks to find a decision boundary of the form:

\begin{equation}
    f(x) = w^T x + b = 0
\end{equation}

where $w$ is the weight vector and $b$ is the bias. The optimal hyperplane is determined by solving the following optimization problem:

\begin{equation}
    \min_{w, b} \frac{1}{2} ||w||^2
\end{equation}

subject to the constraints:

\begin{equation}
    y_i (w^T x_i + b) \geq 1, \quad \forall i
\end{equation}

This ensures that all data points are correctly classified with the maximum possible margin.

\subsubsection{Kernels and why we need them}

For many real-world problems, data is not linearly separable in its original feature space. To address this, SVMs use a technique called the kernel trick, which maps input features into a higher-dimensional space where they may become linearly separable \cite{boser1992training}.

A kernel function $K(x_i, x_j)$ computes the inner product in this higher-dimensional space, without explicitly computing the transformation, making SVMs computationally efficient.

\begin{equation}
    K(x_i, x_j) = \phi(x_i) \cdot \phi(x_j)
\end{equation}

where $\phi(x)$ is the mapping function to the high-dimensional space.

\subsubsection{Types of Kernels}

Different kernel functions are used depending on the data structure and application:

\begin{itemize}
    \item \textbf{Linear Kernel}: Used when the data is approximately linearly separable.
    \begin{equation}
        K(x_i, x_j) = x_i^T x_j
    \end{equation}
    
    \item \textbf{Polynomial Kernel}: Useful for capturing polynomial relationships in the data.
    \begin{equation}
        K(x_i, x_j) = (x_i^T x_j + c)^d
    \end{equation}
    where $d$ is the degree of the polynomial and $c$ is a constant.
    
    \item \textbf{Radial Basis Function (RBF) Kernel}: The most widely used kernel, effective for non-linear data.
    \begin{equation}
        K(x_i, x_j) = \exp\left(-\gamma ||x_i - x_j||^2\right)
    \end{equation}
    where $\gamma$ controls the width of the Gaussian function.
\end{itemize}

Although deep learning methods have largely replaced SVMs in modern audio processing tasks, SVMs remain competitive in cases with limited data such as in medical audio analysis, as we will see in chapter \ref{chap:contribution1}.

\section{Deep neural networks}

A deep neural network (DNN) is a computational model inspired by the structure and function of the human brain \cite{mcculloch1943logical, rosenblatt1958perceptron, hopfield1982neural}. It consists of layers of interconnected nodes (neurons) that process data by applying weighted connections and activation functions. Neural networks are the foundation of most of the advancements in artificial intelligence over the last decade \cite{khan2022comprehensive}.

\subsection{Feedforward neural networks}

A feedforward neural network (FFNN) is the simplest type of DNN, where information moves in one direction—from the input layer through hidden layers to the output layer—without cycles or loops. Each neuron in a layer takes a weighted sum of inputs from the previous layer and applies an activation function to introduce non-linearity. Mathematically, a neuron in layer $ l $ is computed as:

\begin{equation}
    h^l = f(W^l h^{l-1} + b^l)
\end{equation}

where:
 $ h^l $ is the activation of the current layer,
 $ W^l $ is the weight matrix,
 $ b^l $ is the bias term,
 $ f(\cdot) $ is the activation function (such as ReLU \cite{glorot2011deep}, Sigmoid \cite{cybenko1989approximation}, Tanh \cite{lecun1998gradient}).

FFNNs are effective in processing structured numerical features but struggle with sequential and spatially structured data, making them suboptimal for raw audio analysis. Instead, specialized architectures like convolutional neural networks and recurrent neural networks are used for audio processing.

\subsection{Backpropagation}

Backpropagation is the fundamental algorithm used for training neural networks by adjusting weights to minimize error. It works by propagating the gradient of the loss function backward through the network using the chain rule of differentiation\cite{rumelhart1986learning}. The key steps of backpropagation include:

\begin{itemize}
    \item Forward pass: Compute the output of the network by passing input through each layer.
    \item Compute loss: Measure the difference between the predicted output and the actual target.
    \item Backward pass: Compute gradients of the loss with respect to network parameters using the chain rule.
    \item Weight updates: Adjust weights using gradient descent or other optimization techniques.
\end{itemize}

Backpropagation enables deep neural networks to learn from large datasets by iteratively refining weights to improve accuracy.

\subsection{Convolutional neural networks}

A convolutional neural network (CNN) is a type of neural network designed for spatial feature extraction \cite{fukushima1980neocognitron}. It uses convolutional layers that apply filters to local regions of input data, preserving spatial relationships and capturing patterns such as edges in images \cite{lecun1989backpropagation}.

Mathematically, a convolution operation is given by:

\begin{equation}
    y(i, j) = \sum_m \sum_n x(i-m, j-n) w(m, n)
\end{equation}

where:
 $ x(i, j) $ is the input spectrogram,
 $ w(m, n) $ is the convolutional filter,
 $ y(i, j) $ is the feature map output.

While CNNs are widely used in image processing, they also work well for audio signals, particularly spectrogram-based representations. Since a spectrogram is a 2D image-like representation of sound, CNNs can detect frequency-time patterns similar to how they recognize shapes in images. CNNs are commonly used across a wide-range of sound classification tasks including speech recognition, speaker identification, music genre classification , cry classification and more.

\subsection{Recurrent neural networks}

A recurrent neural network (RNN) is a neural network designed for sequential data, where the current output depends on previous inputs \cite{elman1990finding, williams1989learning}. Unlike feedforward networks, RNNs have loops (recurrent connections), allowing them to retain memory of past information.

Mathematically, an RNN updates its hidden state at time $ t $ as:

\begin{equation}
    h_t = f(W_h h_{t-1} + W_x x_t + b)
\end{equation}

where:
 $ h_t $ is the hidden state at time $ t $,
 $ x_t $ is the input at time $ t $,
 $ W_h $ and $ W_x $ are weight matrices,
 $ b $ is the bias term.

Audio signals are temporal in nature making RNNs particularly suited for them. RNNs take a sequence of audio feature vectors like MFCCs as input. Each time step processes a frame of audio, and the hidden state stores information from previous frames. This allows the network to better capture dependencies across time.
\\
\paragraph{Advanced RNN architectures}
The standard RNN architecture has difficulties in capturing long-rang dependencies due to the vanishing gradient problem -- a situation in which gradients shrink exponentially during backpropagation, causing earlier layers to receive near-zero updates. To address this, more advanced RNN architectures have been proposed, such as: Long short-term memory (LSTM) \cite{Hochreiter1997LongMemory} and Gated recurrent units (GRUs) \cite{cho2014}, which we adapt to create compressed models in chapter \ref{sec:rnn-architectures}.




\section{Tensors and tensor decomposition}

\subsection{Tensors}

A tensor is a generalization of matrices to higher dimensions. While a matrix is a two-dimensional array of numbers (rows and columns), a tensor extends this concept to multiple dimensions (also called "modes" or "axes"). Scalars and vectors can also be seen as a subset of tensors:

\begin{itemize}
    \item A scalar is a 0-dimensional tensor: $ x \in \mathbb{R} $.
    \item A vector is a 1-dimensional tensor: $ \mathbf{x} \in \mathbb{R}^n $.
    \item A matrix is a 2-dimensional tensor: $ \mathbf{X} \in \mathbb{R}^{m \times n} $.
    \item A tensor (order $ k $) extends this concept: $ \mathcal{X} \in \mathbb{R}^{d_1 \times d_2 \times ... \times d_k} $.
\end{itemize}

\subsection{Tensor decomposition}

Tensor decomposition is the process of breaking down a tensor into simpler components \cite{kolda2009tensor}, similar to matrix factorization using techniques like singular value decomposition \cite{eckart1936approximation}. Tensor decomposition can be used to reduce complexity, extract latent structures, and compress multi-dimensional data \cite{delathauwer2000multilinear, anandkumar2014tensor, rambhatla2019tensormap}, making it essential in machine learning and signal processing.

Given a tensor $ \mathcal{X} $, the goal of tensor decomposition is to approximate it using a sum of simpler components, often in the form of factor matrices and core tensors.

\subsection{Methods of tensor decomposition}

\subsubsection{Canonical polyadic (CP) decomposition}

CP decomposition \cite{kolda2009tensor} expresses a tensor as a sum of rank-one tensors:

\begin{equation}
\mathcal{X} \approx \sum_{r=1}^{R} \mathbf{a}_r \circ \mathbf{b}_r \circ \mathbf{c}_r
\end{equation}

where $ \mathbf{a}_r, \mathbf{b}_r, \mathbf{c}_r $ are vectors corresponding to each mode, and $ \circ $ represents the outer product.

\subsubsection{Tucker decomposition}

Tucker decomposition \cite{tucker1966some} generalizes CP by introducing a core tensor and factor matrices:

\begin{equation}
\mathcal{X} \approx \mathcal{G} \times_1 \mathbf{A} \times_2 \mathbf{B} \times_3 \mathbf{C}
\end{equation}

where $ \mathcal{G} $ is a smaller core tensor and $ \mathbf{A}, \mathbf{B}, \mathbf{C} $ are factor matrices along each mode.

\subsubsection{Tensor-train decomposition}

Tensor-train (TT) decomposition \cite{oseledets2011} breaks down high-dimensional tensors into a sequence of lower-rank 3D tensors, reducing computational complexity.
\begin{equation}
\mathcal{X} \approx \mathcal{X}_1 \times_1 \mathcal{X}_2 \times_2 \mathcal{X}_3 \times_3 ... \times_{d-1} \mathcal{X}_d
\end{equation}

where each $ \mathcal{X}_i $ is a core tensor of significantly reduced size, making computations more efficient. We develop TT decomposition for end-to-end model compression of neural networks in this dissertation (See \ref{sec:tensorizing_rnns}).

\section{Pre-training, transfer learning, and domain adaptation}
Pre-training, transfer learning, and domain adaptation are inter-related methodologies for leveraging large-scale datasets and improving model performance on new tasks. Pre-training helps models learn general representations, transfer learning enables knowledge reuse across tasks, and domain adaptation ensures robustness on the same task across different environments. In this thesis, we build on these ideas in the context of medical audio to develop robust models that transfer into different clinical tasks and contexts. First, we provide here an overview of these concepts.

\subsection{Pre-training and transfer learning}

Transfer learning is a technique in deep learning where a model is trained on a large dataset before being fine-tuned for a specific task \cite{erhan2010does}. Instead of training from scratch, knowledge learned from a source domain is transferred to a target domain, reducing the need for large labeled datasets. This allows the model to learn generalized representations, which can be adapted for different downstream applications.

Pre-training on a large dataset is typically the first step in transfer learning. Pre-training can be categorized based on how the initial training is conducted:

\paragraph{Unsupervised Pre-training \cite{hinton2006fast, bengio2007greedy}}
This is one of the earliest proposed forms of pre-training. The model is trained layer-by-layer on unlabeled data using techniques like autoencoders. Unsupervised pre-training helps the network learn useful feature representations without requiring human-labeled annotations.

\paragraph{Supervised Pre-training \cite{yosinski2014transferable}}
The model is trained on a large labeled dataset for a broad task before being fine-tuned on a smaller dataset for a more specific task. For example a model is trained to classify generic real-world images then fine-tuned for medical imaging classification.

\paragraph{Self-supervised Pre-training \cite{mao2020survey}}
A hybrid between supervised and unsupervised learning where the model generates pseudo-labels from the data itself. Examples include contrastive learning methods like SimCLR \cite{chen2020simple} and MoCo \cite{he2020momentum}, which train models by maximizing the similarity between augmented views of the same sample.

\subsection{Domain Adaptation}
Domain adaptation is a specialized form of transfer learning where a model trained in a source domain is adapted to a target domain that has a different data distribution \cite{shimodaira2000improving, gretton2009covariate}. Unlike standard transfer learning, domain adaptation is required when there is a shift in data characteristics between domains.

\subsubsection{Types of Domain Adaptation}

\paragraph{Supervised Domain Adaptation}
Labeled data is available in both the source and target domains. For example, a speech recognition model trained on American English adapted to \textit{labeled} British English data.

\paragraph{Unsupervised Domain Adaptation}
No labeled data is available in the target domain. For example, a model trained on one hospital’s X-ray images being adapted for another hospital’s imaging system without having labels in the new hospital.

\paragraph{Semi-supervised Domain Adaptation}
A small amount of labeled data is available in the target domain. For example, a voice assistant trained on generic speech being adapted to noisy environments with minimal labeled samples.

\subsubsection{Approaches to Domain Adaptation}
Approaches to domain adaptation can be broadly categorized according to how they address the underlying distribution shift between source and target domains. Some key approaches include:

\begin{itemize}
    \item \textbf{Feature alignment}: Here the objective is to ensure that feature distributions from different domains are similar. Successful techniques include minimizing the distribution discrepancy between domains \cite{gretton2012kernel}, using domain adversarial training to make feature representations domain-invariant \cite{ganin2015unsupervised} and batch normalization adaptation where batch statistics are adjusted to match target domain statistics \cite{li2016revisiting}.
    
    \item \textbf{Instance-based adaptation}: Here individual examples from the source domain are adjusted to resemble those in the target domain. This can be achieved by importance weighting, where source samples are reweighted based on their similarity to the target distribution \cite{huang2007correcting}.

    \item \textbf{Parameter-based adaptation}: Borrowing from the core idea of transfer learning, model parameters trained on the source domain can be fine-tuned for applicability in the target domain, using target data \cite{YosinskiHowNetworks}.
    
    \item \textbf{Self-training}: In unsupervised domain adaptation problems, generating pseudo-labels for the target domain and iteratively refining them has been shown to be effective \cite{xie2020self, sun2019unsupervised}. 
\end{itemize}

\chapter{Building rich audio representations for improved downstream accuracy and robustness}
\label{chap:contribution1}
\section{Overview}
Despite continuing medical advances, the rate of newborn morbidity and mortality globally remains high, with over 6 million casualties every year. The prediction of pathologies affecting newborns based on their cry is thus of significant clinical interest, as it would facilitate the development of accessible, low-cost diagnostic tools\cut{ based on wearables and smartphones}. However, the inadequacy of clinically annotated datasets of infant cries limits progress on this task. This study explores a neural transfer learning approach to developing accurate and robust models for identifying infants that have suffered from perinatal asphyxia. In particular, we explore the hypothesis that representations learned from adult speech could inform and improve performance of models developed on infant speech. Our experiments show that models based on such representation transfer are resilient to different types and degrees of noise, as well as to signal loss in time and frequency domains.

\section{Problem}
We consider the case study of detecting perinatal asphyxia from infant cry sounds. The physiological interconnectedness of crying, respiration and the central nervous system has been long appreciated. Crying presupposes functioning of the respiratory muscles \cite{lagasse2005assessment}. In addition, cry generation and respiration are both coordinated by the same regions of the brain \cite{lester1990colic, zeskind2001analysis}. The study of how pathologies affect infant crying dates back to the 1970s and 1980s with the work of Michelsson et al \cite{michelsson1977pain, michelsson1977sound, michelsson2002cry}. Using spectrographic analysis, it was found that the cries of asphyxiated newborns showed shorter duration, lower amplitude, increased higher fundamental frequency, and significant increase in ``rising'' melody type.

The unavailability of reasonably-sized clinically-annotated datasets has limited progress in developing effective models for disease diagnosis from cry. The Baby Chillanto Infant Cry database \cite{Reyes-Galaviz2004}, a relatively small database of 69 babies, at the time of this work, remains the only known available database for pursuing this problem. Previous work using this data has mainly focused on classical machine learning methods or very limited capacity feed-forward neural networks \cite{Reyes-Galaviz2004,onu2014harnessing}.

We adopt a transfer learning approach to leverage larger neural networks and also a larger array freely available speech datasets. In numerous domains (e.g., speech, vision, and text) transfer learning has led to substantial performance improvements by pre-training deep neural networks on some different but related task \cite{howard2018universal,oquab2014learning,karpathy2014large}. 
In our setting, we seek to transfer models trained on adult speech to improve performance on the relatively small Baby Chillanto Infant Cry dataset. 
Unlike newborns, adults have voluntary control of their vocal organs and their speech patterns have been influenced, over time, by the predominant language spoken in their environment. We nevertheless explore the hypothesis that there exists some underlying similarity in the mechanism of the vocal production between adults and infants, and that model weights learned from adult speech could serve as better initialization (than random) for training models on infant speech.

We hypothesize that the choice of source task to pre-train the neural network matters. Ideally, the task on which the model is pre-trained should capture variations that are relevant to those in the target task.
For instance, a model pre-trained on a speaker identification task would likely learn embeddings that identify individuals, whereas a word recognition model would likely discover an embedding space that characterizes the content of utterances.
What kind of embedding space would transfer well to diagnosing perinatal asphyxia is not clear a priori.
For this reason, we evaluate and compare 3 different (source) tasks on adult speech: speaker identification, gender classification and word recognition. For completeness, we also evaluate 2 tasks not based on human speech: audio event recognition and environmental sound classification. We study how different source tasks affect the performance, robustness and nature of the learned representations for detecting perinatal asphyxia.

\textbf{Key results.} On the target task of predicting perinatal asphyxia, we find that a classical approach using support vector machines (SVM) represents a hard-to-beat baseline. Of the 3 neural transfer models, one (the word recognition task) surpassed the SVM's performance, achieving the highest unweighted average recall (UAR) of 86.5\%. By observing the response of each model to different degrees and types of noise, and signal loss in time- and frequency-domain, we find that all neural models show better {\em robustness} than the SVM.

\section{Related work}

\xhdr{Detecting pathologies from infant cry}
The physiological interconnectedness of crying and respiration has been long appreciated. Crying presupposes functioning of the respiratory muscles \cite{lagasse2005assessment}. In addition, cry generation and respiration are both coordinated by the same regions of the brain \cite{lester1990colic, zeskind2001analysis}. The study of how pathologies affect infant crying dates back to the 1970s and 1980s with the work of Michelsson et al \cite{michelsson1977pain, michelsson1977sound, michelsson2002cry}. Using spectrographic analysis, it was found that the cries of asphyxiated newborns showed shorter duration, lower amplitude, increased higher fundamental frequency, and significant increase in ``rising'' melody type.

In \cite{Orlandi2016}, four classification methods -- logistic regression, multilayer perceptron, support vector machine, and random forest -- were trained for the task of identifying preterm and full-term infant from their cry. Dataset included 38 infants in total. Random forest performed best giving an accuracy of 87\%. Further analysis showed that the relevant features were a set of 10 prosodic measure including the $F0$, $F1$, $F2$, indicating that the difference concern not only the vocal folds, but the anatomical and physiological characteristics of the tract. For the evaluation of infants who had suffered asphxyia, \cite{Reyes-Galaviz2004} collected the Chillanto Infant Cry database. The database which contained 1389 cry recordings of 69 deaf and asphyxiated infants. Their best model was a neural network trained on features extracted as mel-frequency cepstral coefficients (MFCC). This model achieved a precision and recall of 72.7\% and 68\%. \cite{sahak2013optimization} have also worked on the problem of asphyxia diagnosis from cry. They first applied principal component analysis \cite{mackiewicz1993principal}, a linear dimensionality reduction technique, to compress sound features, then trained SVMs for the classification task.

Other clinical diagnostics such as hearing impairment and hypoxia in neonates have also been studied. In \cite{Etz2012APalate}, authors applied decision tree classifiers \cite{quinlan1986induction} analysed 128 spontaneous cries of infants who had normal development, hearing impairment (HI) and unilateral cleft lip and palate (UCLP) to examine if acoustic features had marked differences. The models applied on prosodic features achieved an accuracy of 89.2\%. Features with significant differences included: durtion, $F0$, $F2$, $F4$, intensity, jitter, shimmer and harmonics-to-noise ratio. Support vector machines have also been applied for detecting hypoxia, a situation in which there is deficiency in the amount of oxygen reaching the tissues, from infant crying \cite{poel2006analyzing}, achieving up to 85\% accuracy.

\xhdr{The Baby Chillanto Database}
Amongst efforts geared at solving the problem pathology detection from cry sounds, the work of Reyes-Galaviz and Reyes-Garcia \cite{reyes2004system} is notable. They emphasized the crucial importance of early diagnosis of pathologies like asphyxia in newly born babies and went ahead to develop a system that processes infant cry to automatically recognize babies born with asphyxia using very shallow feed-forward neural networks. Their work was based on the fact that “crying in babies is a primary communication function, governed directly by the brain, and any alteration on the
normal functioning of the babies' body is reflected in the cry.” \cite{reyes2004system}. In developing the system, they collected cry samples of normal, deaf and asphyxiating babies into a
corpus (the Baby Chillanto Database) and applied the techniques of automatic speech recognition to create a pattern recognition model. The authors experimented with audio representations as linear predictive coefficients (LPC) and mel-frequency cepstral coefficients (MFCC), training a time delay neural network as the classifier. They achieved a precision and recall of 72.7\% and 68\%. For the work described in this chapter, access to the Baby Chillanto Database was obtained courtesy of the National Institute of Astrophysics and Optical Electronics, CONACYT, Mexico.

\xhdr{Weight initialization and neural transfer learning}
Modern neural networks often contain millions of parameters, leading to highly non-linear decision surfaces with many local optima. The careful initialization of the weights of these parameters has been a subject of continuous research, with the goal of increasing the probability of reaching a favorable optimum \cite{glorot2010understanding, he2015delving}. 
 Initialization-based transfer learning is based on the idea that instead of hand-designing a choice of random initialization, the weights from a neural network trained on similar data or task could offer better initialization. This pre-training could be done in an unsupervised \cite{erhan2010does}, self-supervised \cite{wang2022towards} or supervised \cite{yosinski2014transferable, bengio2007greedy} manner.

Across the domains of vision, speech and language, the use of a pre-trained model's parameters to initialize a network before finetuning has led to state-of-the-art performance in many (sometimes unrelated) tasks \cite{donahue2014decaf, sharif2014cnn}. 
Models pre-trained on large image corpora achieve representations in intermediate layers of the network that extract semantically salient aspects of the input image such as corners, edges, colors, as well as more complex elements such as mesh patterns \cite{zeiler2014visualizing}.  In medical imaging, models transfered from even unrelated non-medical vision tasks consistently outperform models trained on analytical (or "hand-engineered") features \cite{morid2021scoping}.

In audio and music, supervised transfer learning from models pre-trained on large data sources consistently improved accuracy on a variety of tasks and datasets \cite{van2014transfer, diment2017transfer, kong2020panns}. In a similar fashion to pre-training for medical imaging tasks, weights learned by pre-trained audio models have also been found to provide good representations for training medical audio models. For example, \cite{koike2020audio} surpassed the state-of-the-art on heart sounds classification by leveraging an extensive environment sounds database.

\section{Cry signals}
The infant's cry is one of the first signs of life. The mechanism of cry production is interlinked with the central nervous system, such that the cry of a neonate communicates  also its social, psychological and neurological status. In essence, the cry can tell us about pathologies that affect the newborn or if thee baby is hungry, in pain, sleepy, happy, etc.

The development of computational tools to accurately analyze and classify infant cry sounds is therefore of significant interest, since such tools could facilitate the deployment of low-cost medical diagnostic devices for newborns through wearable devices and smartphones. Moreover, the ability to recognise an infant's psychological or social state from its cry will empower parents to better attend to its needs.

One of the earliest research programs to systematically analyze infant crying was started by a Finnish research group in the 1960s \cite{WaszHockert1968TheAnalysis, wasz1985twenty}. Through auditory listening and sound spectography, they associated specific changes in cry characteristics to different medical conditions. For instance, a high fundamental frequency $F0$ (or "high-pitched shrill cry") was observed among infants with insults to the central nervous system. 

\begin{figure}[t]
  \centering 
  \includegraphics[width=1\textwidth]{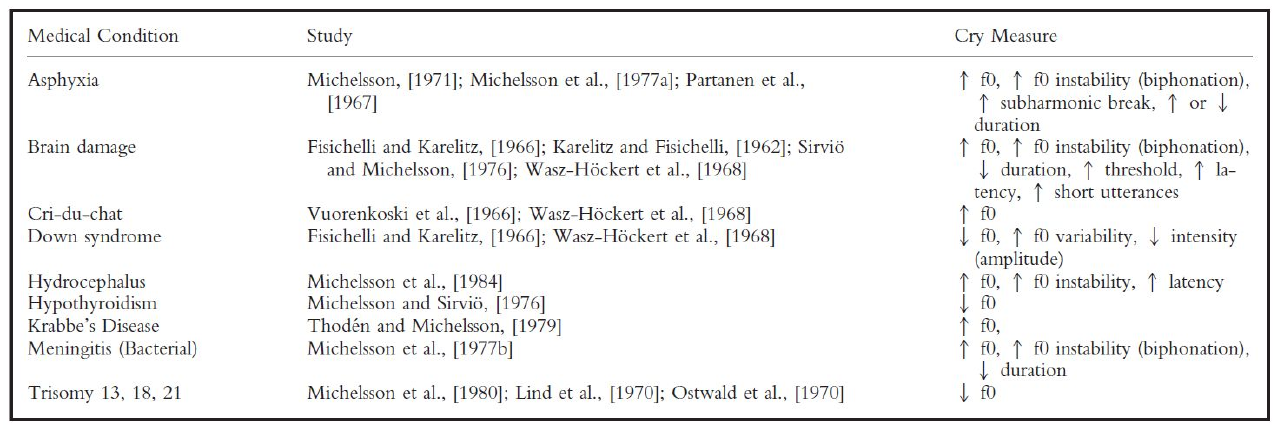} 
  \caption{Conditions affecting infants and cry measures that each alters.}
  \label{fig:cry-measures-conditions}
\end{figure}

Other cry features such as the melody type, duration, biphonation (double series of $F0$s) and glide (rapid changes in $F0$) were found to exhibit notable differences between the populations of normal and ill infants \cite{Michelsson1971, Michelsson1977}. These findings were corroborated by other researchers and are summarised in Fig. \ref{fig:cry-measures-conditions}.

\subsection{Physiology of cry production}
\label{sec:cry-physiology}

\begin{figure}[htbp]
  \centering 
  \includegraphics[width=1\textwidth]{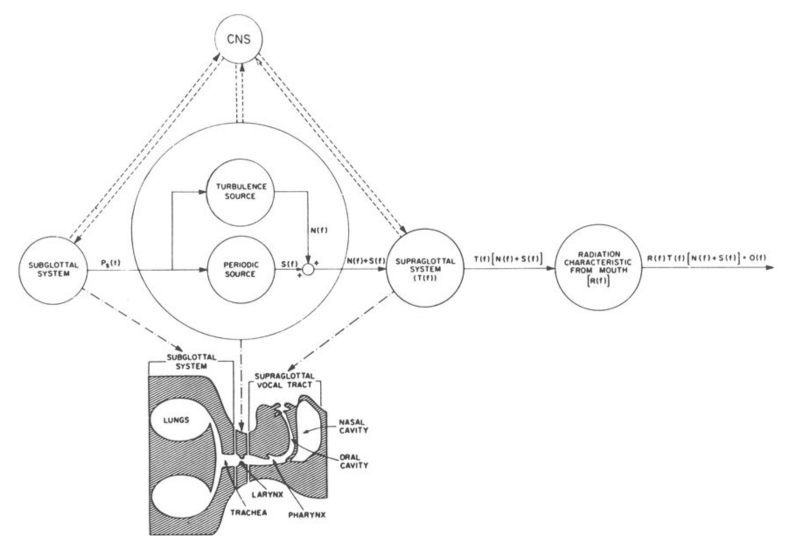} 
  \caption{A physioacoustic model of the infant cry illustration how the central nervous system (CNS) coordinates all the muscles in the upper and lower respiratory tract which collaborate to create cry sounds. Image source: \cite{golub1985physioacoustic}}
  \label{fig:cry-physiology}
\end{figure}

Crying in infants is a reflexlike action in which the central nervous system coordinates several muscles in the subglottal, lower and upper respiratory tract to produce sound \cite{golub1985physioacoustic}. See Figure \ref{fig:cry-physiology}. It occurs as part of the expiratory phase of respiration. Air pressure is generated in the subglottal respiratory system and as it leaves the lungs is shaped in 2 stages: first by the lower and then the upper vocal tract.

In the lower vocal tract, the rate of vibration of the vocal cords (within the larynx) determines the fundamental frequency $F0$ of the cry. $F0$ is perceived by the human ear as the \emph{pitch} of the sound; the higher the $F0$ the more shrill the sound. $F0$ is typically in the 250 - 600 Hz range for healthy infants during normal crying \cite{Michelsson1999PhonationCry} and could be as high as 2000 Hz in abnormal cries \cite{LaGasse2005AssessmentPerception}. This period signal is accompanied by noise.

The upper vocal tract modifies the travelling sound, producing resonant frequencies or formants over the fundamental frequency of the vocal folds. Formants are distinctive frequencies, above the fundamental, that are induced by the size and contour of the vocal tract. These formants further characterize the sound that is produced.

This acoustic model of cry production is important because depending on the anomalies seen in the cry it could be connected to the malfunction, either of one of the 3 muscles above or of the part of the CNS that controls them. In other words, different medical conditions will cause different alterations to the characteristics of the sound produced.\cite{Michelsson1999PhonationCry}.


\subsection{Challenges in acquiring cry data}
Unlike for adult speech, there is only one publicly available database of annotated infant cries - the Chillanto Infant Cry Database~\cite{Reyes-Galaviz2004} which contains 2,267 1-sec cry samples from 88 infants in different states (normal, asphyxia, deaf, hunger, pain). Several factors may have led to this lack of data including the fact that the infant speech is:
\begin{enumerate}
    \item \textbf{Harder to collect ethically}: When acquiring speech data from adults, the individual from whom data is being collected is the giver of consent. However, in infants consent has to be given by proxy (e.g., parents).

    \item \textbf{More difficult to label objectively}: Adult speech can be readily transcribed to text with very low inter-observer variability, whereas parents need time to understand the peculiarities of their baby's cry, and this knowledge, when acquired, is not easily transferable to understand another infant.

    \item \textbf{More costly to obtain clinically}: As a consequence of the previous point, the objective annotation of infant cries, e.g. for anomalies, needs to be done in a clinical setting. This, in general, requires clinical studies which cost significant time and money.
\end{enumerate}

Other research groups have acquired infant cries for different purposes but there is no indication that the data were collected into reusable corpora at the end of their studies. See~\cite{Michelsson1971}, \cite{WaszHockert1968TheAnalysis}, \cite{lind1970spectographic}, \cite{thoden1985sound}, \cite{Orlandi2016}, and \cite{abou2017fully}.

\subsection{Representations}
\label{sec:cry-representation}
Most recording devices today produce speech as a digital waveform over time, where the value at each time point is the amplitude of the signal (i.e., compression or rarefaction of air measured in pascals, Pa). This analog-to-digital conversion is achieved through 2 steps: sampling and quantization, both of which characterize the resulting digital signal. 

The sampling rate is the number of samples taken per second (in hertz, Hz). This must be at least the Nyquist rate, i.e., twice the maximum component frequency of the signal, else the sampled digital signal will be aliased. During quantization, the bit rate (in bits per sample) specifies the number of bits used to store each amplitude measurement: the larger the bit rate, the higher the resolution of the stored signal.

The digital waveform produced from sampling and quantization is typically the input to a computational system. In this section, we discuss different ways of expressing salient information in the infant speech signal for downstream analysis.

\subsubsection{Acoustic Parameters}

At the start of infant cry research in 1960s, no nomenclature had been defined on what to measure. Specific acoustic parameters which define the cry, its intonation, stress and rhythm, were first proposed by~\cite{WaszHockert1968TheAnalysis}, and standardised over the years. These parameters (presented in Table~\ref{table:acoustic-parameters}) can be roughly divided into two: time-domain and frequency-domain parameters.

\begin{table}[t]
\caption{Acoustic parameters of cry. Adapted from \cite{Kheddache2013FrequentialCries}}
\label{table:acoustic-parameters}
\centering
\begin{small}
\begin{tabularx}{\columnwidth}{p{5cm}p{8cm}}
\toprule
CRY CHARACTERISTIC & DEFINITION \\ 
\midrule
\textbf{Fundamental Frequency (F0)} & The average vibratory frequency of the vocal folds (Hz) \\
\textbf{Hyper-phonation} & The average percentage of 25 ms blocks having an F0 \textgreater 1000 Hz. \\
\textbf{Phonation} & The average percentage of 25 ms blocks having an F0 in the range 350 - 750 Hz. \\
\textbf{F0 Irregularity} & Sudden change in F0 \textgreater 100 Hz within 20 ms. \\
\textbf{Dysphonation} & The average percentage of 25 ms blocks containing noise or aperiodic sound. \\
\textbf{Utterances} & The number of vocal sounds produced by exhaling during the cry. \\
\textbf{No of cry mode changes} & The number of blocks that change between phonation and dysphonation. \\
\textbf{Shift} & Sudden change in F0. \textgreater 100 Hz \\
\textbf{Glide} & Very fast increase or decrease in F0 of 600 Hz or more during a time of 0.1 s. \\
\textbf{First and second formant (F1, F2)} & The average resonance frequencies produced by filtering the upper vocal passage. \\
\textbf{Duration of inhalation} & The interval in seconds between the first and second vocalisation. \\
\textbf{Intensity} & The average energy in dB during a vocalization. \\
\textbf{Bi-phonation} & Characterized by the presence of two F0. \\
\textbf{Variability in F0, F1, F2 \& Intensity} & Interquartile spread of each parameter. \\
\bottomrule
\end{tabularx}
\end{small}
\vskip -0.1in
\end{table}

\subsubsection{Intensity, utterances and duration of inhalation/inspiratory time}
The features that can be extracted from the time-domain signal include intensity, number of utterances and duration of inhalation (or inspiration time). Intensity is the average energy, in decibels (dB), during a cry vocalization. This is essentially a measure of how "loud" the infant was. For an audio signal $x$ with $N$ samples, intensity is estimated as:
\begin{equation}
    \text{Intensity} = 10\log_{10}\frac{1}{N{P_0}} \sum_{n=1}^N x^2_n,
\end{equation}
where $P_0$ is the auditory threshold pressure ($2 \times 10^{-5} Pa$)~\cite{martin2009speech} and $x_n$ is the signal value at the $n^{th}$ of $N$ samples. 

\begin{figure}[t]
  \centering 
  \includegraphics[width=1\textwidth]{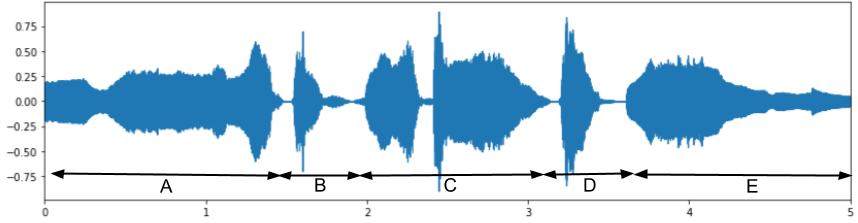} 
  \caption{Time-domain representation of a 5-sec cry signal. The y-axis represents amplitude. (A) A voiced expiratory cry, (B) A voiced inspiration between 2 non-voiced inspirations (C) A voiced expiratory cry with a hiccup near the middle (D) Another voiced inspiration between 2 non-voiced inspirations (E) Another voiced expiratory cry.}
  \label{fig:tagged-waveform}
\end{figure}

The number of utterances refers to the number of voiced expiratory cries produced during a vocalization, while the inspiration time, which could be voiced or non-voiced, is the average interval from one expiration and another. Fig.~\ref{fig:tagged-waveform} illustrates these properties using an example cry vocalization from a healthy infant.

\subsubsection{$F0$, formants and derived features}
\label{sec:f0_and_formants}
Another set of acoustic parameters are based on the (quasi) periodic nature of the cry signal. The fundamental frequency ($F0$) measures the rate of vibration of the vocal folds, and the formants characterise resonant behaviour above $F0$. The first 2 formants, $F1$ and $F2$, in conjunction with $F0$ are known to capture the most salient periodic information in cry signals. Several other features are derived from them including hyper-phonation, phonation, $F0$ irregularity, dysphonation, mode change, shift, glide, formants, bi-phonation. We refer the reader to Table \ref{table:acoustic-parameters} for definitions of these features.

\textbf{Autocorrelation Method for Estimating $\textbf{F0}$} \\
Several algorithms have been proposed for estimating $F0$. These are generally called {\em pitch extraction} algorithms. The autocorrelation method aims to first estimate the {\em fundamental period} - the inverse of $F0$~\cite{miller1956measurement, Markel1972TheEstimation}. The period is obtained by correlating the signal with itself at various lags or offsets. The offset at which correlation is highest is the fundamental period. The autocorrelation method is capable of handling the range of infant cry $F0$s~\cite{Petroni}. The autocorrelation of a signal is defined as:
\begin{equation}
   \rho_k = \sum_{n=0}^{N-1-|k|} x_n x_{n + |k|} \qquad k=0,1,...,N-1,
\end{equation}
where $\{x_0, x_1, ..., x_{N-1}\}$ is a length-$N$ window or frame of speech and $k$ is the lag.

\begin{figure}[t]
    \centering
    \includegraphics[width=1\textwidth]{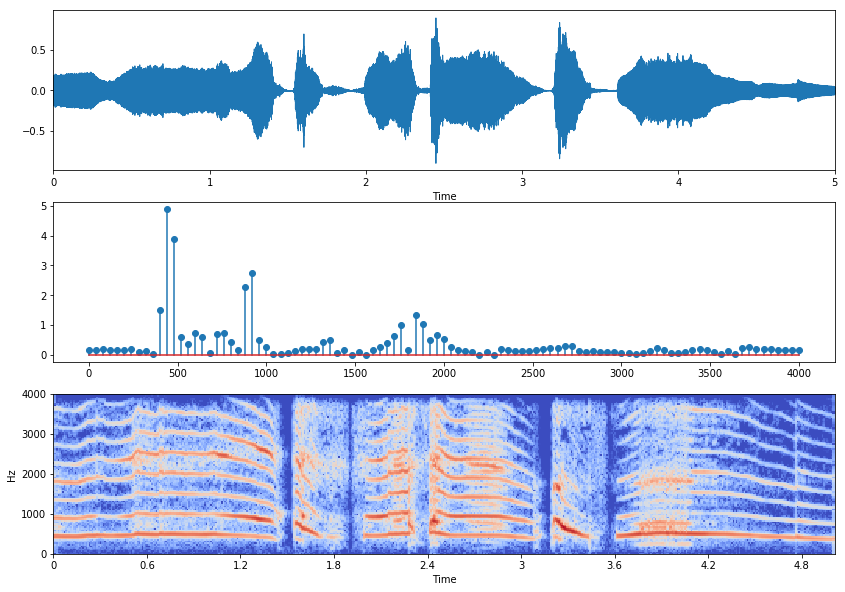}
    \caption{(\textbf{Top}) The raw waveform of a 5-second cry signal. (\textbf{Middle}) The spectrum of of the first 25ms of the cry signal. The x-axis indicates the frequencies present and the y-axis shows the magnitudes of each. (\textbf{Bottom}) The spectrogram of the cry signal. The  x-axis is time, the y-axis is frequency and colour represents the magnitude of the frequency at a given time (the brighter the colour, the larger the magnitude).}
    \label{fig:raw-spectrum-spectogram}
\end{figure}

\textbf{Cepstral Method for Estimating $\textbf{F0}$} \\
The cepstral method for estimating $F0$ relies on the discrete Fourier transform (DFT). The DFT is a decomposition of a signal into its frequency components and their respective magnitudes - a representation typically known as the (frequency) spectrum of the signal. Fig.~\ref{fig:raw-spectrum-spectogram} (middle) shows the spectrum of the first 25ms window of the signal in Fig.~\ref{fig:raw-spectrum-spectogram} (top). Formally, the DFT is a sequence of complex numbers given as:
\begin{equation}
    X_k = \sum_{n=0}^{N-1} x_n e^{-j\frac{2\pi}{N}kn} \qquad k=0,...,N-1,
\end{equation}
where $x_n$ is the $n$th element of the length-$N$ signal whose DFT is to be estimated. The DFT requires $O(N^2)$ operations. However, fast Fourier transform (FFT) algorithms exist which compute the DFT in $O(N\log N)$ time. FFT algorithms achieve this speed up by factorizing the DFT matrix into a product of sparse factors.

Cepstral analysis estimates a double spectrum to obtain the cepstrum\footnote{ The word "cepstrum" was formed by reversing the first 4 letters of "spectrum"~\cite{noll1964short}.}, a representation that separates the $F0$ from its formants. In practice, this implies computing the inverse DFT of the log magnitude of the DFT of a signal ~\cite{noll1964short}, as follows:
\begin{equation}
    c_k = \sum_{n=0}^{N-1} \log \left (\left|\sum_{n=0}^{N-1} x_n e^{-j\frac{2\pi}{N}kn} \right| \right)e^{j\frac{2\pi}{N}kn}
\end{equation}

\begin{figure}[t]
    \centering
    \includegraphics[width=\textwidth]{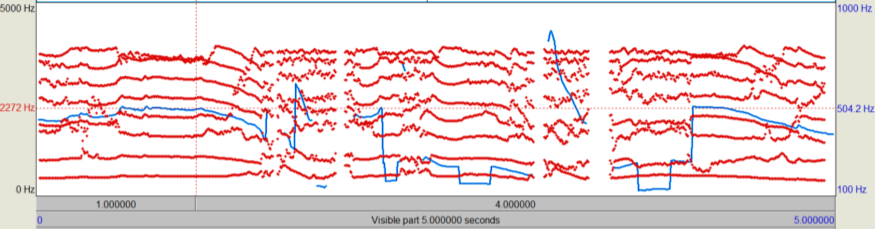}
    \caption{$F0$ or "Pitch" (blue line) and formants (red lines) of cry signal of a healthy baby in Fig. \ref{fig:tagged-waveform}. Generated using Praat software \cite{Praat}.}
    \label{fig:pitch-tracking}
\end{figure}

Fig.~\ref{fig:pitch-tracking} shows $F0$ and the first 8 formants, i.e., $F1$ to $F8$, estimated for the cry signal in Fig. \ref{fig:tagged-waveform}.

These acoustic features are useful from a domain knowledge perspective. For instance in the case of extracting infant speech features for clinical purposes, they may represent parameters that are altered when a particular organ in the vocal system is malfunctioning. For example, conditions affecting the lungs could lead to shorter breaths and consequently shorter {\em duration of utterances}. A diseased larynx could cause {\em fundamental frequency} to deviate from normal ranges.

\subsubsection{Spectrogram}
Spectograms are a more general way of representing audio signals. They indicate how the frequency spectrum evolves over the duration of an audio signal. Recall that the spectrum (Fig. \ref{fig:raw-spectrum-spectogram} (middle)) is a decomposition of a single window/frame of an audio signal into its frequency components.

A spectrogram, on the other hand, is a representation of a spectrum as a function of time. The spectrogram is computed using overlapping windows in the time domain. Each window is chosen to be small enough (typically 20 to 30 ms) so that we can safely assume statistical stationarity within it. An example spectogram is shown in Fig.~\ref{fig:raw-spectrum-spectogram} (bottom).

\subsubsection{Log Mel-Spectogram}
The human ear does not perceive sound equally across the frequency spectrum. The inner ear (cochlea) has evolved to filter audio in a specific way that facilitates interpretation in the presence of adversarial factors. Two such transformations have played an important role in speech recognition systems for adults and have been applied in infant speech research as well. 

First, it has been observed that human hearing is more sensitive to lower frequency ($<$ 500 Hz) sounds than those at higher frequencies. In particular, increasingly larger intervals in frequencies (above 500 Hz) are perceived to be of equal pitch increments. This effective spectrum of the human ear has been described as the Mel-frequency scale~\cite{stevens1937scale}. A given frequency $f$ can be converted to Mel-frequency $m$ as follows:
\begin{equation}
m = 2595 \log_{10} \left (1 + \frac{f}{700} \right)
\end{equation}

\begin{figure}[h]
    \centering
    \includegraphics[width=.85\textwidth]{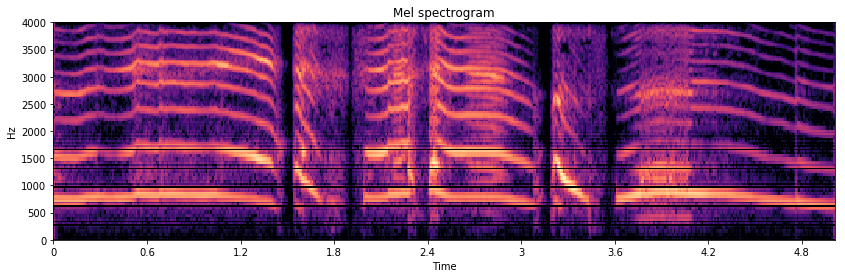}
    \caption{Log mel-spectogram of the cry signal in Fig. \ref{fig:raw-spectrum-spectogram} (top). Notice how the formants at higher frequencies are closer to each other than in the spectogram in Fig. \ref{fig:raw-spectrum-spectogram} (bottom).}
    \label{fig:mel-scale}
\end{figure}

Secondly, the human ear perceives the intensity of audio on a log-scale with respect to the frequency; meaning that audio at higher frequencies (\textgreater 1000 Hz) does not sound as loud as its actual magnitude. Hence it is common to not only transform the frequency spectrum of an audio signal to the mel-scale but to also take the logarithm of the magnitudes. The resulting spectogram after applying these 2 transformations is called a {\em log mel-spectogram}. The log mel-spectogram of the signal in Fig.~\ref{fig:raw-spectrum-spectogram} (top) is shown in Fig.~\ref{fig:mel-scale}.

\subsubsection{Mel-Frequency Cepstral Coefficients}
Mel-frequency cepstral coefficients (MFCCs) are a popular choice of features in adult speech recognition and have been used in infant speech tasks as well~\cite{Reyes-Galaviz2004}. They were first developed as a way of decorrelating $F0$ from higher formants. Similar to the cepstral method for computing $F0$, MFCCs are based on the coefficients of the cepstrum described earlier in sec.~\ref{sec:f0_and_formants}. They are computed by estimating the DFT of the log-mel-spectogram. Typically, only the top 12 cepstral coefficients (containing the formants) of the transform are used, in conjunction with the energy of the signal and measures of the rate of change of the coefficients, known as delta and delta-delta features~\cite{martin2009speech}. 

\begin{figure}[t]
    \centering
    \includegraphics[width=.85\textwidth]{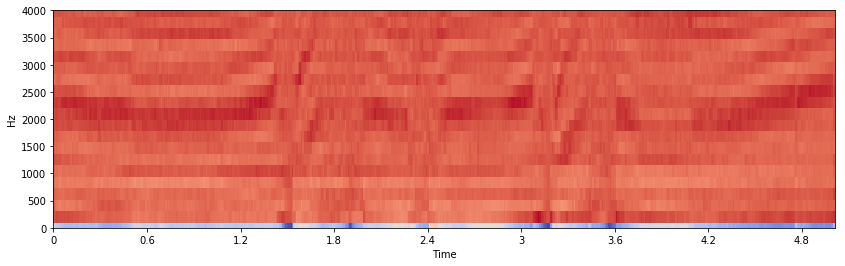}
    \caption{Mel-frequency cepstral coefficients (MFCC) of the cry signal in Fig.~\ref{fig:raw-spectrum-spectogram} (top)}
    \label{fig:mfcc}
\end{figure}

\begin{figure}[h]
    \centering
    \includegraphics[width=.8\textwidth]{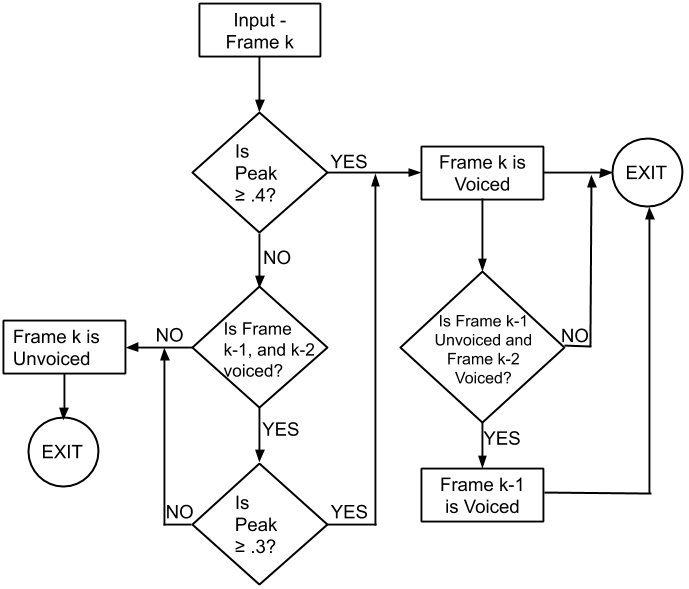}
    \caption{Decision algorithm for voiced-unvoiced detection, proposed in \cite{Markel1972TheEstimation}.}
    \label{fig:detection-flowchart}
\end{figure}

\subsection{Speech to cry transfer}
This work explores the potential benefits of transferring models trained on adult speech signals to tasks based on infant cry signals. As seen in the related work section above, such transfer between tasks is more successful when the source and target tasks share similarities such as being of the same modality (images, audio, text, etc) and being similarly distributed.

In the case of speech and cry the underlying physiology of sound production described in section \ref{sec:cry-physiology} is largely unchanged, begging the question of whether in the absence of large corpuses of infant crying data, can speech be an effective substitute for pre-training. Precisely, the generation of a single vowel by an adult goes through the same process: air pressure generated in the lung, periodic characteristic plus noise added by the vocal chord vibration, formants created by the filtering of the upper vocal tract. However, we note that there are differences caused by 3 main factors. First is adults have developed voluntary control over their speech organs, so it is no more reflexlike as in infancy. Second is adults generally utter a different kind of language than infants, suggesting that speech data could be distributed differently. Lastly, growth and the environment an adult has spent the most time in alters the prosodic characteristics of their voice including intonation, accent, timbre and pitch.

For these reasons, we aim to discover through a careful set of experiments how useful speech data could be for learning models on infant cry signals. And if ultimately large, freely available databases of adult speech are a great asset for building useful cry solutions.

\section{Methods}
In this section, we describe our approach to designing and evaluating transfer learning models for the detection of perinatal asphyxia in infant cry. We present the source tasks selected along with representative datasets. We further describe pre-processing steps, choice of model architectures as well as analysis of trained models.

\subsection{SVM Baselines}
The baseline models were based on support vector machines (SVMs). Instead of a simple boundary, SVMs aim to learn the maximum margin between classes of interest. It achieves non-linear functions using kernels to implicitly map feature vectors to a higher-dimensional spaces. A kernel function is a similarity function that defines the basis for measurement of the proximity of two or a combination of samples from a dataset. Not all similarity functions make valid kernels. A valid kernel must satisfy Mercer’s theorem \cite{mercer1909functions}. We experiment with 2 kernels here: Radial Basis Function and Polynomial kernels.

Radial basis (RBF) \cite{boser1992training} is one of the most popular kernels in use and is very suited for majority of applications. Given the feature vector representations of two samples: $x$ and $x^{\prime}$ and a free parameter $\sigma$, the RBF kernel is:

\begin{equation}
    K(x, x^{\prime}) = \exp (-\frac{\|x - x^{\prime}\|^2}{2\sigma^2})
\end{equation}

The polynomial kernel on the other hand projects feature vectors to higher orders. For polynomials of degree $d$ and free parameter $c$ the kernel is given as:

\begin{equation}
    K(x, x^{\prime} = (x^Tx^{\prime} + c)^d
\end{equation}

For our baseline models, the input cry samples are transformed to feature vectors represented as mean-normalized MFCCs. First, each audio file was downsampled to a sampling rate of 8kHz, then the log-mel-spectrogram of the audio signal is computed, after which the discrete Fourier transform is applied to obtain a $16 \times 12$ representation of the MFCCs which is unrolled into a $168 \times 1$ feature vector. Mean normalization is applied at this point, across the training data to ensure that each feature is between $-1$ and $1$.

The data splits is given in Table \ref{tab:chillanto-splits}. Splits were made such that all recordings from the same patient remained in the same set.

\begin{table}[h]
  \caption{Data split of audio recordings in the Baby Chillanto Database. Files belonging to the same patient were guaranteed to reside in the same set.}
  \label{tab:chillanto-splits}
  \centering 
  {
  \small
  \begin{tabular}{l p{3cm}p{3cm}p{3cm}} 
    \toprule
       & \textbf{Training}& \textbf{Cross Validation}&\textbf{Test}\\
    \midrule
    \textbf{Healthy}& 630& 209&210\\
    \midrule
    \textbf{Asphyxia}& 207& 68&68\\
    \midrule
    &  834& 277&278\\
  \end{tabular}
  }
\end{table}

The cross-validation set was used to select hyperparameters when experimenting with both kernels. The best hyperparameters were then used to evaluate the model on the left out test set. The polynomial kernel SVM achieved the best performance with a sensitivity and specificity of 81.6\% and 87.2\% respectively.

\subsection{Source tasks} We choose a total of 5 source tasks which includes 3 speech (speaker identification, gender classification, word recognition) and 2 non-speech source tasks (audio event recognition and environmental sound classification) with corresponding audio datasets: Voice Cloning Toolkit (VCTK) \cite{veaux2017}, Speakers in the Wild (SITW)  \cite{mitchell2016}, Speech Commands  \cite{warden2018}, AudioSet \cite{gemmeke2017}, and ESC-50 \cite{piczak2015}. We briefly describe each database below:

\noindent\textbf{Voice Cloning Toolkit (VCTK)}. 
The VCTK (Voice Cloning Toolkit) database is a widely used collection of English speech data designed for research and development in the field of text-to-speech (TTS) synthesis, voice cloning and speaker adaptation. It consists of speech recordings from 109 different native English speakers, encompassing both male and female voices uttering phonetically balanced sentences, which are carefully designed to cover a wide range of phonemes and linguistic contexts. As a source task in this work, VCTK was used for training a speaker identification model.

\noindent\textbf{Speakers in the Wild (SITW)}. The "Speakers in the Wild" (SITW) dataset is a speech dataset of 2,000 audio clips originally designed for speaker recognition, voice biometrics and related fields. The version used includes 299 speakers from various ethnic backgrounds. Since the SITW dataset encompasses a wide range of ages, genders, and accents, it was used in this work for training gender classification models. 

\noindent\textbf{Speech Commands}. The Speech Commands dataset is a widely used collection of audio recordings of spoken English words and phrases. It consists of over 65,000 one-second audio clips of individuals speaking single-word commands such as "yes," "no," "stop," "go," "up," "down," and many others. The dataset includes contributions from thousands of different speakers, which introduces speaker variability into the recordings. In accordance with its design, Speech Commands dataset was used for training word recognition models in this work.

\noindent\textbf{AudioSet}. AudioSet is a large-scale, publicly available database of labeled audio data designed for various audio-related machine learning tasks. AudioSet consists of approximately 2,084,320 audio clips, each lasting 10 seconds. These clips are extracted from a wide range of sources and covers 632 different audio event classes of everyday sounds, including musical instruments, animal sounds, human activities, environmental sounds, and more. Corresponding to its design, AudioSet was applied for training a source model for audio event recognition.

\noindent\textbf{Environmental Sound Classification 50 (ESC-50)}. The ESC-50 dataset is a widely used collection of audio recordings designed for the task of environmental sound classification. The dataset contains a total of 2,000 audio clips, across 50 sound categories, each category represented by 40 clips. It is different from AudioSet in that the latter includes a vast collection of audio clips from various sources, encompassing not only environmental sounds but also speech, music, and other audio types. We trained an environmental sound classification model using ESC-50.

All source tasks and corresponding databases are summarized in Table \ref{table:source_task}.

\begin{table}[h]
  \caption{Source tasks and corresponding datasets used in pre-training neural network. Size: number of audio files.}
  \label{table:source_task}
  \centering {\small
  \begin{tabular}{p{4cm}p{8.5cm}p{0.7cm}} 
    \toprule
    Dataset   & Description & Size \\
    \midrule
    VCTK  & Speaker Identification. 109 English speakers reading sentences from newspapers.  & 44K \\
    \midrule
    SITW    & Gender classification. Speech samples from media of 299 speakers. & 2K \\
    \midrule
    Speech commands    &  Word recognition. Utterances from 1,881 speakers of a set of 30 words.  & 65K \\
    \midrule
    AudioSet & Audio event recognition. Over weakly labelled clips of 527 audio classes. &  2M\\
    \midrule
    ESC-50 & Environmental sound classification. Environment audio recordings organized into 50 semantic classes & 2K \\
    \bottomrule
  \end{tabular}}
\end{table}

\subsection{Target task: Perinatal asphyxia detection} Our target task is the detection of perinatal asphyxia from newborn cries. We develop and evaluate our models using the Chillanto Infant Cry Database \cite{reyes2004system}. The database contains 1,049 recordings of normal infants and 340 cry recordings of infants clinically confirmed to have perinatal asphyxia. Audio recordings were 1-second long and sampled at frequencies between 8kHz to 16kHz with 16-bit PCM encoding. The pre-processing steps applied are described in section \ref{sec:neur-trans-preproc}.

We compare a benchmark SVM model trained on the Chillanto database with neural transfer models pre-trained on the different source tasks then fine-tuned on Chillanto database. We assess the impact of pre-training on the accuracy and robustness of the resulting model. While accuracy is measured using the combination of sensitivity and specificity as well as their arithmetic mean, robustness is evaluated via a set of analyses.

To assess robustness, we first conduct ablation experiments simulating channel disruptions that could truncate portions of a recorded audio signals. Ablation experiments were of two kinds: time-domain, in which random (milli)seconds of audio are missing and frequency-domain, in which different portions of the frequency spectrum are attenuated. In both cases we estimate the ability of the models to still make correct predictions. 

A second set of analyses involved assessing the models' resilience to environmental noise. When capturing signals, and audio in particular, other environmental sounds are inevitably recorded alongside our signal of interest. We conducted noise experiments to evaluate how well the neural transfer models handle noise relative to each other and to the baseline randomly initialized model.

Lastly, we examine the linear independence of the dimensions of the embedding vectors output by the model, taking this as a proxy for the robustness of the learned representation. Using the unsupervised dimensionality reduction technique, principal component analysis (PCA) \cite{jolliffe2011principal}, we determine correlations between dimensions of our embeddings. A model that has learned a more robust function will output embeddings with a higher number of principal components.

\subsection{Model architecture and transfer learning}
 We adopt a residual network (ResNet) \cite{he2016deep} architecture with average pooling, for training. Consider a convolutional layer that learns a mapping function $F(x)$ of the input, parameterized by some weights. A residual block adds a shortcut or skip connection such that the output of the layer is the sum of $F(x)$ and the input $x$, i.e., $y = F(x) + x$. This structure helps control overfitting by allowing the network to learn the identity mapping $y=x$ as necessary and facilitates the training of even deeper networks.

\begin{figure*}[t]
  \centering
  \includegraphics[width=.8\textwidth]{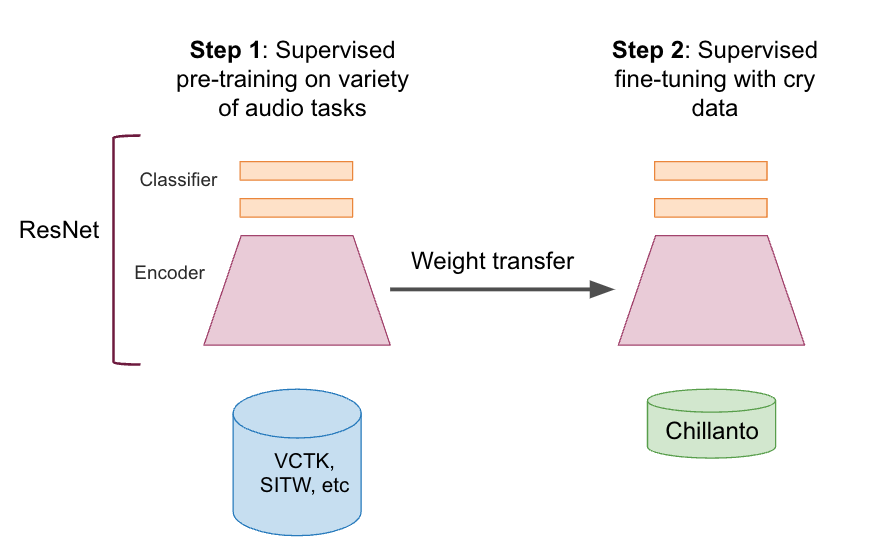}
  \caption{Neural transfer learning methodology. Weights from a pre-trained encoder are used to initialize model on target task.}
\label{fig:model_diagram}
\end{figure*}

ResNets represent an effective architecture for speech, achieving several state-of-the-art results in recent years \cite{tang2018deep}. To assure even comparison across source tasks, and to facilitate transfer learning, we adopt a single network architecture: the {\em res8} as in Tang et al.\@ \cite{tang2018deep}. The model takes as input a 2D MFCC of an audio signal, transforms it through a collection of 6 residual blocks (flanked on either side by a convolutional layer), employs average pooling to extract a fixed dimension embedding, and computes a k-way softmax to predict the classes of interest. Fig \ref{fig:model_diagram} shows the overall structure of our system. Each convolutional layer consists of 45, $3 \times 3$ kernels.



We train the {\em res8} on each source task to achieve performance comparable with the state of the art. The learned model weights (except those of the softmax layer) are used as initialization for training the network on the Chillanto dataset. During this post-training, the entire network is tuned.

\subsection{Pre-processing and baselines}
\label{sec:neur-trans-preproc}
All audio samples are pre-processed similarly, to allow for even comparison between source tasks and compatibility with target task. Raw audio recordings are downsampled to 8kHz and converted to mel-frequency cepstral coefficients (MFCC). To do this, spectrograms were computed for overlapping frame sizes of 30 ms with a 10 ms shift, and across 40 mel bands. For each frame, only frequency components between 20 and 4000 Hz are considered. The discrete cosine transform is then applied to the spectrogram output to compute the MFCCs. The resulting coefficients from each frame are stacked in time to form a spatial ($40 \times 101$), 2D representation of the input audio.

We implement and compare the performance of our transfer models on 2 baselines. One is a model based on a radial basis function Support Vector Machine (SVM), similar to \cite{onu2014harnessing}. The other is a {\em res8} model whose initial weights are drawn randomly from a uniform Glorot distribution \cite{glorot2010understanding} i.e., according to $U(-k, k)$ where $k = \frac{\sqrt{6}}{n_i + n_o}$, and $n_i$ and $n_o$ are number of units in the input and output layers, respectively. This initialization scheme scales the weights in such a way that they are not too small to diminish or too large to explode through the network's layers during training.

\section{Experiments}

\begin{table}[htbp]
  \caption{Performance -- mean (standard error) -  of different models in predicting perinatal asphyxia. 
  The models are Support Vector Machine (SVM), a res8 network trained from random initialization (no-transfer), and res8 networks pretrained on speaker identification (vctk-transfer), gender identification (sitw-transfer), word recognition (sc-transfer), audio event recognition (auds-transfer) and environmental sound classification (esc-transfer) tasks.}
    \label{table:performance}
  \centering {\small 
  \begin{tabular}{llllll}
    \toprule
      & Model    &  UAR \%       &     Sensitivity \%            &   Specificity \%       \\
    \midrule
    Baselines & SVM & 84.4 (0.4)  & 81.6 (0.7) & 87.2 (0.2) \\
    & no-transfer   &    80.0 (2.5) &            71.8 (5.8) &            88.1 (0.8) \\
    \midrule
    Speech pre-training & sc-transfer   &    \textbf{86.5 (1.1)} &            \textbf{84.1 (2.2)} &            \textbf{88.9 (0.4)} \\
    & sitw-transfer &    81.1 (1.7) &            72.7 (3.5) &            89.5 (0.2) \\
    & vctk-transfer &    80.7 (1.0) &            72.2 (2.1) &            89.1 (0.3) \\
    \midrule
    Non-speech pre-training & auds-transfer &     78.6 (5.1) &            82.8 (5.8) &           74.3 (5.2) \\
    & esc-transfer &      75.0 (1.5) &                 61.9 (2.9) &            88.1 (0.2) \\
    \bottomrule
    \end{tabular}}
\end{table}

\subsection{Training details}
There were a total of 1,389 infant cry samples (1,049 normal and 340 asphyxiated) in the Chillanto dataset. The samples were split into training, validation and test sets, with a 60:20:20 ratio, and under the constraint that samples from the same patients were placed in the same set. We evaluate the performance of our models on the target task by tracking the following metrics: sensitivity (recall on asphyxia class), specificity (recall on normal class), and the unweighted average recall (UAR). We use the UAR on the validation set for choosing best hyperparameter settings. The UAR is a preferred choice over accuracy since the classes in the Chillanto dataset are imbalanced.

Each source task was trained, fine-tuning hyperparameters as necessary to obtain performance comparable with the literature. For transfer learning on the target task, models were trained for 50 epochs using stochastic gradient descent with an initial learning rate of 0.001 (decreasing to 0.0001 after 15 epochs), a fixed momentum of 0.9, batch size of 50, and hinge loss function. We used a weighted balanced sampling procedure for mini-batches to account for class imbalance. We also applied data augmentation via random time-shifting of the audio recordings. Both led to up to 7\% better UAR scores when training source and target models.


\subsection{Performance on target task}
Table \ref{table:performance} summarizes the performance of all models on the target task. The best performing model was pre-trained on the word recognition task (sc-transfer) and attained a UAR of 86.5\%. This model also achieves the highest sensitivity and specificity 84.1\% and 88.9\% respectively. All other speech-based transfer models performed better than {\em no-transfer }, suggesting that transfer learning resulted in better or at least as good an initialization. The SVM was the next best performing model and had the lowest standard error among all models. The worst performing models were the ones transferred from non-speech tasks (auds-transfer and esc-transfer). Comparing the UARs, it was notable to find that these models performed even worse than random initialization. We suspect that the domain of non-speech audio is simply too different from that of infant cry sounds, and that the difference between these models and the no-transfer might be accounted for by the relatively high standard errors observed.

For subsequent analysis, we pick the best performing neural transfer model, {\em sc-transfer}, the randomly initialized neurl model, {\em no-transfer}, and the SVM baseline.

\begin{figure}[htbp]
\centering
\includegraphics[width=1\textwidth]{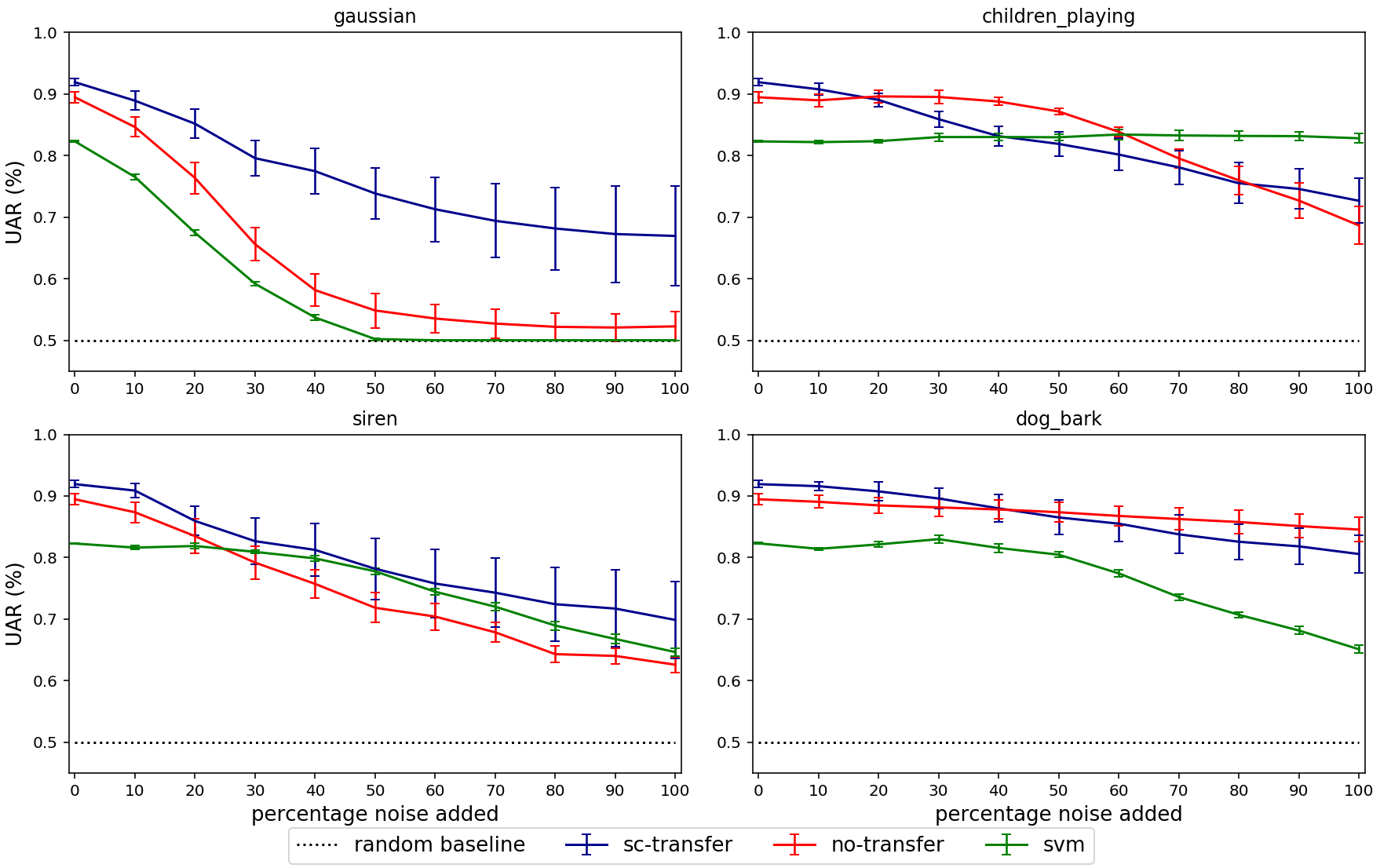}
\caption{Performance of models under different noise conditions.}
\label{fig:noise_analysis}
\end{figure}

\subsection{Robustness analysis}
\begin{figure}[htbp]
\centering
\includegraphics[width=.8\textwidth]{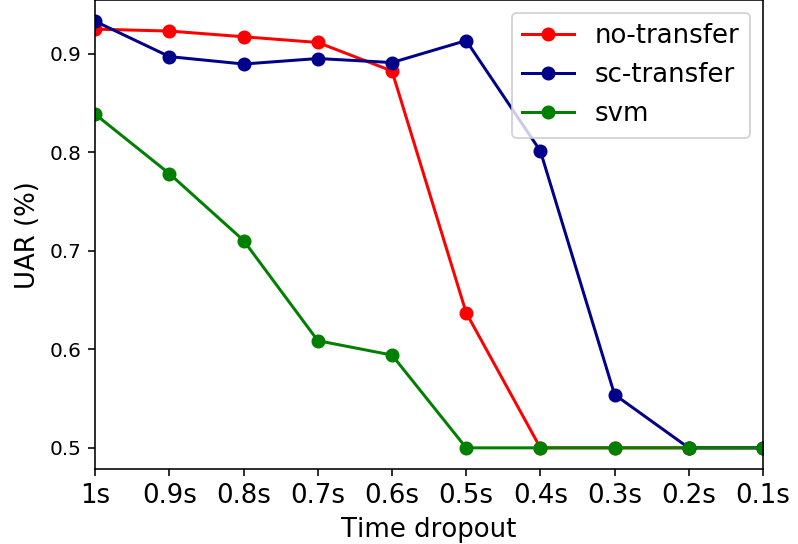}
\caption{Audio length analysis highlighting the impact of using shorter amounts of input audio on UAR performance.}
\label{fig:time_analysis}
\end{figure}

In most cases, our results suggest that neural models have overall increased robustness. We focused on the top transfer model {\em sc-transfer}, {\em no-transfer} and the SVM. Figure \ref{fig:noise_analysis}, shows the response of the models to different types of noise, revealing that in all but one case the neural models degrade slower than the SVM.  Results from Figure \ref{fig:time_analysis} suggest that the neural models are also capable of high UAR scores for short audio lengths, with {\em sc-transfer} maintaining peak performance when evaluated on only half (0.5s) of the test signals.

\begin{figure}[htbp]
\centering
  \includegraphics[width=.8\textwidth]{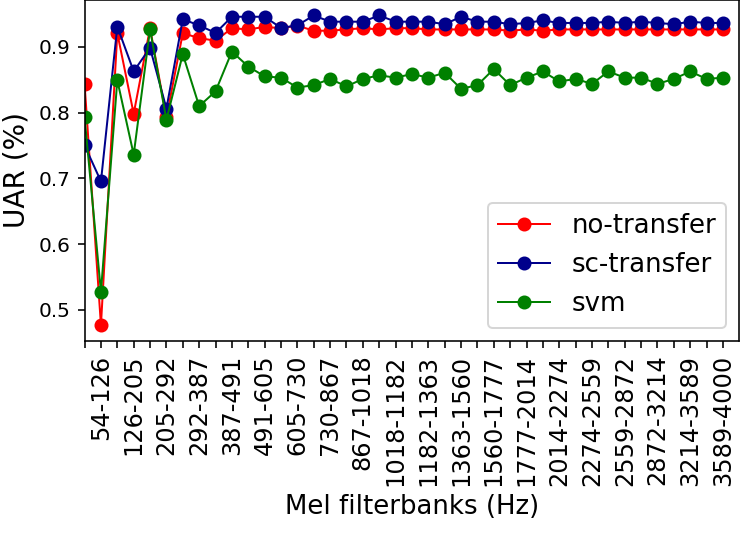}
  \caption{Frequency response analysis of the relative importance of different Mel filterbanks on UAR performance. Each point represents the performance after removing the corresponding Mel filterbank.}
\label{fig:freq_analysis}
\end{figure}

From our analysis of the models' responses to filterbank frequencies (Figure \ref{fig:freq_analysis}), we observe that (i) the performance of all models (unsurprisingly) only drops in the range of the fundamental frequency of infant cries, i.e. up to 500Hz \cite{daga2011acoustical} and (ii) {\em sc-transfer} again is the most resilient model across the frequency spectrum.

\subsection{Visualization of MFCC feature embeddings} \label{appendix:pca_analysis}
In order to further investigate the nature of the embedding learned by each model, we apply principal component analysis (PCA) to the learned final-layer embeddings for all models \cite{jolliffe2011principal}.
By applying PCA, we hope to gain insight on the extent to which the embedding space captures unique information.

Figure 5 shows cumulative variance explained by the principal components (PC) of the neural model embeddings. Whereas in {\em no-transfer}, the top 2 PCs explain nearly all variance in the data (91\%), in {\em sc-transfer} they represent only 52\%---suggesting that the neural transfer leads to an embedding that is intrinsically higher dimensional and richer than the {\em no-transfer} counterpart.

\begin{figure}[htbp]
\label{fig:pca_cum_var}
  \centering
  \includegraphics[width=1\textwidth]{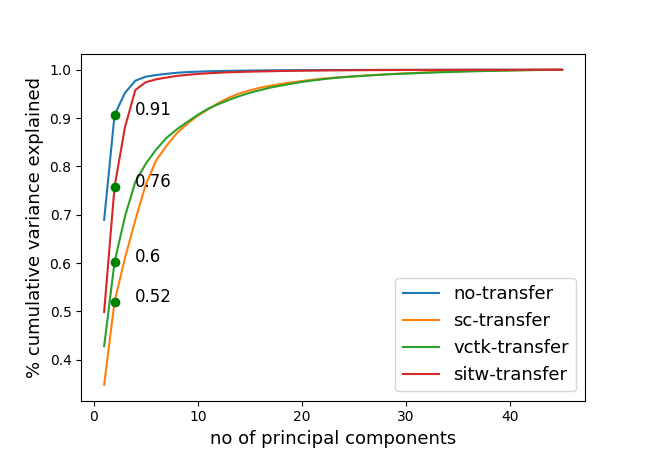}
  \caption{Cumulative variance explained by all principal components (left) and the top 2 principal components on the Chillanto test data (right) based on embeddings of {\em no-transfer} model.}
\end{figure}

\section{Discussion}
This work investigates neural transfer learning from adult speech tasks to the challenge of detecting perinatal asphyxia from infant cry sounds. Given the physiological connection between crying, respiration, and the central nervous system, previous research has established the potency of the infant cry as a diagnostic input and we sought to explore that in the context of analyzing the cries of asphyxiated newborns.

Clinically-annotated datasets for this problem are however scarce, with the "Baby Chillanto Infant Cry" database being one of the few available resources. Consequently, previous work in this area has primarily used classical machine learning methods or limited-capacity neural networks.

In adopting a transfer learning approach, we simultaneously aimed to leverage larger neural networks and freely available speech datasets. Given the parallels between adults and infants in vocal production mechanisms, it begs the question of whether models pre-trained on adult speech can improve performance on infant speech. Various source tasks for pre-training, including speaker identification, gender classification, word recognition, etc were evaluated to understand their impact on diagnosing perinatal asphyxia.

We found that one neural transfer model (pre-trained on word recognition) outperforms the classical support vector machines (SVM) baseline in predicting perinatal asphyxia. The neural models also exhibit enhanced robustness compared to SVM when exposed to noise and signal loss.

In summary, this study validates the transferrability of neural network representations learned from adult speech to target problems in infant cry space for which we may not have sufficient data. Through the use of transfer learning via different source tasks we develop accurate and robust models for the detection perinatal asphyxia from infant cry sounds. This is a notable result given the abundance of free and public databases of adult speech, and in stark contrast, the lack of databases of infant cry sounds. We believe that this work will kick-start a renewed interest in infant cry research, which could be crucial for the early detection of critical conditions. The methods developed here can also be applied to any medical audio domain for which labelled clinical data is scarce.

\chapter{End-to-end compression for efficient modeling}
\label{chap:contribution2}
\section{Overview}
The text discusses the application of tensor-train (TT) formalism to achieve significant compression of model parameters in recurrent neural networks (RNNs). While RNNs have excelled in sequential data tasks, they often require large hidden states, leading to memory and computational challenges, especially on mobile and embedded devices. Various compression techniques have been used to address this issue. Unlike previous methods that applied TT formalism separately to weight matrices, the authors introduce a "fully tensorized" form of weight sharing using TT formalism, applying it jointly to all weight matrices within RNNs. This approach results in lightweight, end-to-end trainable models that maintain or improve performance compared to uncompressed counterparts, even when dealing with high-dimensional hidden states or input representations. Experiments demonstrate state-of-the-art performance on an image classification test case. In speaker verification experiments, a useful first step in an infant cry analysis system, we observe a reduction in speaker verification error by 16\% while achieving a remarkable 200-fold compression in model parameters, effectively addressing the memory and computational challenges posed by large RNNs.

\section{Introduction}
Recurrent neural networks (RNNs) represent a model family that is well-suited for tasks involving sequential data. Although early RNNs were limited by the problem of vanishing gradients during training, this was largely solved by the development of gated RNNs such as long short-term memory (LSTM) and gated recurrent unit (GRU) models~\cite{hochreiter1997,cho2014}, which employ a collection of independent weight matrices to control the propagation of gradients. Such models have allowed RNNs to attain impressive performance in tasks such as speech recognition, language modeling, time series forecasting, and video classification.

RNNs typically employ large hidden states to achieve better performance in difficult modeling tasks, which in turn leads to a significant increase in parameters used to specify large weight matrices. The memory and compute issues associated with running such models, particularly in the limited setting of mobile and embedded devices, has led to the use of various techniques for model compression, including model distillation~\cite{hinton2015}, alternate matrix decompositions~\cite{sainath2013}, and quantization of network weights~\cite{he2019}. The use of such compression strategies is supported by the observation that standard representations of neural networks contain significant amounts of redundancy~\cite{denil2013,cheng2015}.

In this work we use the tensor-train (TT) formalism, a means of efficiently representing multi-modal tensors, to achieve significant compression of the model parameters associated with various RNN architectures. In contrast to previous work~\cite{tjandra2017,yang2017tensor}, we apply the TT formalism jointly to all weight matrices within the RNN, leading to a ``fully tensorized'' form of weight sharing, where various gate matrices are encoded within a single TT format. This permits the development of extremely lightweight, end-to-end trainable models, even in the presence of high-dimensional hidden states or input representations. 

Experiments on image classification and speaker verification show that fully tensorized TT-RNNs give comparable or better performance relative to their uncompressed counterparts. We demonstrate that our method leads to state-of-the-art performance on the LibriSpeech dataset, producing a 16\% reduction in speaker verification error while simultaneously allowing for a 200-fold compression in model parameters.

\section{Related work}
The compression of deep neural networks (DNNs) has been of interest for a long time. It has been shown that DNNs are typically parameterized in a redundant fashion, allowing the prediction of values of some parameters of a trained model given knowledge of the others~\cite{denil2013}.

Several approaches depend on some kind of post-processing after a large model has been trained. Model distillation~\cite{ba2014,hinton2015} for example is a successful technique which retrains a smaller model by using the output activations of the trained large model as labels, instead of the actual data labels. This was found to result in smaller models that are fast to train and match the performance of the larger models from which they were distilled. Quantization is another post-processing technique which uses a more coarse-grained representation for each parameter value, thereby reducing the memory needed to store a trained model's parameters. One complication with these post-processing methods is that they are not end-to-end; the process of pruning the DNN is separate from training.

Matrix and tensor factorization techniques provide an alternative that is end-to-end trainable. A natural first step is to decompose parameter weight matrices in a low-rank matrix factorized format. This was done in~\cite{sainath2013} to compress the last fully-connected layer of a convolutional neural network (CNN). Restricting to the last layer is limited in the compression achieved, since the other layers of the network themselves contain many parameters. However, utilizing this approach in internal layers results in a lower effective number of hidden units, ultimately hurting accuracy.

Tensor factorization methods, such as that employed here, can generally be used to decompose matrices in higher-dimension space, and were used in~\cite{yu2017} to capture higher order interactions in dynamical processes. The idea of tensorizing neural networks in an end-to-end trainable manner using tensor-train decomposition was first introduced in~\cite{novikov2015}, where a fully-connected layer was reshaped and factorized as a tensor in TT format to achieve impressive compression. The extension to convolutional layers was later given in~\cite{garipov2016ultimate}.

Different aspects of these ideas were extended to recurrent neural networks (RNN) in~\cite{yang2017tensor,tjandra2017}. The work of~\cite{yang2017tensor} applied tensor-train layers to the large encoding matrices used for high-dimensional video input, allowing for simultaneous compression and improved performance in video classification. This was later followed by~\cite{yin2020}, which reported further gains through the use of the more complex Hierarchical Tucker decomposition in place of tensor trains. 

By contrast, \cite{tjandra2017} tensorized RNNs by assigning a separate TT matrix to each of the separate weight matrices in a recurrent cell, with a focus on GRU models. This allowed significant compression to be achieved not only with high-dimensional inputs, but also with  high-dimensional hidden states. Our work is similar to \cite{tjandra2017}, but achieves further compression by jointly tensorizing the weights within each RNN cell. We show how this process leads to a novel form of weight sharing, which is verified experimentally to have tangible benefits for performance and compression. The use of a tensor-train parameterization is shown to represent an implicit regularization capable of improving training and generalization. Our TT-RNN model is available as open-source code, and can be used as a drop-in replacement for standard RNN models.

\section{Recurrent neural networks and the tensor-train decomposition}
In this section we first give an overview of common RNN architectures, before introducing the tensor-train decomposition and describing its use for ``tensorizing'' large weight matrices within neural networks. \\

\noindent\textbf{Notation.}
We use bold lower-case letters $\va$ to denote vectors, bold upper-case letters $\mW$ to denote matrices, and bold calligraphic letters to denote tensors $\tT$. Tensor elements are indexed as $\va(i)$, $\mW(i,j)$ and $\tT(i_1, i_2,..,i_d)$, for the respective cases of vectors, matrices, and more general $d$'th-order tensors. The notation $\va * \vb$ represents the element-wise Hadamard product between vectors of equal size. The collection of integers $\{1, 2, \ldots, n\}$ is denoted as $[n]$.

\subsection{Architectures}
\label{sec:rnn-architectures}
Recurrent neural networks (RNN) define a paradigm for learning from sequential data. The recurrent unit of an RNN defines an iterative procedure whose outputs and hidden state at each time step $t$ are a non-linear function of $\vx_{t}$, the input at $t$, and $\vh_{t-1}$, the hidden state at time $t-1$. Many different functions have been proposed for this nonlinear recurrent unit, and we describe two representative choices, long short-term memory (LSTM) and gated recurrent unit (GRU).

\subsubsection{Long short-term memory}
The LSTM cell uses three ``gates'' to control the flow of information, and divides its hidden state into a memory cell state $\vc$ and regular hidden state $\vh$, of identical dimension $D$. These are jointly updated as
\begin{equation}
\begin{split}
    \vc_{t} &= \vu_{t} * \Tilde{\vc}_{t} + \vf_{t} * \vc_{t-1} \\
    \vh_{t} &= \vo_{t} * \tanh(\vc_{t}),
\end{split}
\end{equation}

\noindent where the candidate cell state ($\Tilde{\vc}_{t}$), update gate ($\vu_{t}$), forget gate ($\vf_{t}$), and output gate ($\vo_{t}$) vectors are given by
\begin{equation}
\begin{split}
\label{eq:lstm_gates}
    \Tilde{\vc}_{t} &= \tanh(\mW^{(c)} \vx_{t} + \mU^{(c)} \vh_{t-1} + \vb^{(c)})  \\
    \vu_{t} &= \sigma(\mW^{(u)} \vx_{t} + \mU^{(u)} \vh_{t-1} + \vb^{(u)}) \\
    \vf_{t} &= \sigma(\mW^{(f)} \vx_{t} + \mU^{(f)} \vh_{t-1} + \vb^{(f)}) \\
    \vo_{t} &= \sigma(\mW^{(o)} \vx_{t} + \mU^{(o)} \vh_{t-1} + \vb^{(o)}).
\end{split}
\end{equation}

In the above, $\vx_{t} \in \mathbb{R}^M$ and $\vh_{t} \in \mathbb{R}^D$ are the input and hidden state vectors respectively, while  $\mW^{(c)}, \mW^{(u)}, \mW^{(f)}, \mW^{(o)} \in \mathbb{R}^{D \times M}$ are the input-hidden transition matrices, and $\mU^{(c)}, \mU^{(u)}, \mU^{(f)}, \mU^{(o)} \in \mathbb{R}^{D \times D}$ are the hidden-hidden transition matrices.

\subsubsection{Gated recurrent unit}
The GRU is defined by two (update and relevance) gates and a single hidden state
\begin{equation}
    \vh_{t} = \vu_{t} * \Tilde{\vh}_{t} + (1 - \vu_{t}) * \vh_{t-1},
\end{equation}

\noindent where
\begin{equation}
\begin{split}
\label{eq:gru_gates}
    \Tilde{\vh}_{t} &= \tanh(\mW^{(h)} \vx_{t} + \mU^{(h)} (\vr_{t} * \vh_{t-1}) + \vb^{(h)})  \\
    \vu_{t} &= \sigma(\mW^{(u)} \vx_{t} + \mU^{(u)} \vh_{t-1} + \vb^{(u)}) \\
    \vr_{t} &= \sigma(\mW^{(r)} \vx_{t} + \mU^{(r)} \vh_{t-1} + \vb^{(r)}).
\end{split}
\end{equation}

The number of parameters for either of the above RNN units is $g D (M + D)$, where $g$ is the number of distinct gates, which is 4 for an LSTM and 3 for a GRU. Given any factorization of the input and hidden dimensions into positive integers as $D = \prod_{k=1}^n d_k$ and $M = \prod_{k=1}^n m_k$ (where $d_k, m_k \geq 1$), this parameter count can be expressed as
\begin{align}
\label{eq:params_rnn}
    N_{dense} &= g D (M + D) = g \left(\prod_{k=1}^n d_k m_k + \prod_{k=1}^n d_k^2 \right) \nonumber \\
    &= \bigO(d^n (m^n + d^n)),
\end{align}

\noindent where $d = \max_k d_k$ and $m = \max_k m_k$. This version of the parameter count will allow for an easier comparison of typical RNN models with the tensorized RNNs introduced below.

\subsection{Tensor-train decomposition}
The tensor-train (TT) decomposition, introduced in~\cite{oseledets2011} and equivalent to the earlier matrix product state model of many-body physics~\cite{vidal2003}, gives a method for representing higher-order tensors as a type of iterated low-rank factorization. A TT representation of an $n$th-order tensor $\tT \in \R^{p_1 \times p_2 \times \cdots \times p_n}$ is a tuple of $n$ tensors $\tGn = (\tG_1,\tG_2,...,\tG_n)$, called the TT cores. Each core has dimension $\tG_k \in \R^{p_k \times r_{k-1} \times r_k}$, where the $r_k$ for $k \in \{1, \ldots, n-1\}$ are hyperparameters called the TT ranks of the model. Given a collection of TT cores, the tensor $\tT$ associated with these cores has elements given by the following vector-matrix-vector products
\begin{equation}
\begin{split}
\label{eqn:tt_format}
    \tT(i_1, i_2, \ldots, i_n) &= \tG_1(i_1) \tG_2(i_2) \cdots \tG_d(i_n),
\end{split}
\end{equation}
where $\tG_k(i_k) = \tG_k(i_k, :, :) \in \R^{r_{k-1} \times r_k}$ indicates an index-dependent matrix associated with the $k$th core, with each $i_k \in [p_k]$ and $r_0, r_n$ each taken to be 1. We will refer to $\tT$ as the ``global'' tensor encoded by the TT cores, which constitute a ``local'' representation of $\tT$.

The TT decomposition is capable of exactly representing any $n$th-order tensor given sufficiently large TT ranks using the TT-SVD procedure of~\cite{oseledets2011}, but a more common practice is to fix the TT ranks at small values and use the core tensors as a compact parameterization which is optimized to minimize some loss function defined on the global tensor. This approach is not limited to cases where higher-order tensors are already present, as any vector $\vv$ with dimension $P = \prod_{k=1}^n p_k$ can be reshaped into an $n$th order tensor $\tV \in \R^{p_1 \times \cdots \times p_n}$. Such ``TT vectors'' provide an efficient description requiring only $\sum_{k=1}^n p_k r_{k-1} r_k = \bigO(\log(P))$ parameters when all TT ranks $r_k$ and core dimensions $p_k$ are bounded, compared with $P$ parameters for a dense representation.

The same procedure can be applied to matrices of shape $D \times M$ when $D = \prod_{k=1}^n d_k$ and $M = \prod_{k=1}^n m_k$, yielding a TT matrix defined by $n$ tensor cores. In this case we choose each TT core $\tG$ to have four indices with respective dimensions $d_k$, $m_k$, $r_{k-1}$, and $r_k$, and denote the associated index-dependent matrices by $\tG_k(i_k, j_k) = \tG_k(i_k, j_k, :, :) \in \R^{r_{k-1} \times r_k}$, for $i_k \in [d_k]$ and $j_k \in [m_k]$.


\subsection{Tensorizing neural networks}
\label{sec:tensorizing_nn}
The bulk of the parameters in a neural network consist of large weight matrices represented in dense format. It was shown in~\cite{novikov2015} that the representation of these matrices as TT matrices allowed for a significant reduction in parameter count, while introducing little or no additional error in the performance of the network.

Given a weight matrix $\mW$ of shape $D \times M$, where $D = \prod_{k=1}^n d_k$ and $M = \prod_{k=1}^n m_k$, then the affine transformation implemented as part of a typical neural network layer takes the form $\vy = \mW \vx + \vb$. In a tensorized neural network, $\vx, \vy, \vb$ are represented normally as dense vectors, while the weight matrix $\mW$ is represented in TT form. The affine transformation is carried out by first using multilinear tensor contractions to perform the multiplication $\mW \vx$, with $\vx$ reshaped into a dense $n$th order tensor $\tX$, and then using standard dense addition for the bias vector $\vb$. The output vector $\vy$ can be described in reshaped form as the tensor $\tY$ with elements
\begin{multline}
\label{eq:tt_layer}
    \tY(i_1, \cdots\!, i_n) = \tB(i_1, \cdots\!, i_d)\ + \\
    \sum_{j_1, \cdots, j_n}\!\!\! \left(\tG_1(i_1,j_1) \cdots \tG_d(i_d,j_d)\right) \tX(j_1, \cdots\!, j_d).
\end{multline}

By carrying out the above summations (including those implicit in the matrix-vector products) in an optimal order,~\eqref{eq:tt_layer} can be evaluated with a total cost of $\bigO(n r^2 d M)$, where $r = \max_k r_k$. In the typical setting where $r$, $m$, and $d$ remain bounded as $D$ and $M$ are increased, this cost is $\bigO(\log(\max(D, M)) M)$, compared to $\bigO(D M)$ for the usual affine map. This representation is also compact, requiring only $\bigO(n r^2 d m) = \bigO(\log(\max(D, M)))$ parameters, compared to $\bigO(D M)$ parameters for a dense representation.

For clarity, we refer to a fully-connected layer represented in tensor-train form as a tensor-train layer (TTL), and denote the linear portion of the operation implemented in~\eqref{eq:tt_layer} as $\TTL(\vx; \tGn)$.

\section{Fully tensorized RNNs}

\subsection{Tensorizing RNNs}
\label{sec:tensorizing_rnns}
We describe a straightforward application of the tensorization procedure described in section \ref{sec:tensorizing_nn} to LSTM models which allows for a significant reduction in the models' parameter count~\cite{tjandra2017,yang2017tensor}. In our work, we propose an extension of this procedure which permits an even greater degree of compression to be attained. 


An LSTM recurrent unit contains 8 weight matrices, each providing contributions to one of the four independent gate vectors coming from an input vector $\vx_{t}$ or previous hidden vector $\vh_{t}$. When these matrices are replaced by tensor-train matrices,~\eqref{eq:lstm_gates} can be re-written as
\begin{equation}
\begin{split}
\label{eq:vanilla_ttlstm}
    &\Tilde{\vc}_{t} = \tanh(\TTL(\vx_{t}; \tGn^{(Wc)}) + \TTL(\vh_{t-1}; \tGn^{(Uc)}) + \vb^{(c)})  \\
    &\vu_{t} = \sigma(\TTL(\vx_{t}; \tGn^{(Wu)}) + \TTL(\vh_{t-1}; \tGn^{(Uu)}) + \vb^{(u)}) \\
    &\vf_{t} = \sigma(\TTL(\vx_{t}; \tGn^{(Wf)}) + \TTL(\vh_{t-1}; \tGn^{(Uf)}) + \vb^{(f)}) \\
    &\vo_{t} = \sigma(\TTL(\vx_{t}; \tGn^{(Wo)}) + \TTL(\vh_{t-1}; \tGn^{(Uo)}) + \vb^{(o)}).
\end{split}
\end{equation}

Each of the 8 weight matrices $V_e$ (where $V$ is one of $W$ or $U$, and $e$ is one of $c$, $u$, $f$, or $o$) is replaced by its own collection of tensor-train cores $\tGn^{Ve}$, and we assume for simplicity that the same factorization of $D = \prod_{k=1}^n d_k$ and $M = \prod_{k=1}^n m_k$ is used for each of the 8 tensor-train matrices.

For a tensorized gated RNN with $g$ gates and an identical factorization for each tensor-train matrix, such as the LSTM above, the total parameter count is
\begin{align}
\label{eq:params_vanilla_ttrnn}
    N_{TT1} &= g \sum_{k=1}^n r_{k-1} r_k d_k (m_k + d_k) \nonumber \\
            &= \bigO(g n r^2 d (m + d)).
\end{align}

Although the exact comparison of this count to~\eqref{eq:params_rnn} depends on the TT ranks $r_k$ and the number of cores $n$ employed, it is clear that for the typical case where $r, d, m \ll \min(M, D)$, a tensorized RNN will require significantly fewer parameters. However, the use of a separate TT matrix for each gate in the RNN unit still leads to a multiplicative factor of $g$ in~\eqref{eq:params_vanilla_ttrnn}.

We now introduce a different tensorization method, where a tensor-train factorization is applied to entire collections of concatenated weight matrices, rather than to individual matrices. The efficient nature of the tensor-train decomposition leads to a further reduction in model parameters, with LSTMs requiring approximately four times fewer parameters compared to the tensorization above. We show more generally that gated RNNs with $g$ gates exhibit a roughly $g$-fold reduction in the parameter count with this method, on top of the already sizable reduction coming from the use of tensor-train matrices.

\subsection{Gate concatenation}
We achieve further compression of our tensorized RNN by jointly tensorizing the input-hidden weights, as well as the hidden-hidden weights. Taking the LSTM as an example, we first take the row-wise concatenation of the four input-hidden matrices $\mW^{(c)}, \mW^{(u)}, \mW^{(f)}, \mW^{(o)} \in \R^{D \times M}$, which gives a single input-hidden matrix $\mW \in \R^{4D \times M}$. More concretely, the concatenated weight matrices utilized are
%
%
\begin{equation}
\begin{split}
    &\mW = [ \mW^{(c)}, \mW^{(u)}, \mW^{(f)}, \mW^{(o)}]^T, \\
    &\mU = [ \mU^{(c)}, \mU^{(u)}, \mU^{(f)}, \mU^{(o)}]^T.
\end{split}
\end{equation}

For regular LSTMs with dense weight matrices, this concatenation gives a means of replacing four separate matrix-vector multiplications by a single larger multiplication, permitting greater parallelism. After the single vector $\mW \vx$ is computed it can be split into four equal-sized pieces, each holding the value of one of the gate vectors.

\begin{figure}[t]
    \centering
    \includegraphics[width=.7\columnwidth]{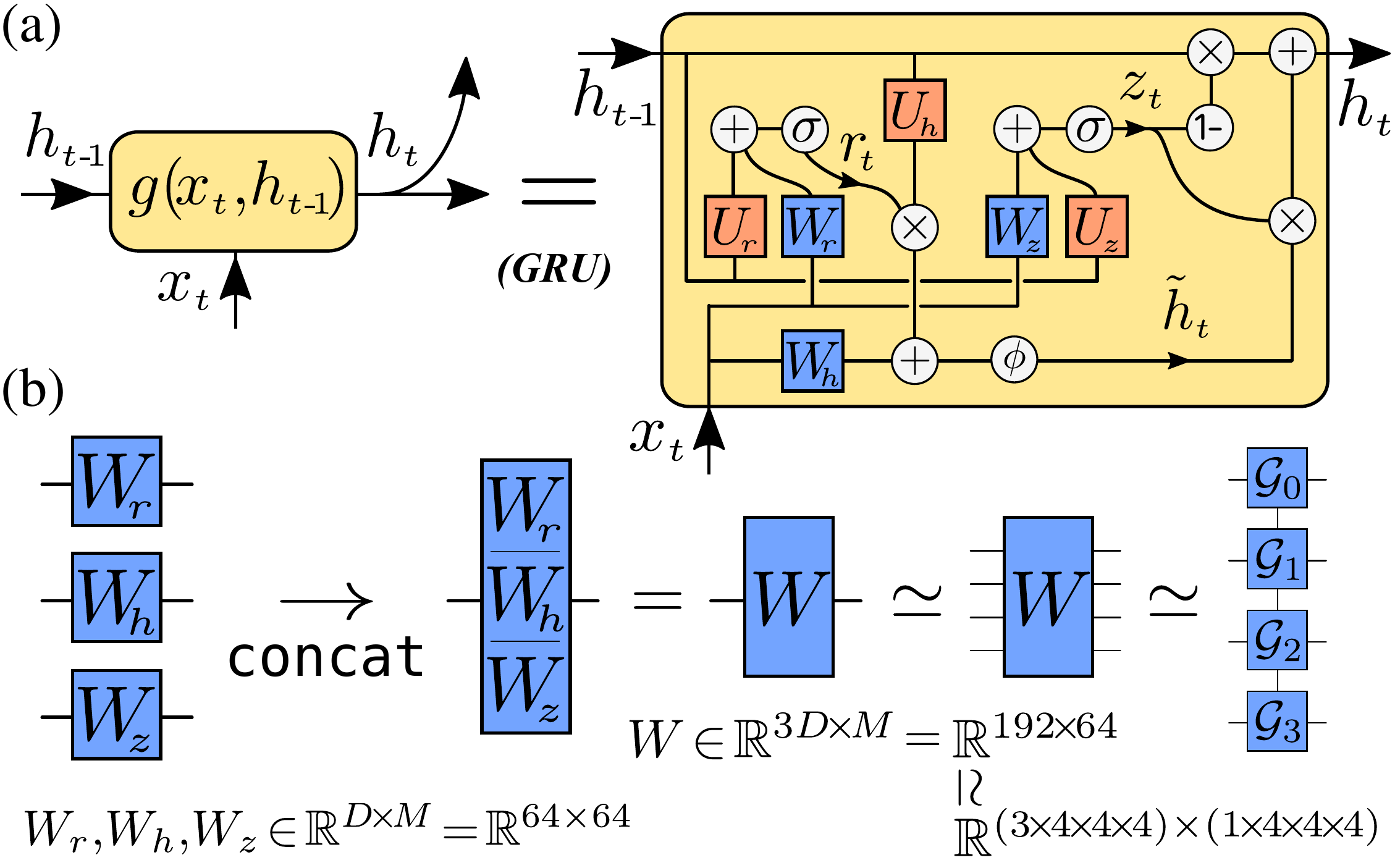}
    \caption{Illustrated of tensorization process on GRU cell with hidden and input dimensions $D = M = 64$.\\ \textbf{(a)} Layout of the recurrent update function $g(x_{t}, h_{t-1})$ for a GRU, with biases omitted for simplicity. Weight matrices are shown in blue and orange, with matrices of the same color having the same shape. In traditional RNNs, these weight matrices are parameterized as separate dense matrices. \textbf{(b)} Our compression process involves first concatenating all matrices of the same type, then tensorizing this composite matrix by parameterizing it as a TT matrix. For the given case, the stacked matrix $W \in \R^{192 \times 64}$ is represented as a tensor $\tT \in \R^{(3\times4\times4\times4) \times (1\times4\times4\times4)}$, which in turn is represented by the contraction of four TT cores $\tG_k \in \R^{d_k \times m_k \times r_{k-1} \times r_k}$. In the particular case shown, the bottom cores $\tG_k$ for $k = 1, 2, 3$ give a family of $r_0$ matrices $\mM_\alpha \in \R^{64 \times 64}$ jointly represented in TT format, while the top core $\tG_0$ acts as a matrix assigning each of the GRU gate matrices to a linear mixture of the TT matrices $\mM_\alpha$).
    }
    \label{fig:tensorized_rnn_cell}
\end{figure}

When the concatenated weight matrices are represented as a tensor-train layer, this leads to the revised LSTM gate equations,
\begin{equation}
\label{eq:ttlstm}
\begin{split}
    \Tilde{\vc}_{t} &= \tanh(\TTL(\vx_{t}; \tG^{W})_{1} + \TTL(\vh_{t-1}; \tG^{U})_{1} + \vb^{(c)})  \\
    \vu_{t} &= \sigma(\TTL(\vx_{t}; \tG^{W})_{2} + \TTL(\vh_{t-1}; \tG^{U})_{2} + \vb^{(u)}) \\
    \vf_{t} &= \sigma(\TTL(\vx_{t}; \tG^{W})_{3} + \TTL(\vh_{t-1}; \tG^{U})_{3} + \vb^{(f)}) \\
    \vo_{t} &= \sigma(\TTL(\vx_{t}; \tG^{W})_{4} + \TTL(\vh_{t-1}; \tG^{U})_{4} + \vb^{(o)}),
\end{split}
\end{equation}

\noindent where $\TTL(\vx_{t}; \tG^{W})_i$ and $\TTL(\vh_{t-1}; \tG^{U})_i$ are the $i$th equally-sized vectors in the TT matrix-vector products associated with $W$ and $U$, which contribute to the $c$, $u$, $f$, and $o$ gates. This process can be carried out analogously for a gated RNN with $g$ gates, where the matrices $\mW \in \R^{gD \times M}$, $\mU \in \R^{gD \times D}$ are each concatenations of $g$ separate matrices. An example of this process for a GRU model is given in Figure~\ref{fig:tensorized_rnn_cell}.


\subsection{Compression and runtime}

When tensorizing the individual weight matrices of an RNN in Section~\ref{sec:tensorizing_rnns}, the hidden and input dimensions were factored into $n$ smaller terms, as $D = \prod_{k=1}^n d_k$ and $M = \prod_{k=1}^n m_k$. For the case of concatenated weight matrices $\mW$ and $\mU$, a closely related factorization can be employed, namely $gD = \prod_{k=0}^n d_k$ and $M = \prod_{k=0}^n m_k$, where we take $d_0 = g$ and $m_0 = 1$, along with identical $d_k, m_k$ for all $k \geq 1$. 

Taking $\mW$ as an example, a tensor-train decomposition relative to this augmented factorization will give the collection of $n+1$ cores $\tGno^{(W)} = (\tG^{(W)}_0, \tG^{(W)}_1, \ldots, \tG^{(W)}_n)$, where the cores $\tG^{(W)}_k$ for $k > 1$ are shaped identically to a tensor-train factorization of any one of the single-gate weight matrices. The single new core appearing in this decomposition has a shape of $\tG^{(W)}_0 \in \R^{g\times 1\times 1\times r_0}$, for a new TT rank parameter $r_0$, and removing the singleton indices gives a matrix $\mV^{(W)} \in \R^{g \times r_0}$. This leads to a revised parameter count of
\begin{align}
\label{eq:params_ttrnn}
    N_{TT2} &= g r_0 + \sum_{k=1}^n r_{k-1} r_k d_k (m_k + d_k) \nonumber \\
            &= \bigO(n r^2 d (m + d)),
\end{align}

\noindent giving a compression ratio approximately $g$ times greater than~\eqref{eq:params_vanilla_ttrnn}. Using an example model in Table~\ref{tab:numerical_example}, we illustrate the level of compression and speedup in inference time that can be obtained for different configurations of our fully tensorized RNNs. This shows particular promise for the application of RNN models in settings with limited resources, such as edge devices. Finally, the training time for TT-RNNs is comparable to untensorized RNNs, although with a clear dependence on the TT rank.

Some intuition for this parameter reduction can be gained by interpreting the concatenated global matrix $\mW$ encoded by the TT cores $\tGno^{(W)}$ in terms of the small matrix $\mV^{(W)}$ coming from the first core $\tG^{(W)}_0$. Seen this way, the contraction of the remaining TT cores $(\tG^{(W)}_1, \ldots, \tG^{(W)}_n)$ gives a tensor which encodes a family of $r_0$ matrices $\{\mM_\alpha \in \R^{D \times M}\}_{\alpha=1}^{r_0}$. Contracting all of the TT cores (including $\tG^{(W)}_0$) and selecting the $i$th subspace then gives a single-gate weight matrix $\mW_i$, which corresponds to the linear mixture of matrices
\begin{equation}
\label{eq:weight_sharing}
    \mW_i = \sum_{\alpha=1}^{r_0} \mV^{(W)}_{i, \alpha} \mM_\alpha.
\end{equation}

Since all of the matrices $\mW_i$ are jointly encoded as a collection of $n$ TT cores whose matrix dimensions are identical to those of a single tensorized gate matrix, specifying the weight matrices for all $g$ gates in this manner requires a comparable number of parameters to specifying a single weight matrix in TT format.

\begin{table}[t]
\caption{Comparison of model size and per-step training and inference times of RNNs and TT-RNNs. Each model has a single recurrent layer with hidden size of 512, a linear projection layer of embedding size 256, and input dimension of 4,096. Each TT-RNNs has 2 cores, and $r$ denotes the TT rank. For both LSTM and GRU models, the tensorized versions achieve significant compression of model parameters while reducing the inference time and, for smaller values of $r$, decreasing the training time. All reported times were obtained on an Intel(R) Xeon(R) CPU E5-1650 v3 @ 3.50GHz with 128GB of RAM, and averaged over 100 runs.} \smallskip
\label{tab:numerical_example}
\centering
\resizebox{.75\columnwidth}{!}{
\smallskip\begin{tabular}{|c|crrr|}
\hline
Model & $r$ & \# params & Train time (s) & Eval. time (s) \\
 \hline
LSTM & $-$ & 9,570,560 & $12.84 \pm .17$ & $3.70 \pm .19$  \\
\hline
& 2 & 21,248 & $9.37 \pm .11$ & $2.13 \pm .13$ \\
TT-LSTM & 3 & 30,720 & $11.92 \pm .22$  & $2.23 \pm .13$ \\
& 4 & 40,192 & $15.55 \pm .37$  & $2.48 \pm .25$  \\
\hline
GRU & $-$ & 7,212,288 & $10.12 \pm .26$ &  $2.53 \pm .07$ \\
\hline
& 2 & 19,200 & $8.09 \pm .21$ & $1.43 \pm .09$ \\
TT-GRU & 3 & 27,136 & $9.18 \pm .15$  & $1.59 \pm .08$  \\
& 4 & 35,072 & $11.23 \pm .30$ & $1.80 \pm .10$  \\
 \hline
\end{tabular}}
\end{table}

\section{Experiments and results}
We consider the task of speaker verification for evaluating the proposed factorized RNNs. Only results for TT-LSTM are discussed. Those for TT-GRU are identical. Beyond assessing the accuracy in these tasks, we characterize trade-offs between compression and accuracy arising from different choices of TT rank and core layout. In the process, we find that the tensor-train parameterization acts as a form of regularization, leading to improved stability and generalization during training.

For simplicity and ease of comparison, all models in the following are trained without explicit regularization such as dropout, weight decay, or gradient clipping. The tensorized models were written in PyTorch~\cite{paszke2019} using the tensor-train implementation from~\cite{khrulkov2019}, and are available on GitHub\footnote{https://github.com/onucharles/tensorized-rnn}.

\subsection{Sequential digit classification}
We first evaluate the TT-LSTMs on the permuted sequential MNIST task~\cite{lecun1998mnist} in which the $28 \times 28$ pixel images of handwritten digits are randomly rearranged using a fixed permutation into sequences of length 784. These are split into 50k training, 10k validation, and 10k test images, with the validation dataset used to determine the end of training by early stopping. 

The LSTM and TT-LSTM were each chosen as single-layer models with 256 hidden units. Training was performed with a batch size of 256 and Adam optimizer, using a piecewise constant learning rate starting at 0.001.

Table~\ref{tab:permuted-mnist-perf} reports the digit classification accuracy, where the hidden dimensions of the TT-LSTM are factored into either 2 or 3 TT cores using TT ranks of 2, 4, or 6 to connect adjacent cores. Although a clear tradeoff is present between compression and accuracy, even the largest TT-LSTM utilizes 46 times fewer parameter in total, while achieving comparable performance to the LSTM baseline ($-0.28\%$ classification accuracy).

\begin{table}[h]
\caption{Comparison of TT-RNN and standard RNN models on the permuted pixel MNIST task. The models use a single-layer containing $D = 256$ hidden units, and are trained identically. The performance of the TT-RNNs varies with the parameter count, but achieves comparable accuracy to a standard RNNs while maintaining a compression ratio of 46 times and 25 times fewer parameters, in the TT-LSTM and TT-GRU respectively.}\smallskip
\label{tab:permuted-mnist-perf}
\centering
\resizebox{.7\columnwidth}{!}{
\smallskip\begin{tabular}{|c|c|crrr|}
\hline
Model   & Cores & $r$ & \#Params & Compr. & Acc. (\%) \\
\hline
LSTM    & $-$   & $-$ & 266,762 & $-$ & 89.77     \\
\cline{1-6}
\multirow{6}{*}{TT-LSTM} & \multirow{3}{*}{2} & 2   &  3,434 &  78  &  87.98    \\
 &     & 4   & 5,834   &  46  &   89.49   \\
 &  & 6  &  8,234  &  32  &   89.22  \\
\cline{2-6}
 & \multirow{3}{*}{3} & 2   & 1,842   & 145   &   85.36   \\
 &     & 4   &  3,354  &  80  &  87.18   \\
 &  & 6  &  5,570  &  48  & 89.30    \\
\hline
GRU    & $-$   & $-$ & 201,482 & $-$ & 91.49     \\
\cline{1-6}
\multirow{6}{*}{TT-GRU} & \multirow{3}{*}{2} & 2   &  3,674 &  55  &  87.94    \\
 &     & 4   & 5,802   &  35  &   89.29   \\
 &  & 6  &  7,930  &  25  &   90.26  \\
\cline{2-6}
 & \multirow{3}{*}{3} & 2   & 2,282   &  88  &   87.62   \\
 &     & 4   &  3,722  &  54  &  88.90   \\
 &  & 6  & 5,866  &  34  & 89.80    \\
\hline
\end{tabular}}
\end{table}

\subsection{Speaker verification}

In the speaker verification problem, the objective is to ascertain if an utterance of speech belongs to a given individual, based on a collection of utterances labeled by individuals. We use the LibriSpeech dataset, containing around 1,000 hours of English language audiobook recordings~\cite{panayotov2015librispeech}, where training, validation, and testing are carried out on the {\em train-clean-100}, {\em dev-clean}, and {\em test-clean} partitions. 

Our model for speaker verification contains two main components, an utterance encoder and a similarity function, as in~\cite{heigold2016end,xie2019utterance}. The utterance encoder consists of an RNN which computes fixed-dimensional embeddings from spectograms of input utterances, while the similarity function assigns similarity scores to pairs of embeddings. 

We use the generalized end-to-end (GE2E) loss function~\cite{wan2018generalized} to train the model, which encourages embeddings of utterances to cluster based on the associated speaker. Given an embedding vector $\ve_{ji}$ for the $i$th utterance by the $j$th speaker, the GE2E loss is
\begin{equation}
\begin{split}
    L(\ve_{ji}) &= -\mS_{ji,j} + \log \sum_{k=1}^N \exp(\mS_{ji,k}), \\
\end{split}
\end{equation}
where $\mS_{ji,j} = w \cdot \cos(\ve_{ji}, \vc_k) + b$ is the scaled cosine similarity between the embedding $\ve_{ji}$ and the centroid of the embeddings of speaker $j$, denoted $\vc_j$. The scaling coefficients $w$ and $b$ are initialized to $10$ and $-5$ respectively. The full loss is then the sum of all utterance-specific losses, $L = \sum_{j,i} L(\ve_{ji})$. 

We report performance in the speaker verification task using the equal error rate (EER) metric, which is the error rate on the receiver-operating characteristic (ROC) curve when the false positive rate and false negative rates are equal.

\subsection{Performance}

Our utterance encoder consists of a single-layer LSTM with hidden size of 768, whose output is converted to an embedding of dimension 256 using a fully-connected linear layer. The input to this encoder is 40-bin $\times$ 160-frame Mel spectograms of utterances. We compare regular LSTMs and TT-LSTMs for these identical input, hidden, and embedding dimensions, as given in Table~\ref{tab:speaker_verif_perf}.

Using a standard LSTM in the encoder gives an EER of 7.33\%, similar to the performance found in~\cite{zhou2019training}. By contrast a TT-LSTM encoder led to significantly better EERs, with the best configuration achieving an EER of 4.34\%. This increased accuracy was accompanied by a reduction in the total parameter count, from 2.6M parameters to only 13K. By reducing the TT rank, this parameter count can be further reduced while still maintaining higher accuracy than the LSTM baseline.

\begin{table}[tbp]
\caption{Performance of RNNs and TT-RNNs on the task of speaker verification. Models have a single layer with 768 hidden units and a linear projection layer of 256. The lowest ranked TT-RNNs outperform the RNNs on this more challenging task of speaker verification, achieving larger compression ratios of 653 (TT-LSTM) and 369 (TT-GRU). \cut{The TT-RNNs have a hidden shape of ($16 \times 16$) when 2 cores and ($4 \times 8 \times 8$) when 3 cores.} EER is the equal error rate (lower is better).}\smallskip
\label{tab:speaker_verif_perf}
\centering
\resizebox{.7\columnwidth}{!}{
\smallskip\begin{tabular}{|c|c|crrr|}
\hline
Model   & Cores & $r$ & \#Params & Compr. & EER (\%) \\
\hline
LSTM    & $-$   & $-$ & 2,682,114 & $-$   & 7.33     \\
\cline{1-6}
\multirow{6}{*}{TT-LSTM} & \multirow{3}{*}{2} & 1   &  8,178 &  328  &  4.71    \\
 &      & 2   & 13,026  &  206  &  4.34    \\
 &     & 4   &  22,722  & 118   & 6.21  \\
\cline{2-6}
 & \multirow{3}{*}{3} & 1   &  4,106 & 653   &  6.09    \\
 &      & 2   & 5,394  &  497  & 5.31     \\
 &     & 4   &  9,506  &  282  & 5.38     \\
\hline
GRU    & $-$   & $-$ & 2,063,106 & $-$   & 7.87     \\
\cline{1-6}
\multirow{6}{*}{TT-GRU} & \multirow{3}{*}{2} & 1   &  9,074 &  227  &  5.31    \\
 &      & 2   & 13,282  & 155   &  6.72    \\
 &     & 4   &  21,698  &  95  & 5.36  \\
\cline{2-6}
 & \multirow{3}{*}{3} & 1   &  5,594 & 369   &  6.46    \\
 &      & 2   & 6,738  &  306  & 6.39     \\
 &     & 4   &  10,274  &  201  & 4.48     \\
\hline
\end{tabular}}
\end{table}

\begin{figure}[htbp]
\centering
\includegraphics[width=.7\columnwidth]{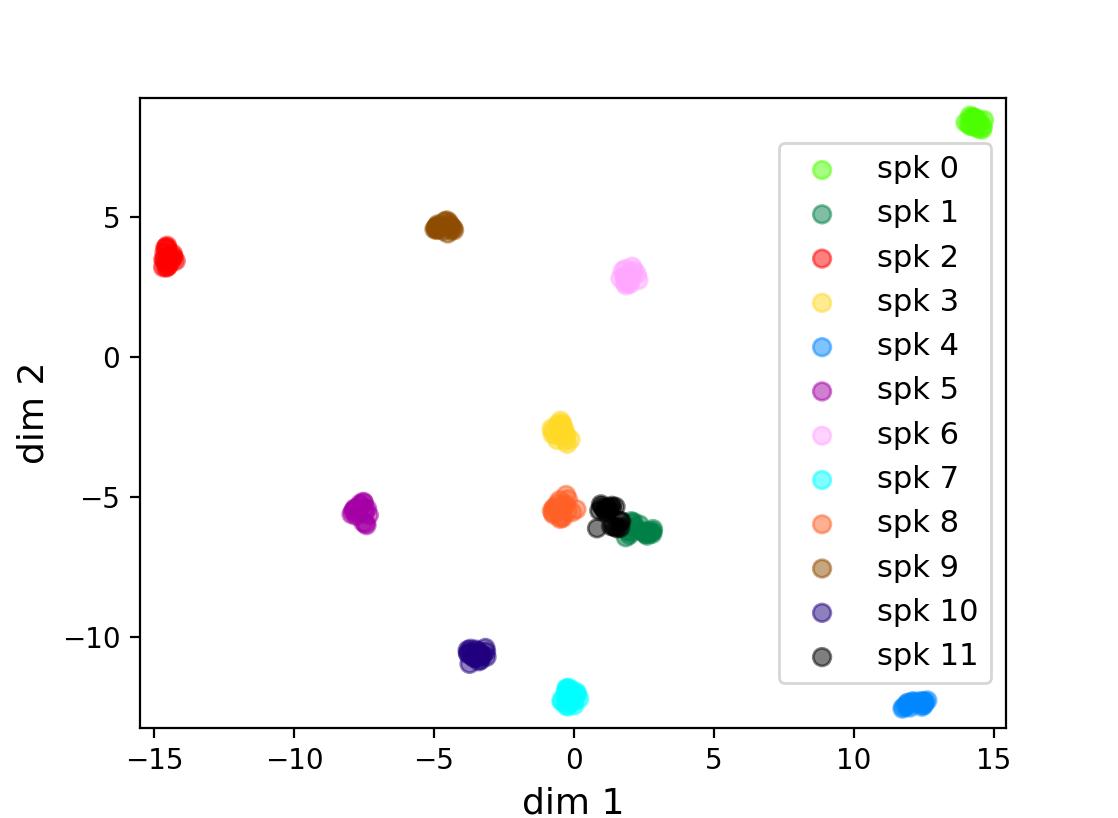}
\caption{Low dimensional UMAP visualization of embeddings from the TT-LSTM. Each datapoint corresponds to the 256-dimensional embedding of an utterance, where colors reflect the identity of different speakers. A clear clustering pattern is seen amongst the utterances from each speaker.}
\label{fig:umap_embeds}
\end{figure}

Analyzing the embeddings learned by the TT-LSTM further demonstrates the performance of the model in speaker verification. We use uniform manifold approximation and projection (UMAP)~\cite{mcinnes2018umap} to project the 256-dimension embedding vectors into 2D space (Figure~\ref{fig:umap_embeds}), which shows that the embeddings learned by the TT-LSTM effectively cluster the utterances from each speaker.

\subsection{Regularization}

\begin{figure*}[h]
\centering
\includegraphics[width=1\textwidth]{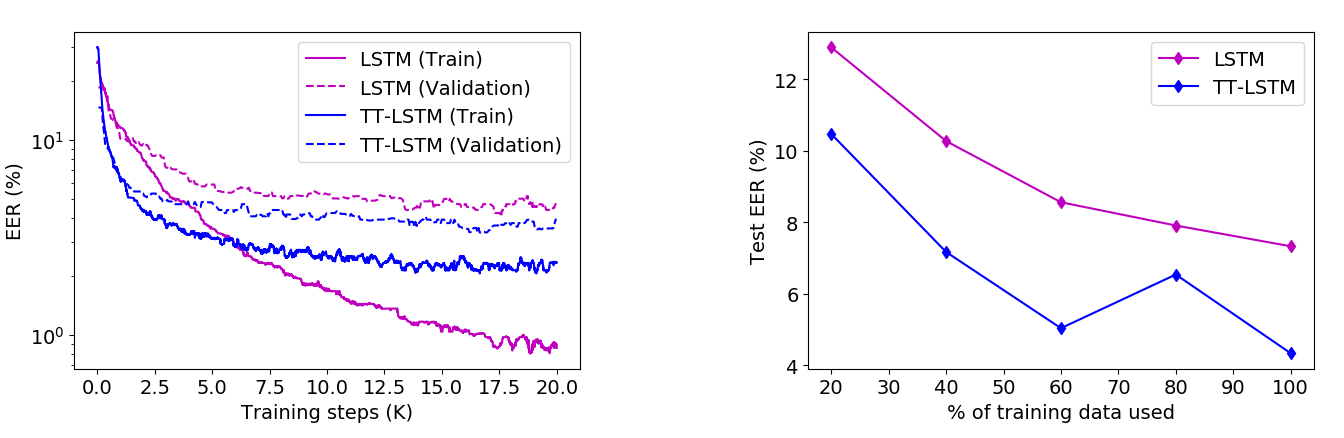}
\caption{Illustration of regularization benefits of TT-RNNs. \textbf{Left}: Learning curves for the best LSTM and TT-LSTM models. The use of tensor-train weights acts as an implicit regularizer, raising the training error while reducing the margin between training and validation EERs of the TT-LSTM compared to the LSTM. \textbf{Right}: Performance of the models using different fractions of the LibriSpeech training set. Each datapoint gives the test EER of the corresponding model after training, and we see the TT-LSTM consistently generalizing better than the standard LSTM.}
\label{fig:regularisation}
\end{figure*}

TT-LSTMs utilize a more compact set of weight parameters, which can be expressed as a low-dimensional family of weight matrices. To assess if this low-dimensional parameterization has benefits for regularization, we first examine the learning curves of TT-LSTMs and standard LSTMs during training (Figure~\ref{fig:regularisation}, left). We observe that while LSTM encoders achieve lower loss during training, this loss is not reflected in the validation loss, likely due to overfitting. By contrast, the TT-LSTM shows better generalization, giving a smaller discrepancy between training and validation loss, and ultimately a lower validation EER.

To further test this generalization, we conduct the speaker verification experiments in a more data-limited setting, using between 20\% and 100\% of the training data. TT-LSTMs consistently performed better than the LSTM baseline when trained with small amounts of data (Figure~\ref{fig:regularisation}, right).

\subsection{Training stability}

\begin{figure}[t]
\centering
\includegraphics[width=.8\columnwidth]{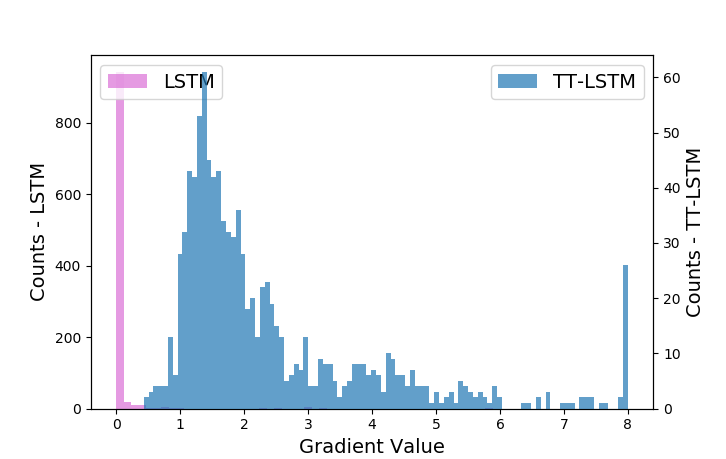}
\caption{Distribution of the norm of gradients of model parameters across 1000 training steps. Model is same configuration as before, but learning rate is increased to 0.01 from 0.001. The LSTM succumbs to the vanishing gradient problem, while the gradients of the TT-LSTM remain distributed over a wide range.}
\label{fig:grad_hist}
\end{figure}

 We observed during the initial hyperparameter search an increased robustness in the performance of TT-LSTMs relative to changes in the learning rate. Both LSTMs and TT-LSTM models were trained at a learning rate of 0.001, but increasing this to 0.01 led to an instability in the former and no noticeable impact on the latter. The distribution of gradients for this case is given in Figure~\ref{fig:grad_hist}. The standard LSTM exhibits vanishing gradients, effectively saturating at 0, while the gradients for TT-LSTM are distributed over a reasonable range. 

\section{Discussion}
The goal of this work was to develop some understanding, using both theoretical and experimental analysis, of what happens when tensor-train (TT) layers are employed in a recurrent speech model. We apply the TT formalism jointly to all weight matrices within the RNN, leading to a ``fully tensorized'' form of weight sharing, where various gate matrices are encoded within a single TT format. We find that:
\begin{enumerate}
    \item TTL facilitates a distinctive form of weight sharing that allows it to learn complex function with far fewer weights (Equation \ref{eq:weight_sharing}).
    \item TTL encourages gradients to be distributed over a wider range of values, thereby reducing the chances of vanishing gradients. (Figure \ref{fig:grad_hist}).
    \item TTL provides clear regularisation benefits, resulting in smaller generalisation error even when using only a fraction of the training data. (Figure \ref{fig:regularisation}).
\end{enumerate}

The findings presented in this chapter highlight the potential of tensor decomposition methods, specifically tensor-train representations, to revolutionize recurrent neural network (RNN) architectures. By enabling significant compression without sacrificing accuracy, the proposed fully tensorized RNN framework demonstrates the viability of deploying efficient and lightweight models in resource-constrained environments. This capability is particularly crucial for real-time applications in medical audio analysis, where portability and processing efficiency are often as important as model accuracy.

Despite these promising advancements, some challenges remain. The training stability of tensorized RNNs under extreme compression, particularly at very high tensor-rank reductions, requires further investigation. Additionally, while this work achieves remarkable parameter reductions, its integration with non-recurrent architectures, such as convolutional or transformer-based models, remains an open avenue for exploration. The comparative scalability of tensor-train RNNs relative to alternative model compression techniques, such as pruning or quantization, also warrants a deeper examination, especially in dynamic settings with varying computational constraints.

Future research should build on these findings to explore hybrid strategies that combine tensorization with other efficiency-enhancing techniques, thereby creating a robust ecosystem of medical AI tools tailored for diverse operational environments.

Our TT-RNN model is available as open-source code, and can be used as a drop-in replacement for standard RNN models\footnote{https://github.com/onucharles/tensorized-rnn}.



\chapter{Self-supervised pre-training for infant cry analysis}
\label{chap:ssl-infantcry}
\section{Overview}
In this paper, we explore self-supervised learning (SSL) for analyzing a first-of-its-kind database of cry recordings containing clinical indications of more than a thousand newborns, recruited as part of this study. Specifically, we target cry-based detection of neurological injury as well as identification of cry triggers such as pain, hunger, and discomfort. 
Annotating a large database in the medical setting is expensive and time-consuming, typically requiring the collaboration of several experts over years. Leveraging large amounts of unlabeled audio data to learn useful representations can lower the cost of building robust models and, ultimately, clinical solutions.
In this work, we experiment with self-supervised pre-training of a convolutional neural network on large audio datasets. 
We show that pre-training with SSL contrastive loss (SimCLR) performs significantly better than supervised pre-training for both neuro injury and cry triggers. In addition, we demonstrate further performance gains through SSL-based domain adaptation using unlabeled infant cries. We also show that using such SSL-based pre-training for adaptation to cry sounds decreases the need for labeled data of the overall system.

\section{Introduction}
\label{sec:intro}

Crying is the primary means by which babies communicate with the world.
Researchers have been interested in infant cry analysis since the early 1960s~\cite{wasz1964identification}. 
Cry characteristics may help us to understand basic baby needs (hunger, pain, etc.) and, more importantly, can be analyzed for the early and non-invasive detection of various diseases~\cite{ji2021review}. 
 For example, clinical research has reported that certain infant cry characteristics are correlated with birth asphyxia~\cite{michelsson1977pain}. This multi-causal condition frequently leads to severe health problems, including neurological injury and even death. 
Various methods based on signal processing~\cite{liu2019infant}, statistical modeling~\cite{felipe2019identification,parga2020defining} and deep learning~\cite{onu2019neural,al2019vcmnet,ozseven2023infant,patil2022constant} have been explored for finding clinical and other insights using cry recordings. \\ 
\indent One of the main challenges in baby cry analysis is data acquisition. Today, cry sounds are not part of routine medical records, so obtaining a database requires targeted efforts such as a clinical study. These are expensive to conduct and typically require the collaboration of several hospital staff over the years. Most machine learning (ML) research on pathology detection from cry sounds was done using the Baby Chillanto~\cite{reyes2004system} database, which contains only six patients diagnosed with birth asphyxia.

From an ML problem point of view, cry classification is analogous to general audio classification, where deep convolutional neural networks (CNNs) have excelled as the state-of-the-art. 
Recently,~\cite{10.1109/TASLP.2020.3030497} demonstrated that Pre-trained Audio Neural Networks (PANNs) - large CNNs pre-trained on generic audio - transferred to a wide range of audio pattern recognition tasks outperformed several previous state-of-the-art systems. Since then, PANNs have been widely adopted for various audio tasks, including emotion recognition from speech~\cite{triantafyllopoulos2021role} and COVID-19 detection from cough~\cite{casanova2021transfer}.

Another popular paradigm in audio classification state-of-the-art is self-supervised learning (SSL) - a method to obtain high-quality representations by training on unlabeled data. SSL has revolutionized the fields of Natural Language Processing and Computer Vision and is currently widely adopted in audio processing~\cite{mohamed2022self}. 
A neural network (encoder) pre-trained with SSL can be seen as a non-linear mapping of an audio sequence to a hidden representation - an embedding. 
The embeddings can be used as input to a classifier trained on a specific task with a supervised objective (using labeled data and conventional cross-entropy loss).
This approach is common for benchmarking various SSL models on multiple diverse audio tasks~\cite{yang2021superb, wang2022towards}. 
Recently, a similarity-based contrastive learning method called SimCLR introduced in Computer Vision~\cite{chen2020simple,chen2020big} demonstrated good performance in multiple audio tasks~\cite{wang2022towards,wang2022learning}, including music  analysis~\cite{spijkervet2021contrastive,mccallum2022supervised}.
SimCLR maximizes the similarity between modified (distorted) views of the same object. For audio, such distortion can be done, for example, by mixing random audio samples~\cite{wang2022towards}, spectrogram masking~\cite{park2019specaugment} in~\cite{wang2022learning}, or/and reverberation, pitch shifting, etc~\cite{spijkervet2021contrastive}. 

In this paper, we experiment with PANNs using both supervised and self-supervised pre-training to learn representations for two  downstream tasks. The first task is classifying brain injury (resulting from birth asphyxia), and the second is predicting cry triggers (pain, hunger, discomfort). The methods are tested on a unique clinical database of newborn cries collected by Ubenwa Health in collaboration with hospitals across three countries~\cite{onu1711ubenwa}. 

In addition, we evaluate the impact of SimCLR-based adaptation of PANNs using unlabeled cries inspired by self-supervised domain adaptation in Speech~\cite{chen2021self} and Natural Language Processing~\cite{gururangan2020don}. 
It should be noted that speech and audio SSL state-of-the-art frequently uses transformers instead of CNNs and relies on different learning objectives~\cite{mohamed2022self}. However, our preliminary experiments with some popular pre-trained speech and audio transformers (specifically, Wav2Vec2.0~\cite{baevski2020wav2vec}, HuBERT~\cite{hsu2021hubert}, WavLM~\cite{chen2022wavlm} and  SSAST~\cite{gong2022ssast}) have not shown sufficient improvements but generally required many parameters to be adapted and hyperparameters tuned. We, therefore, focus on CNN and SimCLR, which demonstrated a good balance of accuracy and adaptation complexity.


\section{Methodology}
\label{sec:method}

\subsection{Clinical data acquisition}
Over a period of 3 years, we conducted a prospective, multi-center, international clinical study with the aim of acquiring a large database of labeled infant cry sounds to study the relationship between cries and medical issues. The study, dubbed Ubenwa (meaning “cry of a baby” in Igbo language), involved sites at 5 health networks in Brazil, Canada and Nigeria, namely Santa Casa de Misericordia de Sao Paulo (SCDM), McGill University Health Centre (MUHC), Enugu State University Teaching Hospital (ESUTH), Rivers State University Teaching Hospital (RSUTH) and Lagos State University Teaching Hospital (LASUTH).

\subsubsection{Protocol and cry database}
The dataset of cries and associated clinical information was collected by the above five hospitals between 2020 and 2023. The overall raw database consists of 4,312 recordings from 2,631 term babies (i.e., of at least 36 weeks gestational age).

Newborns belonged to one of two cohorts: (a) “asphyxia cohort”, that is those who were admitted to the hospitals’ neonatal intensive care units (NICU) with a history of a hypoxic insult, including sentinel events, Apgar scores, resuscitation requirements, and/or blood gasses where available; (b) “healthy cohort”, that is patients who had no evidence of a hypoxic insult, typically recruited from the normal newborn nurseries. 

\begin{figure}[h]
    \centering
    \includegraphics[width=1\textwidth]{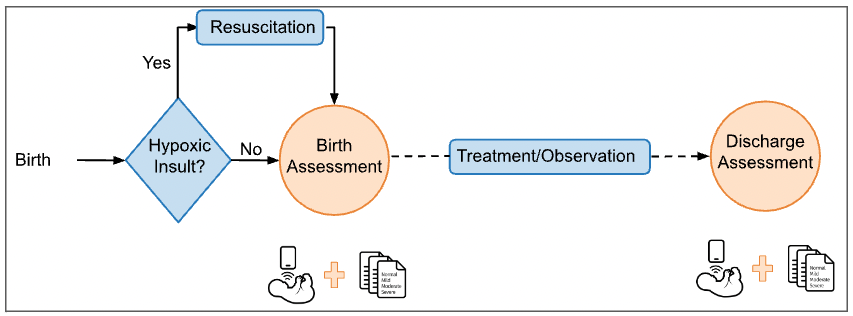}
    \caption{Protocol for the Ubenwa clinical study. Infant cry recordings are taken at 2 time points – birth at birth and at discharge. Each time a 30-sec cry sample is obtained and a Sarnat exam conducted by a neonatologist.}
    \label{fig:study-protocol}
\end{figure}

The study protocol is summarized in Figure \ref{fig:study-protocol}. Patients received a neurological assessment at birth (or admission) and at discharge by a clinician using a modified Sarnat scoring system \cite{sarnat1976neonatal, prempunpong2018prospective} – assigning the level of injury as normal, mild, moderate or severe. In total, 344 infants were admitted with symptoms of neurological injury measured by the modified Sarnat.

\begin{figure}[h]
    \centering
    \includegraphics[width=.8\textwidth]{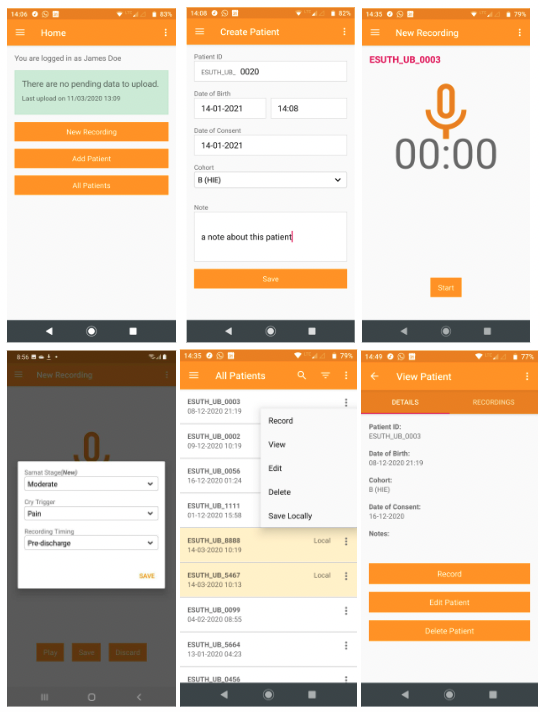}
    \caption{Data collection interface. Patient information is added to the database along with date of birth and cohort information. Recording is done with Samsung A10 smartphone held at 10-15 cm from the newborn's mouth. One or more recordings were collected at birth and before discharge.}
    \label{fig:study-app-pics}
\end{figure}

At the time of Sarnat assessments, a recording of the newborn’s unelicited cry was obtained using a custom-built mobile application (see Figure \ref{fig:study-app-pics}) on a Samsung A10 smartphone held at 10-15 cm from the newborn's mouth. In this work, we set out to utilize a cry sample taken after birth (“birth assessment” in Fig \ref{fig:study-protocol}) to identify the presence of neurological injury, sequel to perinatal asphyxia.

\subsubsection{Curated labeled subset of birth recordings for model development}
To ensure the high quality of data used in model development, a number of data cleaning steps were done. First, we manually segmented the cry expiratory sounds in the audio files recorded prior to November 2022. We then only kept audio files that had at least 3 seconds of cry sound.

In addition, the annotations were cross-referenced with independently collected clinical records (data collection forms) and a number of recordings were removed as potentially mislabeled resulting in 2,174 recordings with 17 hours of cry signal in total. Finally, we excluded discharge recordings as we focus on birth screening in this study. The resulting curated labeled subset of birth recordings consists of 1,108 audio samples - 959 recordings of controls and 149 diagnosed with neuro injury. This dataset is further subdivided into train, validation and test subsets (see Table \ref{tab:data-description}). 

\subsubsection{Unlabeled dataset of cry recordings for self-supervised domain adaptation}
Data cleaning for self-supervised model adaptation did not have to be cleaned as rigorously as the labeled subset. We took all 4,312 recordings and only excluded recordings of patients that appeared in the validation and test sets. Also, instead of manual segmentation of cry, all audios were processed by an automatic cry activity detector - a small convolutional neural network trained to discriminate cry audio segments. The resulting automatically cleaned unlabeled dataset for self-supervised domain adaptation consists of 3,613 audio recordings.

\subsubsection{VGGSound dataset for model pre-training}
We used the VGGSound database \cite{Chen20} for model pre-training. The VGGSound database is a large-scale audiovisual dataset designed for training and evaluating machine learning models in sound classification tasks. Developed by researchers at the Visual Geometry Group (VGG), it comprises over 200,000 video clips spanning 310 classes of human actions, objects, and events that are associated with sound. Each clip is sourced from YouTube and is approximately 10 seconds long, ensuring that the audio and visual components are aligned in time. The dataset stands out for its diversity and scale, covering a wide range of real-world scenarios, such as musical instruments, environmental sounds, and human activities. Importantly, VGGSound includes weakly labeled audio events, where labels are inferred from video metadata and tags, making it suitable for research in self-supervised learning and multimodal analysis. Its broad coverage of sound classes and robust curation process have made it a valuable resource for advancing the state of the art in audio classification, audio-visual learning, and related fields.

\subsection{Proposed SSL-based training pipeline}
We designed a 3-stage training methodology: (1) pre-training a large model on a massive audio data set (2) adaptation using an unlabelled subset of our cry database and (3) fine-tuning using the clinically-annotated cry samples. This approach is illustrated in Figure~\ref{fig:method}. In the subsequent sections, we provide details on each stage.
\newcommand{\SSLtwo}{SSL cry adaptation}
\begin{figure}[th]
    \centering
    \includegraphics[width=.8\textwidth]{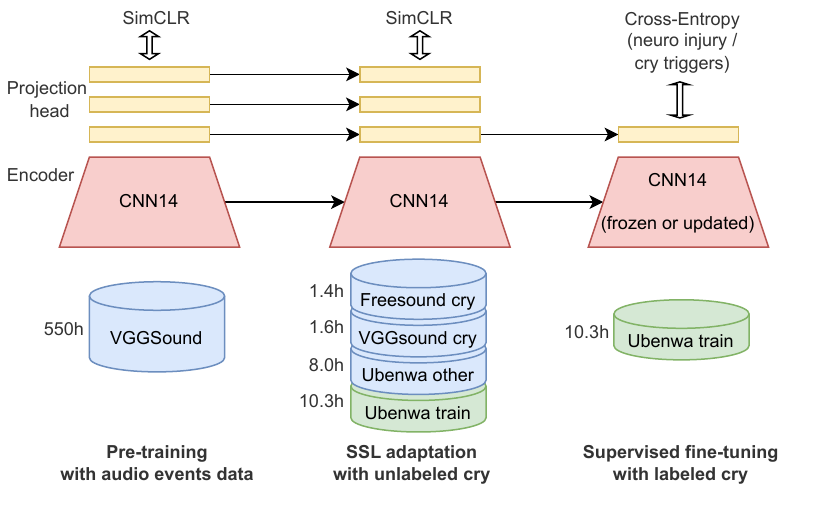}
    \caption{Summary of the proposed SSL-based training pipeline. \cem{\textbf{(Left)} First the CNN14 backbone is pre-trained on the VGGSound dataset using SimCLR. \textbf{(Middle)} The CNN14 backbone is further pre-trained via SSL using cry-specific datasets. We denote this stage \emph{\SSLtwo}. \textbf{(Right)} The model is finally trained with supervision on a labeled cry dataset.}}
    \label{fig:method}
\end{figure}

\subsubsection{Model architecture}
\label{sec:cnn14-desc}
The model is based on CNN14 architecture \cite{kong2020panns} inspired by popular computer vision VGG architecture \cite{simonyan2014very}. CNN14 demonstrated strong performance across various audio classification tasks when pre-trained on large generic audio data.

The audio signal is first processed using short-time Fourier transform (STFT) to create log Mel filterbanks. The resulting log Mel spectrogram is a 2D representation with 10 millisecond time frames on x-axis and 80 frequency bands on y-axis with values corresponding to the corresponding energies.

The spectrogram is further passed to a neural network with 6 layers, where each block is composed of two convolution layers with 3x3 kernel size, batch normalization and ReLU. The network uses 4-second audio clips during training and arbitrary sequences during inference by means of a global pooling operation at the output of the last convolutional layer. 

To summarize, the network contains 76 million parameters which encode arbitrary input audio file into a 2048-dimensional feature vector for classification or contrastive self-supervised learning.


It is schematically shown in Table~\ref{tab:cnn14-v1}. 
\setlength{\tabcolsep}{1pt}
\begin{table}[h]
    \centering
    \begin{tabular}{ll|rcccl}
    \multicolumn{2}{c|}{Blocks} & \multicolumn{5}{c}{Output dimension} \\ \hline
     \multicolumn{2}{c|}{Log-Mel Filterbank} & 1 & x & T & x & 80 \\ 
     2x[Conv(64)+BN+ReLU] &$\rightarrow$Pool 2x2      & 64  & x & T/2 & x  & 40 \\
     2x[Conv(128)+BN+ReLU] &$\rightarrow$Pool 2x2     & 128 & x  & T/4  & x & 20 \\
     2x[Conv(256)+BN+ReLU] &$\rightarrow$Pool 2x2    & 256 & x & T/8 & x & 10 \\
     2x[Conv(512)+BN+ReLU] &$\rightarrow$Pool 2x2    & 512 & x & T/16 & x & 5 \\
     2x[Conv(1024)+BN+ReLU] &$\rightarrow$Pool 2x2   & 1024 & x & T/32 & x & 2  \\
     2x[Conv(2048)+BN+ReLU] &                           & 2048 & x & T/32 & x & 2  \\
     \multicolumn{2}{c|}{Global Average Pooling} & \multicolumn{5}{c}{2048} \\ 
    \end{tabular}
    \caption{CNN14 blocks and output dimensions when processing T audio frames. Conv refers to 3x3 convolutions. T is 400 at training and arbitrary length at inference}
    \label{tab:cnn14-v1}
\end{table}
\setlength{\tabcolsep}{5pt}

\subsection{Supervised and self-supervised pre-training}
We compare two identical CNN14 models pre-trained in a supervised and self-supervised manner on VGGSound. The initial pre-trained models were provided by Wang, Z. et al~\cite{wang2022learning} with the publicly available training code. Supervised pre-training is done over 200 epochs with stochastic gradient descent with batches of 32 4-second random audio segments of VGGSound. In addition, input spectrograms are corrupted using a popular speech data augmentation method called SpecAugment~\cite{park2019specaugment}, which randomly masks parts of the spectrograms thus increasing the diversity of samples and model robustness.

Self-supervised pre-training was done by the same authors in a similar manner to supervised pre-training but using SimCLR method. SimCLR is a contrastive self-supervised learning algorithm originally proposed for computer vision \cite{Chen20} that recently demonstrated good results in audio / music classification tasks \cite{wang2022learning, spijkervet2021contrastive}. It maximizes the similarity between two views of the same audio recording. In our implementation, each view is created by randomly extracting a 4-second audio segment and applying SpecAugment to the spectrogram.

\subsection{Supervised fine-tuning and evaluation}
\label{sec:protocol}
To transfer the pre-trained network to the task of neuro injury prediction we add a randomly initialized classification layer (head) and use labeled data to fully or partially fine-tune the resulting network by minimizing cross-entropy loss. To ensure that we utilize pre-trained models efficiently, for all models we evaluate three fine-tuning strategies that work best in 10-fold cross-validation along with additional hyperparameter tuning.

First, we use \textit{linear probing}, where the frozen encoder  extracts features for a linear classifier. 
Second, in addition to learning the linear classifier, we also update statistics of batch normalization parameters of the encoder while still keeping other parameters frozen. The primary motivation is to compensate for the difference between pre-training and target data characteristics~\cite{frankle2020training,yazdanpanah2022revisiting}. This also allows us to understand what portion of improvement obtained by SSL fine-tuning with cry data may be attributed to a simple update of normalization parameters occurring naturally during SimCLR. 
Finally, aiming to improve classification results further, we perform \textit{end-to-end fine-tuning}, where the encoder parameters are optimized jointly with the classifier on target tasks.

The supervised training is done for 50 epochs in all three settings, and the model with the best validation score is selected. We use weighted random sampling to balance the distribution of classes during training. 
We use Adam optimizer~\cite{kingma2014adam} with a learning rate reduced two times if validation loss does not improve for three epochs. For end-to-end supervised fine-tuning, a much smaller and separately tuned learning rate is used for the encoder. Also, the encoder learning rate is linearly increased from zero to target one over the first ten epochs. 
In all experiments, the learning rates of the classifier and encoder are optimized using grid search. For the model with the best validation score, we repeat the experiment 10 times and report the mean area under the receiver operating characteristic curve (AUC) along with standard error. For multi-class classification (triggers), the macro averaged AUC is computed using a one-versus-rest approach.

Similar to~\cite{chen2020big}, in our preliminary studies, we found that keeping one layer of projection head of the SSL pre-trained model leads to slightly better results. Therefore, we always transfer from layer 1 of the projection head.

\subsection{Self-supervised cry adaptation}

To improve the quality of SSL representations, we further explore a second stage of self-supervised domain adaptation. Our goal is to adapt the encoder from the domain of general audio sounds to the domain of cry sounds, using unlabeled cry data (middle column of Figure \ref{fig:method}). This SSL adaptation is done by reusing the encoder and the projector from CNN14 trained on VGGSound and running 100 more epochs of SimCLR using infant cry data with the same learning rate and schedule as the initial pre-training. The only difference is that we use batch size 200 for faster training and because larger batches performed better for SimCLR in the literature~\cite{spijkervet2021contrastive}. Notably, we did not find a significant difference trying to improve the initial SSL pre-trained model of~\cite{wang2022learning} by using more training epochs and large batches without cry sounds.

Note that during training for cry adaptation, half of the batch is sampled from the unlabeled Ubenwa dataset and another half - randomly sampled from VGGSound. Adding VGGSound to batches can be seen as so called “rehearsal” or “replay" that are common in domain adaptation scenarios (continual learning) for reducing catastrophic forgetting (drop of performance on an initial task when learning new task)~\cite{robins1995catastrophic}. While we are not studying or addressing forgetting in this study, we found in that rehearsal consistently leads to better generalization which is important for supervised transfer learning.
Similar to the initial SSL pre-trained models, the cry-adapted ones are evaluated with linear probing and end-to-end settings described in the previous section. \cut{We expect linear probing of the cry-adapted models to perform significantly better than the non-adapted ones. This is perhaps less obvious for end-to-end fine-tuning as PANN may forget some generic knowledge useful for supervised fine-tuning.}

\section{Experimental setup and results}
\label{sec:exp_setup}

\subsection{Dataset description}

This study is based on a subset of a larger Ubenwa newborn cry clinical database collected from five hospitals in Nigeria, Brazil, and Canada since 2020~\cite{onu1711ubenwa}. For most infants, one 
 recording is done after birth and one before discharge. A neurological exam was conducted on all infants, and the level of neuro injury was recorded using a four-scale measure called Sarnat score~\cite{sarnat1976neonatal}. For classification, we categorize the recordings into two groups: normal (no neuro injury) and injured (mild, moderate, or severe injury). 
We further split the data into train, validation, and test, making sure the recordings of a given patient go to one subset. Table~\ref{tab:data-description} summarizes key statistics of the data.

\setlength\arrayrulewidth{1.0pt}
\begin{table}[h]
    \centering
    \begin{tabular}{c|ccc|ccc}
      & \multicolumn{3}{c|}{\textbf{Healthy}} & \multicolumn{3}{c}{\textbf{Neuro Injury}} \\
      & Train & Val & Test & Train & Val & Test \\
     \textbf{\# recordings} & 1360 & 247 & 238 & 92 & 40 & 45 \\
     \textbf{\# patients} & 885 & 165 & 163 & 75 & 33 & 38 \\
     \textbf{\# hours} & 10.3 & 1.9 & 2.0 & 0.8 & 0.3 & 0.3 \\
    \end{tabular}
    \caption{\cem{The description of our} neurological injury dataset.}
    \label{tab:data-description}
\end{table}

In addition, the recordings are annotated with a trigger - the primary reason for crying determined by the medical or research staff. In this study, we use a subset of three main triggers resulting in 267 recordings of discomfort, 200 hunger, and 682 pain. 

Compared to the commonly used Chillanto database~\cite{reyes2004system}, our dataset has much more patients with neurological injury (146 vs 6) and more annotated cry signals in general (14.2 vs 0.6 hours). 
In our database, cry recording is a segment of arbitrary length (a second to a few minutes). Conversely, in  Chillanto, the recordings correspond to 1-second cry expirations annotated as belonging to the healthy or sick infant. However, there is insufficient evidence to determine whether every cry expiration of a sick infant has distinct characteristics from healthy infants or if only some expirations have them. 
Furthermore, cry expirations of a single infant are generally quite similar, so if recordings are split randomly for training, testing, and validation without considering infant identities (for example, in~\cite{patil2022constant}), the resulting performance may be over-estimated.

For SSL experiments, we also use an additional 8 hours of Ubenwa unlabeled cries along with 1.6 hours available in VGGSound and about 1.4 hours collected from Freesound website\footnote{ \url{https://freesound.org} } using search query ``baby cry''.

\subsection{Baselines}

While our primary focus is on SSL pre-training and fine-tuning, we use two supervised approaches as baselines that do not rely on pre-training. 
The first system is a statistical model using ComParE 2016~\cite{schuller2016interspeech} acoustic features extracted with OpenSmile toolkit~\cite{eyben2010opensmile}. The feature set contains 6373 recording-level derivatives (mean, standard deviation, etc.) of various acoustic descriptors (MFCC, pitch, jitter, etc.). It is commonly used in computational paralinguistics and infant cry classification~\cite{parga2020defining}. The model and hyperparameters are selected using grid search and 10-fold cross-validation, maximizing the average AUC score. 
%
The second baseline is CNN14 described in Section~\ref{sec:method} using random initialization, no pre-training and end-to-end supervised fine-tuning described in Section~\ref{sec:protocol}
The performance of baselines on neurological injury and triggers is summarized in Table~\ref{tab:baselines} for five experiments with different random seeds. 
CNN14 without pre-training does not outperform the statistical baseline, which is not surprising given that our training datasets are quite small for such a model.

\begin{table}[h]
    \centering
    \begin{tabular}{c|c|c}
     \multirow{2}{*}{\textbf{Model}}& \multicolumn{2}{c}{\textbf{AUC \% (mean and standard error)}} \\ 
     & Neuro Injury & Cry Triggers \\ \hline
     Statistical & 75.1 {\scriptsize $\pm$ 0.4} & 71.1 {\scriptsize $\pm$ 0.2} \\
     CNN14 & 74.6 {\scriptsize $\pm$ 1.7} & 59.8 {\scriptsize $\pm$ 0.9} \\
    \end{tabular}
    \caption{Baseline performance obtained without any pre-training}
    \label{tab:baselines}
\end{table}

\subsection{SSL experiments}
\label{sec:exp}

The main results of neurological injury and cry trigger experiments are summarized in Tables~\ref{tab:asphyxia-results} and~\ref{tab:trigger-results}. The first row \cem{in both tables} refers to supervised training with random initialization and is provided to give an idea about the model performance without pre-training.

The last three columns \cem{in Table \ref{tab:asphyxia-results} and Table~\ref{tab:trigger-results}} \cem{summarize the performances obtained after fine-tuning the network with supervised training. From left to right, the results in the tables correspond to the following: 
\begin{enumerate}
    \item Evaluation with linear probing, where a linear layer is fine-tuned on top of the frozen encoder weights (Linear).
    \item Evaluation with linear probing, with batch-norm layers  updated during fine-tuning (Linear+BN).
    \item End-to-end fine-tuning where the linear layer and whole encoder are updated (End-to-end).
\end{enumerate}
}


\setlength{\tabcolsep}{2pt}
\begin{table}[h]
    \centering
    \begin{tabular}{c|c|c|ccc}
   & \textbf{Pre-} & \textbf{SSL cry adapt.}  & 
 \multicolumn{3}{c}{\textbf{\% AUC after fine-tuning}}  \\ 
      & \textbf{training} & \textbf{Dataset} & Linear & Linear+BN & End-to-end \\
    \hline \hline
       1 & -- & --  & 60.6 {\scriptsize $\pm$ 1.5}  & 60.4 {\scriptsize $\pm$ 1.3} & 74.6 {\scriptsize $\pm$ 1.7} \\
       2 & supervised & -- & 75.5 {\scriptsize $\pm$ 0.6}  & 75.9 {\scriptsize $\pm$ 0.7}  & 80.0 {\scriptsize $\pm$ 0.7} \\
       3 & SSL & -- & 71.3 {\scriptsize $\pm$ 0.9} & 78.5 {\scriptsize $\pm$ 1.2}  &  83.9 {\scriptsize $\pm$ 0.6} \\ \hline     
       4 & SSL & train set & 78.8 {\scriptsize $\pm$ 0.5} & 78.0 {\scriptsize $\pm$ 0.7} & 80.8 {\scriptsize $\pm$ 0.8} \\
       5 & SSL & + 11h cry &  79.8 {\scriptsize $\pm$ 0.4} & 81.3 {\scriptsize $\pm$ 0.5} & 81.3 {\scriptsize $\pm$ 0.7} \\
       6 & SSL & + replayVGG &   80.8 {\scriptsize $\pm$ 0.5}  &  83.3 {\scriptsize $\pm$ 0.6} &  {85.0} {\scriptsize $\pm$ 0.9} \\
    \end{tabular}
    
    \caption{\cem{Performance of neuro injury classification under various types of pre-training and fine-tuning strategies.   \textbf{Column 1} indicates the type of pre-training. Note that for the first row, no pre-training is applied. For the second row, supervised pre-training on the VGGSound dataset is applied. The rest of the rows use SSL-based pre-training on the VGGSound. \textbf{Column 2} indicates the datasets used in \SSLtwo. Note that in rows 4-5-6, the datasets used in SSL cry adaptation are cumulative. The 4th row uses neuro injury train, the 5th adds 11h cry to the neuro injury train, and the 6th adds a replay buffer from the VGG Sound dataset to the previous datasets from rows 4-5. \textbf{Columns 3-5} show the \% AUC (with mean and standard error) obtained with different supervised fine-tuning strategies (after the SSL fine-tuning as shown in Figure \ref{fig:method}). \cut{Linear denotes standard linear probing where a linear layer is trained on a frozen encoder, Linear+BN denotes the case where batch-norm parameters are updated in addition to the linear layer. End-to-end denotes the case where the entire encoder is fine-tuned after pre-training.}} }
    \label{tab:asphyxia-results}
\end{table}

\begin{table}[h]
    \centering
    \begin{tabular}{c|c|c|ccc}
    & \textbf{Pre-} & \textbf{SSL cry adapt.}  & 
   \multicolumn{3}{c}{\textbf{\% AUC after fine-tuning}}  \\ 
    & \textbf{training} & \textbf{Dataset} & Linear & Linear+BN & End-to-end \\
    \hline \hline
      1 &  -- & --  &  57.1 {\scriptsize $\pm$ 2.6} & 60.7 {\scriptsize $\pm$ 0.5} & 59.8 {\scriptsize $\pm$ 0.9} \\
       2 & supervised  & -- & 67.9 {\scriptsize $\pm$ 1.9}  & 68.0 {\scriptsize $\pm$ 1.7}  & 68.1 {\scriptsize $\pm$ 1.6} \\
       3 & SSL & -- & 65.9 {\scriptsize $\pm$ 0.8} & 69.5 {\scriptsize $\pm$ 0.7}  & 69.0 {\scriptsize $\pm$ 0.9} \\ \hline

       4& SSL & neuro injury train & 71.7 {\scriptsize $\pm$ 0.5}  & {75.4} {\scriptsize $\pm$ 0.8}  & 72.4 {\scriptsize $\pm$ 1.4} \\
       5 & SSL & + 11h cry &   74.5 {\scriptsize $\pm$ 0.4}  & 74.7 {\scriptsize $\pm$ 0.4}    & 72.0 {\scriptsize $\pm$ 1.8}  \\
      6 & SSL & + replayVGG & 74.2 {\scriptsize $\pm$ 0.4}  &   {75.6} {\scriptsize $\pm$ 0.6}  &   74.4 {\scriptsize $\pm$ 0.7}  \\
    \end{tabular}
    \caption{\cem{Performance of cry trigger classification. We follow the same structure used in Table \ref{tab:asphyxia-results}, therefore the same caption applies.}}

    \label{tab:trigger-results}
\end{table}

The second and third rows \cem{in both tables} compare supervised and self-supervised initial pre-training with VGGSound. In these experiments, the cry database is used only for supervised fine-tuning \cem{(In other words, no-additional \SSLtwo \; stage is applied). We observe that,} while simple linear probing performs better for supervised pre-training, the self-supervised pre-training achieves better results when updating BN statistics. Also, \cem{we see that with the end-to-end fine-tuning strategy} SSL pre-training performs significantly better on neuro injury task.  

Next, \cem{in rows 4-5-6 of Table \ref{tab:asphyxia-results} and Table \ref{tab:trigger-results} we show the results when additional \SSLtwo \;stage is employed.} First, we fine-tune with SimCLR using only the neuro injury training dataset, as shown in row 4 of Table~\ref{tab:asphyxia-results}). \cem{For neuro injury, we observe that} this significantly improves AUC for linear evaluation (71.3 to 78.8), but the improvement vanishes when BN is updated (78.5 and 78.0). We hypothesize that SimCLR in this experiment performs better mostly due to a significant domain mismatch between VGGSound that a simple BN update can compensate for. \cem{For cry triggers, we observe that \SSLtwo \; in general improves the performance obtained after supervised fine-tuning. }

\cem{Next, as shown in the 5th row of Table \ref{tab:asphyxia-results} and Table \ref{tab:trigger-results}} we further double the amount of unlabeled data for SimCLR \SSLtwo \; stage. \cem{This is achieved by adding} an 8-hour portion of previously unused Ubenwa cry along with some \cem{unlabeled} cry sounds from VGGSound and freesound. \cem{In total, these \SSLtwo \; datasets amount to approximately 11 hours of recording (hence it is called 11h cry in Table \ref{tab:asphyxia-results}, and Table \ref{tab:trigger-results})}. This significantly improves the performance of linear evaluation with and without BN update \cem{for neuro injury}. 

This, however, is not the case for end-to-end fine-tuning, where the initial VGGSound pre-training results in better transfer for neuro injury (row 3 of Table~\ref{tab:asphyxia-results}). \cem{We hypothesize that the model loses its generalization properties that are useful for fine-tuning due to catastrophic forgetting \cite{castastrophicforgetting} when we further adapt the model with SSL.} 

To mitigate this forgetting effect and preserve generalization properties that seem to be important for transfer learning, we perform \SSLtwo \; using replay technique from continual learning literature~\cite{doi:10.1080/09540099550039318}. \cem{We show this on the last row of both tables.} This is done by replaying 50\% of the VGGSound dataset when \cem{applying \SSLtwo \;stage}. We, therefore, observe that \SSLtwo \;with replay performs significantly better in all evaluations for neuro injury. We also \cem{obtain} the best results on trigger classification using linear+BN evaluation with this approach.

\subsection{Training with subsets of labeled data}
\label{sec:analysis}

Obtaining labeled, high-quality medical data is extremely time-intensive and expensive. Therefore, in this section, we aim to understand if a small amount of labeled data can still yield decent performance. 
Specifically, we aim to analyze how SSL cry adaptation helps in such a scenario and which supervised fine-tuning method gives the best performance. To this end, we experiment with linear+BN and end-to-end fine-tuning using randomized subsets of labeled neuro injury dataset.

\begin{figure}[h]
    \centering
    \includegraphics[width=.8\textwidth]{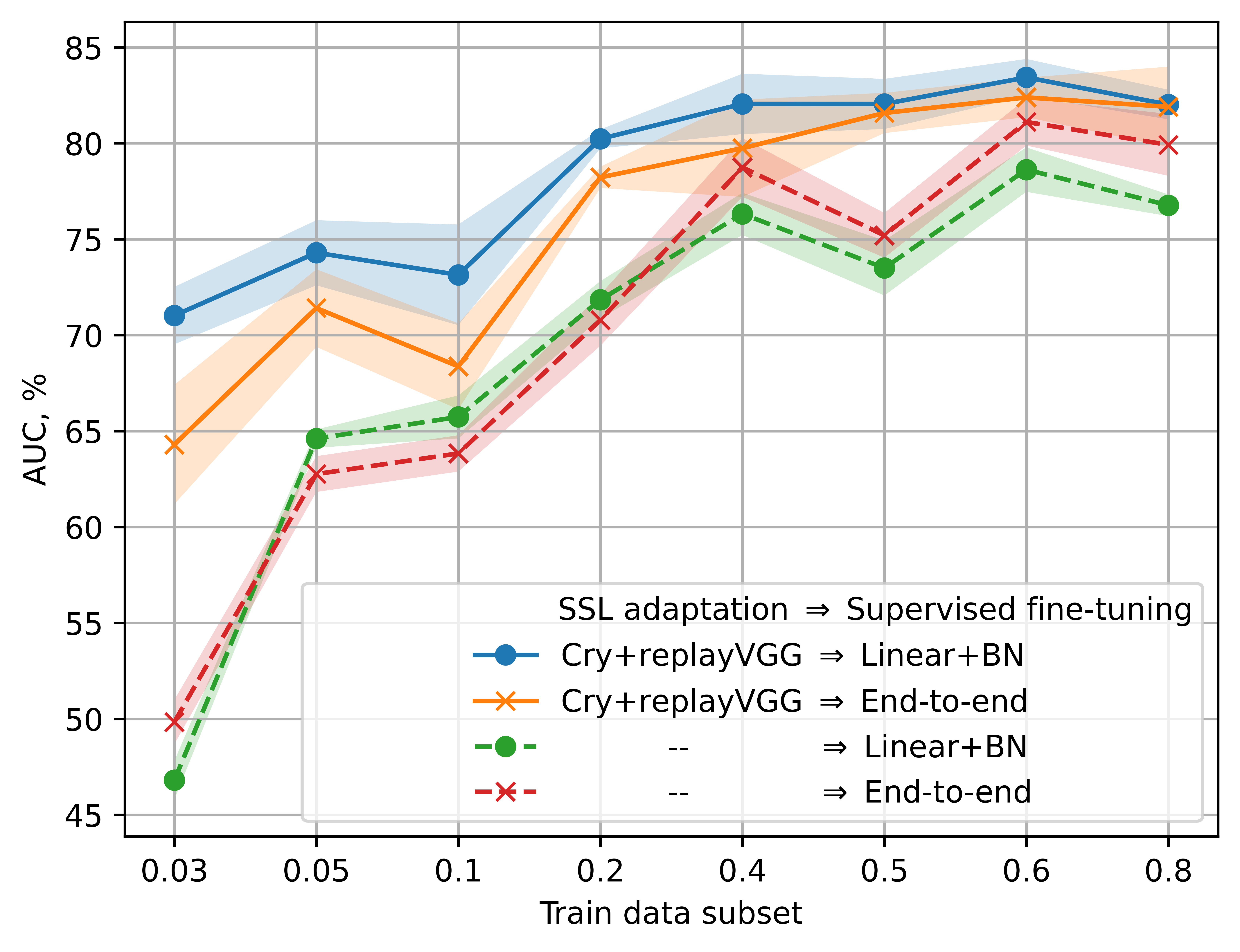}
    \caption{Performance using subsets of labeled neuro injury data in supervised fine-tuning (Linear+BN and End-to-end). Solid lines - model with SimCLR cry adaptation (last row of Table~\ref{tab:asphyxia-results}), dashed - same model without adaptation (row 3 of Table~\ref{tab:asphyxia-results}). The filled areas show the standard error of AUC from five runs with different random seeds.}
    \label{fig:auc-subsets}
\end{figure}

In Figure~\ref{fig:auc-subsets} we show results for neuro injury classification using two models: the model pre-trained with SSL on VGGSound without \SSLtwo \; stage (row 3 of Table~\ref{tab:asphyxia-results}) and the model that obtains the best performance with SSL cry adaptation (last row of Table~\ref{tab:asphyxia-results}).

The SSL cry-adapted model (solid lines) consistently outperforms the not-adapted ones (dashed lines) with a larger margin as the size of the supervised subset is reduced.
%
Another point to note is that we observe that for the not-adapted models, as the amount of labeled fine-tuning data decreases, the end-to-end fine-tuning strategy decreases more rapidly in performance. This could perhaps explain why end-to-end adaptation was performing worse than Linear+BN for cry triggers (Table~\ref{tab:trigger-results}), where less labeled data is available for fine-tuning compared to neuro injury.

\cem{We see that, very i}nterestingly, with only 3\% (a few dozen samples) of labeled data, we can still achieve about 70\% AUC by using a cry-adapted model. 
Also, with only 20\% of data, the adapted model significantly outperforms our supervised baselines.
\cem{This showcases that SSL cry-adaptation has huge potential to obtain satisfactory model performance by only incorporating a small amount of labeled data in the supervised fine-tuning stage.} 

\section{Explainability analysis}
\subsection{Acoustic biomarkers for explainability and clinical decision support}
Given the rich database acquired, we study the opportunity for acoustic biomarkers of the infant cry to deepen our understanding of how pathology alters cry patterns and as a means of model explainability. Such explainability has immense value in AI-based clinical decision support as it advances science in an era of black box predictors, and keeps control in the hands of physicians thereby providing an opportunity for safe and robust decision-making. 

\begin{figure}[h]
    \centering
    \includegraphics[width=1\textwidth]{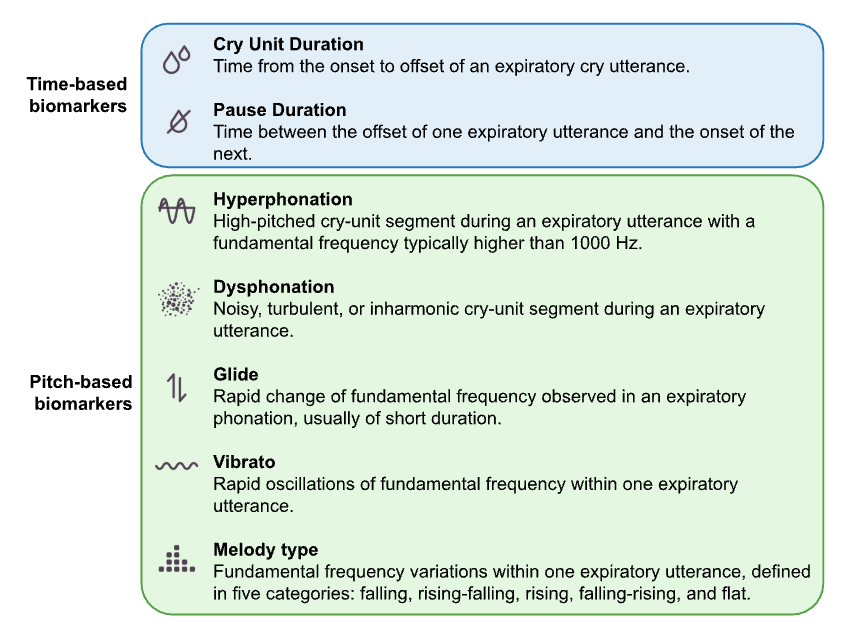}
    \caption{The 7 acoustic biomarkers of infant cry studied in this work.}
    \label{fig:biomarkers-summary}
\end{figure}

To develop these acoustic biomarkers, we studied two kinds of features: \textit{generic voice features} and \textit{cry-specific biomarkers}. Generic voice features include measures commonly used in audio analysis such as fundamental frequency, resonance frequencies and cepstral coefficients. Cry-specific biomarkers are higher-level features that measure specific aspects of the infant’s physiology. See an overview of 7 cry-specific biomarkers in Figure~\ref{fig:biomarkers-summary}.

\subsection{Generic voice features as cry biomarkers}
Generic voice features used in this study were computed using openSMILE (open-source Speech and Music Interpretation by Large-space Extraction), an open-source toolkit for audio feature extraction and audio classification~\cite{eyben2010opensmile}. Specifically, we make use of the extended Geneva Minimalist Acoustic Parameter set (eGeMAPS), which includes 25 low-level descriptors (LLDs) designed for automatic voice analysis tasks~\cite{eyben2015geneva}. The LLDs are time sequences computed from the input audio. They are further aggregated into a total of 88 features by taking various statistics over the sequence.

The motivation for choosing eGeMAPS for our study is twofold. First, the LLDs in eGeMAPS were selected based on their potential to index physiological properties in human voice, thus making it possible to verify physiological hypotheses against feature-analysis findings. Second, eGeMAPS are widely adopted in voice and audio literature~\cite{atmaja2020differences, haider2019assessment, li2020waveform, macary2020multi, haider2021emotion, yang2022improving}. 

To extract features from an input cry recording, we first collect all expiration cry segments from manually annotated recordings. The expiration segments are concatenated into one audio array and processed by the openSMILE feature extractor. At the output, each cry recording is represented by an 88-dimensional feature vector. 

\subsection{Cry-specific biomarkers}
We refer to cry biomarkers as signal processing-based features that measure specific aspects of the infant’s physiology. Some of these cry-specific biomarkers were first introduced by Truby et al in 1960s~\cite{truby1965cry}. We refined the definitions, and developed signal-processing algorithms leveraging our much larger database. The biomarkers are either based on fundamental frequencies ($F_0$) such as hyperphonation, dysphonation, glide, vibrato, melody types, or time-domain durations: cry unit duration and pause duration.

\textit{\textbf{Hyperphonation}} is defined as a high-pitched cry-unit segment during an expiratory utterance with a fundamental frequency typically higher than 1000 Hz~\cite{wasz1985twenty, corwin1996infant, chittora2016spectral}. It results from a "falsetto" like vibration pattern of the vocal folds. It has been reported as an indicator of neural constriction of the vocal tract48 and associated with various pathologies such as laryngomalacia~\cite{chittora2016spectral}, asthma~\cite{chittora2016spectral}, respiratory distress syndrome~\cite{chittora2016spectral}, and prenatal exposure to opiate~\cite{corwin1987cry, lester2002maternal}, cocaine~\cite{corwin1992effects}, and alcohol~\cite{lester2002maternal}. 

\textit{\textbf{Dysphonation}} is defined as a noisy, turbulent, or inharmonic cry-unit segment during an expiratory utterance~\cite{corwin1996infant, lagasse2005assessment, golub1985physioacoustic} and has been reported to indicate unstable respiratory control\cite{lagasse2005assessment}. In existing clinical cry research, it has been associated with depression\cite{lagasse2005assessment}, laryngomalacia~\cite{chittora2016spectral}, congenital heart disease~\cite{chittora2016spectral}, meningitis~\cite{chittora2016spectral}, brain hemorrhage~\cite{chittora2016spectral}, and prenatal exposure to cocaine~\cite{lester2002maternal, golub1985physioacoustic}, marijuana~\cite{lester1989effects}, and alcohol~\cite{nugent1996effects}. 

\textit{\textbf{Glide}} is defined as a rapid change of fundamental frequency observed in an expiratory phonation, usually of short duration~\cite{wasz1985twenty, golub1985physioacoustic}. In existing clinical studies, glide has been associated with birth asphyxia~\cite{Michelsson1971}, meningitis~\cite{golub1985physioacoustic}, and hydrocephalus~\cite{golub1985physioacoustic}. It also occurs more frequently in the cry of premature neonates~\cite{golub1985physioacoustic}. 

\textit{\textbf{Vibrato}} is defined to occur when there are at least four rapid up-and-down movements of fundamental frequency within one expiratory utterance~\cite{Michelsson1971, golub1985physioacoustic} and has been studied in the context of congenital heart disease~\cite{Michelsson1971}, deafness~\cite{chittora2016spectral},  and birth asphyxia~\cite{Michelsson1971, chittora2016spectral}. 

\textit{\textbf{Melody type}} describes the fundamental frequency variations within one expiratory utterance, defined in five categories: falling, rising-falling, rising, falling-rising, and flat~\cite{Michelsson1971, corwin1996infant}. It reflects the trend of fundamental frequency over time. It was reported that a typical cry of the healthy newborn has a falling or rising-falling melody while a significant increase in rising, falling-rising, and flat types of melody was observed in those with central respiratory failure~\cite{Michelsson1971, michelsson1982sound}.

\textit{\textbf{Cry-unit duration}} is defined as the time from the onset to offset of an expiratory utterance \cite{lagasse2005assessment, golub1985physioacoustic}. A deviation from the normal range of cry-unit durations was associated with asphyxia, meningitis, hydrocephalus, peripheral respiratory distress, central respiratory distress~\cite{golub1985physioacoustic}, hyperbilirubinemia~\cite{wasz1971spectrographic}, and prenatal exposure to opiate~\cite{corwin1987cry} and cocaine~\cite{lester2002maternal}. 

\textit{\textbf{Pause duration}} is defined as the time between the offset of one expiratory utterance and the onset of the next. It is typically during this time that the newborn takes in air and prepares for the next expiration. Together with cry-unit duration, it reflects the neural control of the respiratory system. Past studies show that durational biomarkers are dependent on the state of the infant’s respiratory system~\cite{golub1985physioacoustic}. 

\begin{figure}[h]
    \centering
    \includegraphics[width=.8\textwidth]{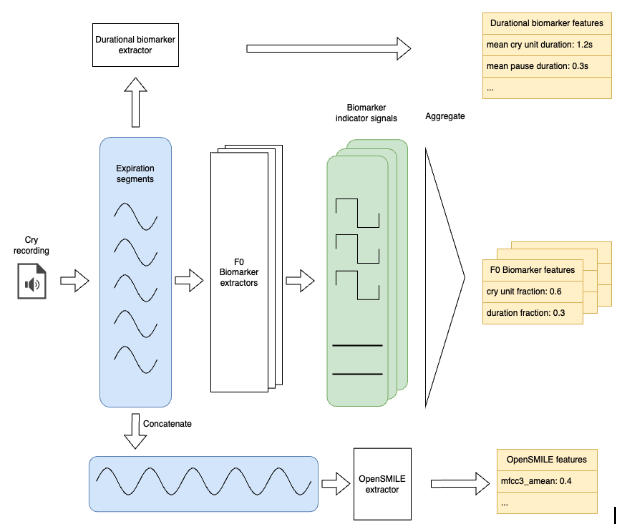}
    \caption{Schematic representation of biomarker extraction.}
    \label{fig:biomarker-extraction-schematic}
\end{figure}

\subsection{Cry-specific biomarkers extraction}
\label{sec:cry-specific-biomarkers-extraction}
We designed extractors for all cry biomarkers. Durational biomarkers (cry unit and pauses) are computed in a straightforward way from manually segmented expiration timestamps and further aggregated for each recording using mean, standard deviation, maximum, and minimum statistics. The schematic in Figure~\ref{fig:biomarker-extraction-schematic} describes the extraction process, discussed in detail below.

Most pitch-based biomarkers are extracted from $F_0$ contour computed with an open-source pitch estimator - CREPE~\cite{kim2018crepe}. The only exception is dysphonation which relies on spectral flatness computed with Librosa~\cite{mcfee2015librosa}.

All pitch-based biomarkers are binary indicators computed for each 10-millisecond audio frame (1 - detected and 0 - otherwise). Hyperphonation is implemented as an indicator of $F_0$ exceeding a pre-defined threshold for a substantial duration. Dysphonation is detected following the same procedure as hyperphonation, except that $F_0$ is substituted by spectral flatness. Vibrato is calculated as an indicator of the oscillating $F_0$ curve detected by measuring the distance between peaks. Glide is implemented as an indicator of a sharp increase or decrease in the $F_0$ sequence. Melody type is represented by five biomarkers that indicate the shape of $F_0$ contour within an expiratory cry unit: falling, rising-falling, rising, falling-rising, or flat.

The thresholds and other parameters of biomarker extractors are tuned using our training database to maximize the classification performance of individual biomarkers. For classification, the indicators of pitch-based biomarkers are aggregated on a per-recording basis as cry unit fraction (number of cry units with biomarker observed in proportion to the total number of cry expiration segments in the recording) and duration fraction (number of frames with biomarker observed in proportion to total number of frames in the recording). 

\subsection{Selecting a compact subset of biomarkers}
First, we calculated Pearson correlation coefficient (PCC) between each feature and the corresponding positive (Sarnat mild, moderate, severe) or negative label (Sarnat Normal) using a subset of train and validation curated birth recordings. 

This was performed on a per-hospital basis, and a feature was selected if its PCC has the same directionality across all hospitals, indicating that this feature has a consistent, non-spurious relationship with the infant's health state and may be a reasonably robust discriminators of neurological injury

Out of 114 total features (88 generic and 26 cry-specific) 18 were selected using this approach. Among these selected features summarized in Table~\ref{tab:sarnat_correlation}. Six come from cry biomarkers while the other 12 are generic voice features.

\begin{table}[h!]
\centering
\begin{tabular}{>{\raggedright\arraybackslash}m{10cm}>{\centering\arraybackslash}m{3cm}}
\hline
\textbf{Feature name} & \textbf{Correlation with Sarnat} \\ 
\hline
\multicolumn{2}{l}{\textbf{Cry-specific biomarkers}} \\ 
Fraction of cry units with rising-falling melody & Negative \\ 
Fraction of cry units with flat melody & Positive \\ 
Fraction of cry units with glide biomarker & Negative \\ 
Fraction of frames with glide biomarker & Negative \\ 
Fraction of cry units with dysphonation & Positive \\ 
Fraction of frames with dysphonation & Positive \\ 
\hline
\multicolumn{2}{l}{\textbf{Generic voice features}} \\ 
slopeUV0-500\_sma3nz\_amean & Negative \\ 
slopeV0-500\_sma3nz\_stddevNorm & Positive \\ 
slopeV0-500\_sma3nz\_amean & Negative \\ 
F2frequency\_sma3nz\_amean & Negative \\ 
F3frequency\_sma3nz\_amean & Negative \\ 
F3frequency\_sma3nz\_stddevNorm & Positive \\ 
mfcc3\_sma3\_amean & Positive \\ 
mfcc3V\_sma3nz\_amean & Positive \\ 
mfcc3V\_sma3nz\_stddevNorm & Positive \\ 
loudness\_sma3\_stddevFallingSlope & Positive \\ 
mfcc2\_sma3\_stddevNorm & Negative \\ 
mfcc4V\_sma3nz\_stddevNorm & Negative \\ 
\hline
\end{tabular}
\caption{Features and their correlation with Sarnat. A feature is only considered positively correlated if its Pearson correlation coefficient is positive across all hospitals.}
\label{tab:sarnat_correlation}
\end{table}

\subsection{Statistical modeling using biomarkers as features}

Finally, in order to validate the utility of the selected features, we conduct linear classification experiments using logistic regression trained on various feature subsets. Linear classifiers are useful for this kind of analysis as they quantify the relative contribution of each feature as percentage weights in the model. Specifically, we consider 88 generic voice features, and 26 cry-specific biomarkers, their combinations and subsets selected based on consistent sign of correlation across three hospitals. 

Logistic regression is built using the scikit-learn package61 with hyperparameters selected using 10-fold cross-validation similar to neural network training setup. The detailed results are summarized in Extended Data Table 3.

By comparing each feature set with its selected counterparts, we conclude that the selected subset of features, although small in size, can achieve classification performance comparable to that achieved by the entire feature set. Although these signal processing-based features are far from neural network performance, they remain valuable due to their interpretability and association with physiological characteristics.

\begin{table}[h!]
\centering
\begin{tabular}{@{}>{\raggedright\arraybackslash}m{4cm}>{\centering\arraybackslash}m{3cm}>{\centering\arraybackslash}m{5cm}>{\centering\arraybackslash}m{3cm}@{}}
\toprule
\textbf{Feature set} & \textbf{Number of features} & \textbf{Mean 10-fold Cross-Validation AUC ± standard error} & \textbf{Test AUC \%} \\ 
\midrule
Voice & 88 & 66.0 ± 2.2 & 59.3 \\ 
Selected voice & 12 & 60.9 ± 2.1 & 59.7 \\ 
Cry & 26 & 65.1 ± 3.5 & 56.9 \\ 
Selected cry & 6 & 61.9 ± 2.6 & 60.5 \\ 
Voice + cry & 114 & 68.5 ± 1.8 & 63.7 \\ 
Selected voice + selected cry & 18 & 66.1 ± 2.7 & 61.6 \\ 
\bottomrule
\end{tabular}
\caption{Comparison of feature sets and their performance. The feature set selected via the procedure in section \ref{sec:cry-specific-biomarkers-extraction} outperforms all others and its AUC is only 2\% less than the full feature set even though it uses 85\% fewer parameters. This indicates the strength of the selected features.}
\label{tab:feature_set_performance}
\end{table}

\section{Conclusions}
\label{sec:conclusion}

In this paper, we explored large-scale SSL pre-training for infant cry analysis, namely for detecting neurological injury and cry triggers. \cem{We observe that} SSL pre-training performs significantly better than the conventional supervised pre-training, and both perform significantly better than training from random initialization.
Furthermore, with limited annotated data, we observe that SSL adaption on cry-specific unlabeled data significantly decreases the need for labeled data in the supervised fine-tuning stage.  
\cem{We show that when we adapt the encoder through SSL using unlabeled cry data, the downstream performance for neurological injury is significantly improved. We therefore believe that, with many unlabeled cry recordings, this opens a promising research direction where it would be possible to train a classifier to detect new diseases using only a small amount of annotated cry sounds from the target population.}

\chapter{Understanding and addressing dataset bias and domain shift}
\label{chap:contribution3}
\section{Overview}
The issue of domain shift is a problem in many real-world datasets and clinical audio is no exception. In this work, we study the nature of domain shift in a clinical database of infant cry sounds acquired across different geographies. We explore methodologies for mitigating the impact of domain shift in a model for identifying neurological injury from cry sounds. Our aim is to learn an audio representation that is domain-invariant to different hospitals (and geographies) and is task discriminative. We adapt unsupervised domain adaptation strategies from computer vision to address these biases, including kernel regularization, entropy minimization, domain-adversarial training, adaptive normalization, and the novel Target Noise Injection (TNI) technique. Unlike images, audio data often requires handling temporal dependencies, varied spectral features, and noise patterns specific to recording environments. By systematically modifying these methods to align with the temporal and frequency-domain properties of audio signals, we demonstrates their effectiveness in creating domain-invariant representations for clinical audio data.

\section{Introduction}
When solving a given task using machine learning, we ideally want to build a single model that performs at similar accuracy when deployed in new settings or domains. For example a model trained to detect asphyxia in a Montreal hospital (the source domain) should detect asphyxia equally in a hospital in Lagos, Nigeria (the target domain). In reality, the distributions of source and target data are rarely the same \cite{candela2009dataset}. Neural nets are quite good at capturing dataset bias in its internal representations, and its been found to leads to significantly lower performance on target domain data \cite{torralba2011unbiased, yosinski2014transferable}.

Here, we are interested in domain adaptation in the context of identifying signs of neurological injury from audio recordings of infant cries. Over a span of 3 years, the Ubenwa clinical study \cite{gorin2023selfsupervised} collected cry recordings across hospitals in 3 countries (Brazil, Canada, and Nigeria) for this problem. Multiple prior work have developed neuro injury detection models from cry sounds using neural transfer learning\cite{onu2020neural} and self-supervised learning\cite{gorin2023selfsupervised}. Although these methods show effectiveness on in-domain test sets, they fail to generalize as well to new hospital data.

In this work, we identify and study patterns of domain shift using this international database of infant cry recordings and explore methods for domain adaptation. We show that DA methods from computer vision can be repurposed and applied to infant cry audio. By experimenting with 5 different methods we illustrate that the best methods not only improve target accuracy but also accuracy in the source domain. Secondly, we validate previous clinical findings about the newborn cry as a universal language -- the pitch of baby cries is similarly distributed regardless of geography. We propose a relatively simple and promising approach for DA in infant cry data. Our method requires no architectural changes nor complex, min-max optimization, employs a simple cross-entropy loss function, and requires neither labels nor cry recordings from the target domain -- only target noise samples \cite{NANNI2020101084}. Experiments show that this is a promising direction worth exploring further in future work.


\begin{figure}[h]
  \centering
  \centerline{\includegraphics[width=.7\linewidth]{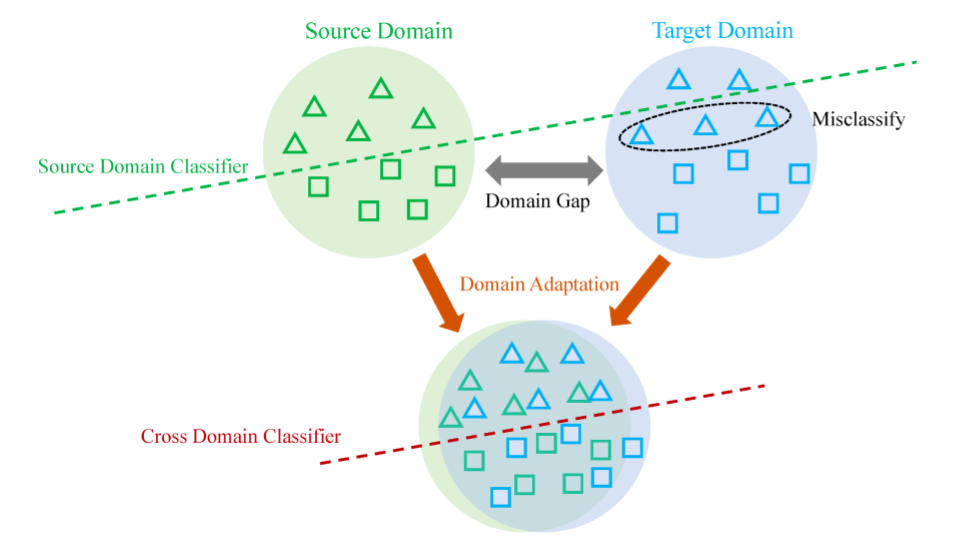}}
\caption{In domain adaptation, we aim to build a cross-domain classifier that generalizes in a consistent fashion regardless of biases in the data samples. Image source \cite{shi2022deep}.}
\label{da-illustration}
\end{figure}


\section{Related work}
Training a neural net for a new task can be expensive. Models typically contain hundreds of millions of parameters requiring immense compute but also large amounts of labelled data which can be costly to obtain in a clinical setting. When solving a given task, we ideally want to build a model that performs at similar accuracy when deployed in new settings or domains. In reality, it is rarely the case that distributions of source and target data are the same \cite{candela2009dataset}, typically resulting in inconsistent model performance \cite{torralba2011unbiased, yosinski2014transferable}. This bias in a dataset can be due to many factors including variations in sensors used to capture data, environmental noise, acquisition protocols, and many more. 

Domain adaptation (DA) aims to address the impact of dataset bias on generalization. In pursuing DA, we want to learn a cross-domain classifier that performs well in both the source and the target domain by mitigating the distributional shift. Domain adaptation can be framed as semi-supervised or unsupervised depending on the availability of labels in the target domain. Given high cost of acquiring labelled medical data, we are mostly concerned with unsupervised domain adaptation, in which labelled data is observed in the source domain and only unlabelled data is available in the target domain during training.

The core idea in most DA algorithms is to simultaneously solve a classification task while learning a domain-invariant representation. This is typically achieved by minimizing a loss consisting of terms for the classification error as well as a measure of the statistical difference between the 2 distributions. Divergence measures used in the latter include mean maximum discrepancy (MMD) \cite{tzeng2014deep},  maximum mean feature norm discrepancy \cite{xu2019larger}, correlation distance \cite{sun2016deep, sun2016return}, $\mathcal{H}$-divergence \cite{ben2007analysis} and $\mathcal{H}\Delta\mathcal{H}$-divergence \cite{ben2007analysis}. 

Another class of domain adaptation methods employ an adversarial objective to learn domain-invariant features...The H-divergence has also been employed in an adversarial objective by using a gradient reversal on a domain classifier \cite{ganin2016domain} or adversarial discriminative domain adaptation \cite{tzeng2017adversarial}. The Wasserstein distance has also proved a useful divergence in adversarial domain adaptation \cite{shen2018wasserstein}.


In the medical domain, dataset bias and domain adaptation have been studied extensively in classifiers built for imaging modalities such as x-rays, CT scans and histopathology \cite{perone2019unsupervised}, \cite{pooch2020can}, \cite{stacke2019closer}. In contrast, dataset bias in medical audio is much less understood.

\section{Methods I: Identifying patterns of domain shift}
\label{sec:domain_shift_id}
In this work, we are interested in addressing the issue of transferability of models models for classifying infant cry sounds across geographically diverse clinical settings. To begin, we investigate the existence and patterns of dataset bias as the likely driver of model degradation across domains.

For experimentation, we select 2 hospitals from the Ubenwa database that had sufficient sufficient samples to be split into their own training and test set. We will refer to these as Hospital A and Hospital B, each of which has its own training and test sets. More details about the data from each hospital is provided in section \ref{sec:da-experiment-data} below.

\subsection{Cross-hospital generalization}
\label{sec:cross-hosp-gen}
The first set of experiments we conducted were to evaluate the generalizability of models trained on one hospital and tested on another. In the previous chapter, all hospitals were treated as a joint training set, and tested as well on their joint test set. But what if a model was trained on only one hospital and tested on another, would it perform at the same accuracy?

We train a classifier to test how well our model generalizes across hospitals. In 2 sets of experiments, using the test sets of hospitals A and hospital B, we compare 3 models: model trained on hospital A (Model-HA), model trained on hospital B (Model-HB) and model trained on both (Model-HA+HB). The expectation is that if there is no bias in the datasets, Model-HA and Model-HB would have approximately the same accuracy regardless of which hospital they were tested on, while Model-HA+HB may have a higher accuracy on the test sets since it had access to more training data. Note that training is always done using the training portion of the hospitals' data, while testing is always carried out on the test portion.

\begin{figure}[htbp]
\label{fig:cross-hosp-gen}
  \centering
  \centerline{\includegraphics[width=.5\linewidth]{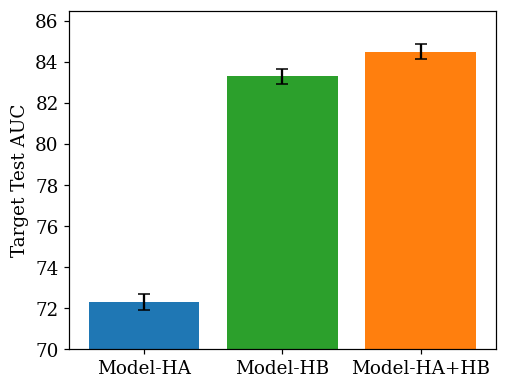}}
\caption{Cross-hospital generalization results showing that model trained on only source data does not generalize well to target domain. "Source" was trained on only source data, "Target" was trained on only target data, while "Source+Target" was trained on source and target data. All models were tested on Target test set.}
\end{figure}

For the experiments, we used the "CNN14" network (see section \ref{sec:cnn14-desc}) as classifier employing the same training parameters and configuration. The results are presented in figure \ref{fig:cross-hosp-gen}. We see clear patterns of bias in the data set. Model-HA performs better on hospital A's test set than it does on hospital B's test set, while Model-HB performs better on hospital B than on hospital A. Model-HA+HB, as expected, is the best performing model in both test sets.

\subsection{Spectral distribution}
\label{sec:spectral-dist}
Here we investigate the possibility that the cry recordings of newborns recorded from different hospitals are truly distributed differently, in spite of the fact that the same recording devices were used across hospitals and that only infant cries were recorded and analyzed (all other non-cry sounds were manually segmented). The poor cross-hospital generalization seen in the previous experiment suggests that the distribution of the data is not quite the same.

To achieve this, we compare at the distribution of energies across the spectrum for each data set. We separate healthy from sick patients since by definition, their spectral patterns would be different. Concretely, we select 100 healthy patients from Hospital A and 100 healthy patients from Hopsital B. For each healthy baby in hospital A, we compute a short-time fourier transform of their recording to get the spectrogram. Then the spectrograms across all patients are averaged through time to get an $N$-dimensional vector representing the average distribution of spectral energies, where $N$ is the number of frequency bins. The $N$-dimensional standard deviation is also computed through time in a similar fashion to get the error bars. We repeat this process for healthy recordings from hospital B, as well as for sick recordings from both hospital A and B. 



\begin{figure}[htbp]
\label{fig:spect-dist-healthy}
  \centering
  \centerline{\includegraphics[width=.6\linewidth]{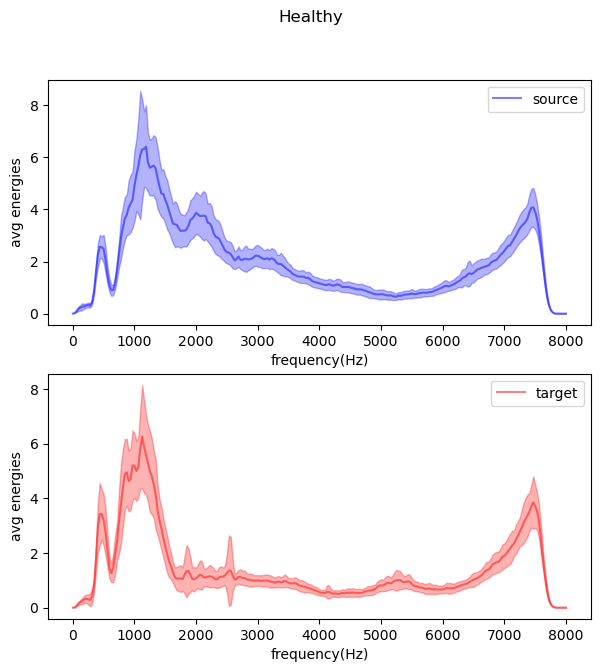}}
\caption{Average distribution of spectral energy for healthy patients in hospital A (top) and hospital B (bottom)}
\end{figure}

\begin{figure}[htbp]
\label{fig:spect-dist-sick}
  \centering
  \centerline{\includegraphics[width=.6\linewidth]{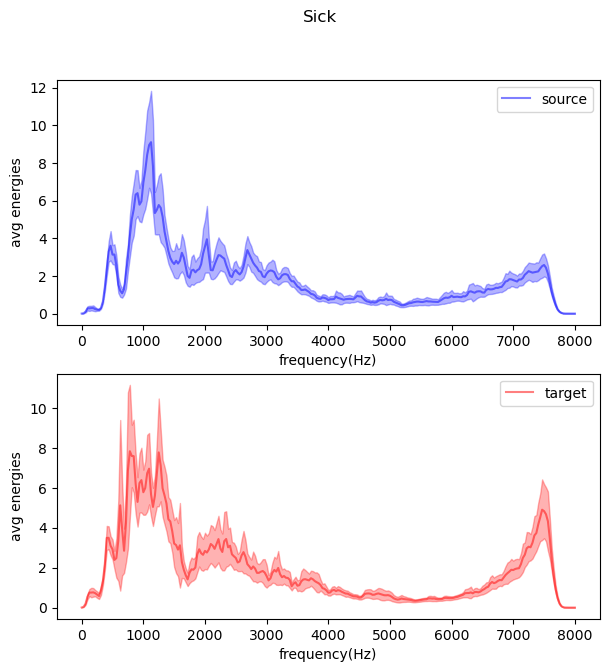}}
\caption{Average distribution of spectral energy for sick patients in hospital A (top) and hospital B (bottom)}
\end{figure}

The resulting 4 graphs of the average distribution of spectral energies are shown in figures \ref{fig:spect-dist-healthy} and \ref{fig:spect-dist-sick}. We find that the distribution of spectral energy between hospital A and hospital B have some notable differences regardless of whether it is healthy or sick patients being compared. Precisely, in the healthy patients we observe higher energies in the $2000 - 3000$Hz range in hospital A than we see in hospital B. Hospital B also attains a small peak around $5,500$Hz, which is not observed in hospital A.

\subsection{Name-the-hospital challenge}
\label{sec:name-the-hosp}
How significant are these differences in spectral energies? We develop a simple but informative measure: name-the-hospital challenge which is inspired by {\em Name That Dataset} in \cite{5995347}. In that work, the authors discovered that not only can humans identify which visual benchmark dataset an image sample came from but that this was a relatively easy task for a linear support vector machine model. The findings supported the theory that inspite of best efforts, human inevitably incorporate biases into collected data.

To play the name-the-hospital challenge, we ask the question: given a cry recording of a baby, can we build a classifier to identify in which hospital it was recorded? Domain shift is characterized by bias in the datasets. If such bias exists, the classifier would be accurate identifying which hospital a sample comes from. If it doesn't, the classifier should struggle to distinguish recordings from different hospitals.

\begin{figure}[htbp]
\label{fig:name-the-hospital}
  \centering
  \centerline{\includegraphics[width=.6\linewidth]{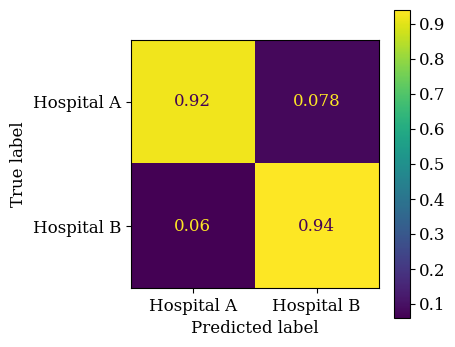}}
\caption{Confusion matrix of a simple classifier trained to predict the originating hospital of a cry recording. Results suggest that audio recordings contain information that uniquely relates to the domain in which they were recorded.}
\end{figure}

We implement a CNN14-based model as a binary classifier using the hospitals as target labels. The confusion matrix in Figure \ref{fig:name-the-hospital} shows that the model achieves greater than 90\% AUC true positive and true negative rates, suggesting the presence of strong biases which effectively makes each hospital data a different dataset. What is the source of this bias? Is it the infant cry sounds, human speech, environmental noises or something else?

\newpage

\subsection{Addressing sources of bias}
\label{sec:pitch-distribution}
To wrap up our investigation on the existence and patterns of domain shift in this clinical database, we zoom in on the possible underlying sources of the observed bias:

\begin{enumerate}
    \item Microphone. The kind of microphone used to record could introduce biases into the data given that difference acoustic sensors behave differently over the frequency spectrum, allowing and modulating certain frequencies. We conclude that the recording microphone is not the source of bias since the same device type, a Samsung Galaxy A10 was used in all hospitals.
    \item Infant cries. Could the bias come from the infant cries themselves? Previous research have indicated that newborn crying is an involuntary action directly coordinated by the central nervous system. This suggests that babies' cries are effectively drawn from the same underlying distribution and thus should not be distributed differently by geography or location.
    \item Adult speech. If adult speech were present in the database, this would add bias that is unique to who is speaking and what was being said, effectively corrupting the spectrograms of cries. In this work, adult speech that occured in the database (eg between infant crying or prior to cry activation) was removed through manual annotation. So this couldn't be the source.
    \item Environmental noise. Lastly, noise would corrupt the cry spectrograms. For example a siren in the background would effectively change the spectral signature. Like speech, non-overlapping noises were removed through manual annotation. However, noises that overlapped with the cry sounds were difficult to remove and may in the database. It is actually useful to have naturally occurring noises as this increases the chances that models learn robust features. As we'll see shortly these noises could also be the source of significant data biases, requiring careful treatment to extract their benefits and minimize their harms on the final model.
\end{enumerate}

Eliminating \#1 and \#3, it appears that the 2 possible sources of bias in this database are the infant cries themselves and environmental noise. Since there is no straight forward approach to visualize or compare these noises which themselves may comprise a variety of sounds, we take the approach of elimination. If we compare infant cries across hospitals and they turn out to be similarly distributed, then it must be that environmental noise is the main source of bias.

In the case of an infant cry, one of the most important features is pitch or fundamental frequency \cite{corwin1996infant}, which corresponds to the rate of vibration of the vocal cords during a cry expiration and defines the harmonic properties of infant cries such as the formants \cite{golub1985physioacoustic}. We compare pitch distributions across the two domains as a way of isolating where bias in the dataset might be coming from. To do this, we use a cry activity detection model to segment clean cry sounds, then run a pitch estimator (CREPE \cite{8461329}) to obtain the pitch tracks per cry utterance.

\begin{figure}[htbp]
\label{fig:pitch-distribution}
  \centering
  \centerline{\includegraphics[width=.6\linewidth]{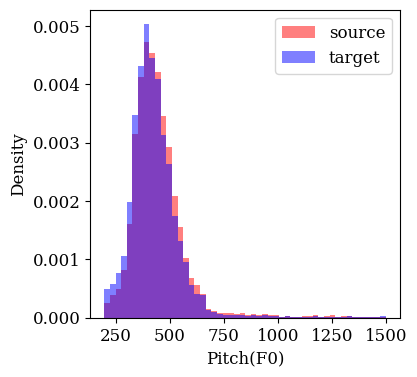}}
\caption{The pitch distribution of cry recordings from source and target domains suggests that dataset bias does not emanate from the cry signals.}
\end{figure}

In figure \ref{fig:pitch-distribution}, we compare the distribution of pitches of cries in hospital A to the cries in hospital B. Both distributions are almost perfectly matched as seen in the overlapping histograms in figure \ref{fig:pitch-distribution}. This finding is consistent with previous research demonstrating that the newborn cry is a pure signal not altered by a baby's genes or birthplace. Furthermore, it strongly suggests that the bias observed in our database (see sections \ref{sec:cross-hosp-gen}, \ref{sec:spectral-dist}, \ref{sec:name-the-hosp}) is driven by the environmental sounds. 

A corollary of these findings is that the most effective domain adaptation methods should address ambient noise and ultimately embed samples into a space where they are indistinguishable by the kind of noise occurring in the background. If we achieve this, the generalization performance of our models should go up.

\section{Methods II: Learning domain-invariant classifiers for sound}
We frame our solution in the context of unsupervised domain adaptation. We study 5 different paradigms for the unsupervised DA, some of which were originally formulated for computer vision but have not been studied extensively in audio and rarely in medical audio. We implement canonical, state-of-the-art methods in each of these categories, adapt these methods to our problem set and study the trade-offs between them. We further propose two new methods for unsupervised DA for medical audio. One of our methods, normalized multi-layer kernel regularization, achieves best performance over all approaches.


\textbf{Preliminaries} \\
In the unsupervised domain adaptation setting, our main assumption is that we have labelled data in the source domain, and unlabelled data in the target domain. Our goal is to use the labeled source samples (infant cry sounds from one hospital plus target - sick or healthy) in conjunction with unlabeled target samples (cry sounds from another hospital) to learn a classifier that generalizes well to the target domain. To conduct this evaluation, we ideally need a small labeled test set from the target domain on which to evaluate our models. We set out the notations used in the following subsections as follows:

Given a source domain $\mathcal{D}_s = \{x_i^s, y_i^s\}_{i=1}^{n_s}$, comprising $n_s$ labelled samples with $|\mathcal{C}_s|$ categories, while the target domain is $\mathcal{D}_t = \{x_i^t\}_{i=1}^{n_t}$ comprising $n_t$ unlabelled samples with $|\mathcal{C}_t|$ categories. In the standard setting of unsupervised DA, the source and target domain share an identical label space $|\mathcal{C}_s| = |\mathcal{C}_t|$. The goal is to learn a feature extractor $F$ and a classifier $C$ to minimize the expected risk in the target domain, $\E_{(x^t, y^t) \sim \mathcal{D}_t} [\mathcal{L}(C(F(x^t)), y_t)]$, where the loss function $\mathcal{L}_c$ is the canonical cross-entropy loss used in classification.

\subsection{Discrepancy-based kernel regularization}
\label{sec:kernel-reg-discrepancy}
The kernel regularization framework involves adding a penalty term to the classification objective to encourage domain confusion. Neural nets typically transition from general to task-specific features as we go from lower to higher layers \cite{yosinski2014transferable}. Thus this penalty or regularization term is computed on kernels at higher layers of the network since these are the most likely to embed domain-specific information.

Consider a neural network with $l$ layers. We denote the output of the representation on the penultimate layer, $(l-1)$th as $\phi(\cdot)$. Given a sample $x$, the embedding vector, $\phi(x)$, is the input to the final layer which computes class probabilities using the softmax operator. To regularize the model, we incorporate a measure of domain discrepancy to the loss function. Our goal is to minimize the distance between domains (or maximize domain confusion), while achieving our classification objective. 

The domain discrepancy statistic estimates how distinguishable a source embedding $\phi(x_s)$ is from a target one $\phi(x_t)$. Several domain discrepancy measures have been proposed including maximum mean discrepancy (MMD) \cite{tzeng2014deep}, maximum mean feature norm discrepancy \cite{xu2019larger}, correlation distance \cite{sun2016deep, sun2016return}, $\mathcal{H}$-divergence \cite{ben2010theory} and $\mathcal{H}\Delta\mathcal{H}$-divergence \cite{ben2010theory}. Here we implement MMFND, which derives from the mean maximum discrepancy, that is, the difference between the kernel means. MMD is computed empirically by:

\begin{equation}
MMD = \| \frac{1}{n_s} \sum_{x_s \in \mathcal{D}_s} \phi(x_s) - \frac{1}{n_t} \sum_{x_t \in \mathcal{D}_t} \phi(x_t)\|
\end{equation}

The maximum mean feature norm discrepancy (MMFND) was first introduced in \cite{xu2019larger}. It improves on the MMD by restricting the mean feature norms of each domain, encouraging them towards a common restrictive scalar $R$, thereby increasing the chance that the domain gap will vanish to zero. The discrepancy loss given by MMFND is:

\begin{equation}
\mathcal{L}_d = d (\frac{1}{n_s} \sum_{x_s \in \mathcal{D}_s} \phi(x_s) , R) + d (\frac{1}{n_t} \sum_{x_t \in \mathcal{D}_t} \phi(x_t), R),
\end{equation}

where $d(\cdot, \cdot)$ represents the $L_2$ distance. Combining the classification loss with the MMFND loss above, we get the hard adaptive feature norm (HAFN) algorithm \cite{xu2019larger} which maximizes domain confusion while training a strong classifier using labeled source domain data:

\begin{equation}
\mathcal{L} = \mathcal{L}_c + \lambda \mathcal{L}_d
\end{equation}

where $\lambda$ is a hyperparameter which controls how much the regularization term contributes to the loss.

The HAFN algorithm is effective in learning domain-invariant, task-discriminative models even at fairly small values of $R$, but has some draw backs. Setting large values of $R$ can cause the gradients generated explode. Given that accuracy can continue to increase for larger norms (larger $R$), stepwise adaptive feature norm (SAFN) was introduced to address this by effectively the norm constraint in a progressive, stepwise fashion throughout learning until saturation. SAFN is given by:

\begin{equation}
\mathcal{L}_d = \frac{1}{n_s + n_t} \sum_{x \in \mathcal{D}_s \cup \mathcal{D}_t}  d (\phi(x; \theta_0) + \Delta r, \phi(x; \theta))
\end{equation}

where $\theta_0$ and $\theta$ represent the model parameters from the previous and current iterations respectively. SAFN encourages larger norms that are more informative and leads to better target performance.


\subsection{Entropy-based kernel regularization}
\label{sec:entropy-min}
We adapt entropy minimization \cite{grandvalet2004advances} for our problem of unsupervised domain adaptation. Entropy minimization was first proposed in the semi-supervised learning context as a way of leveraging unlabelled data to improve generalization ability of learned models \cite{grandvalet2004advances, zhang2018importance}. For our use case, we consider the source hospital as the set of labelled data available for training and the target hospital as the set of unlabeled data available during training. The hypothesis is that by regularizing the supervised model using a measure of the entropy of predictions on the unlabeled data, we ultimately achieve a model whose representations is less able to distinguish source versus target hospital samples, but yet is able to distinguish the classes of interest within each hospital database. In other words, the model is encouraged to make more confident and less uncertain predictions on the target domain.

To achieve this, we optimize an extra term in conjunction with the cross-entropy loss:

\begin{equation}
    \min_{G, C} \mathcal{L}_{em} = -\frac{1}{n_s} \sum_{x_s,y_s \in D_s} \log(p_{y_s} (x_s)) - \lambda \frac{1}{n_t} \sum_{x_t \in \mathcal{D}_t} \sum_{k=1}^K p_k (x_t) \log (p_k (x_t))
\end{equation}

where $p_k (x)$ is the probability that the sample $x$ is of class $k$ as output by the model with feature extractor $G$ and classifier $C$. Not that the first term is the supervised cross-entropy loss computed over all labeled source samples, while the second term is entropy regularizer using only unlabeled samples from the target domain. $\lambda$ is a hyperparameter that trades off the impact of the regularizer during training.

\subsection{Domain-adversarial training}
In the domain-adversarial training framework, we solve the problem of unsupervised domain adaptation by pitting 2 competing optimization objectives against each other. One one hand we are learning a discriminator to distinguish source from target samples, while on the other hand we are learning a feature extractor to output representations that are invariant enough to fool the discriminator. Domain alignment is achieved when the discriminator is unable to accurately classify samples by domain yet the features extracted are task-discriminative. Domain-adversarial training is largely inspired by the literature on generative adversarial networks (GANs) \cite{goodfellow2014generative}.

A key aspect of domain-adversarial training is the loss used to learn the feature extractor given a domain discriminator. Previous work has emplohyed at least three different approaches including minimax \cite{ganin2016domain, ganin2015unsupervised}, domain confusion \cite{tzeng2015simultaneous, long2017unsupervised, zhang2019domainsymmetric} and GAN loss \cite{tzeng2017adversarial, shen2018wasserstein}. In this work, we implement and experiment with domain confusion with a focus on Domain-Symmetric Networks (SymNets)\cite{zhang2019domainsymmetric} to align not only the hospital domains but also the similarity structure between the label categories. Such alignment of label distribution has been shown to facilitate more accurate domain transfer \cite{tzeng2015simultaneous, zhang2019domainsymmetric}.

The overall training objective for SymNets involves iteratively optimizing 2 loss functions given by Eq \ref{eq:symnet-loss}. The first loss updates the source task classifier $C^s$, an explicit target task classifier $C^t$ and a constructed classifier $C^{st}$ which handles domain discrimination. While the second loss represents the category- and domain-level confusion losses and updates the feature extractor accordingly. The objective is given by:

\begin{equation}
\label{eq:symnet-loss}
\begin{aligned}
    \min_{C^s,C^t,C^{st}} \mathcal{L}_s (G, C^s) + \mathcal{L}_t (G, C^t) + \mathcal{L}_{st} (G,C^{st}) \\
    \min_{G} \mathcal{F}_{category} (G,C^{st}) + \lambda (\mathcal{F}_{domain} (G,C^{st}) + \mathcal{M}(G,C^{st})),
\end{aligned}
\end{equation}

where $G$ is the feature extractor, $C^s$ is the source task classifier predicting $K$ classes, $C^t$ is the explicit target task classifier also predicting $K$ classes. Considering our model to be a standard convolutional neural network like the CNN14, the feature extractor is typically the convolutional layers of a network, the task classifiers ($C^s$ and $C^t$) are the fully-connected layers ending with a softmax. As seen above, SymNets have an additional classifier $C^{st}$ which is constructed from a pre-softmax concatenation of the outputs of $C^s$ and $C^t$, with a softmax applied over the $2K$ dimensions to give a single probability distribution.

We now describe briefly each term in the optimization objective given in Eq. \ref{eq:symnet-loss}. The source classifier loss $\mathcal{L}_s (G, C^s)$ is a simple cross-entropy loss between the model's predicted probabilities and the true labels of source samples. Since there exists no labeled target samples to provide supervision to the target task classifier (unsupervised DA), the target classifier loss $\mathcal{L}_t (G, C^t)$ is also computed as a cross-entropy over the labeled source samples, but using probabilities $p^t$ generated by the target classifier. The domain discrimination training via $C^{st}$ ensures that this is not a duplicate of the source classifier, and in fact, this approach helps to facilitate category-level alignment such that classes that are similar in the source subspace remain similar in the target subspace. Formally, both loss are given by:

\begin{equation}
    \begin{aligned}
        \mathcal{L}_s (G, C^s) = - \frac{1}{n_s} \sum_{x_s,y_s \in D_s} \log(p^s_{y_s} (x_s)) \\
        \mathcal{L}_t (G, C^t)= - \frac{1}{n_s} \sum_{x_s,y_s \in D_s} \log(p^t_{y_s}(x_s)) 
    \end{aligned}
\end{equation}


Since $C^s$ and $C^t$ are both trained on source samples, the constructed classifier $C^{st}$ serves to discriminate between them. To achieve this, $C^{st}$ is updated by the following loss term:

\begin{equation}
    \mathcal{L}_{st} (G,C^{st}) = -\frac{1}{n_s} \sum_{x_s \in \mathcal{D}_s} \log (\sum_{k=1}^K p^{st}_k (x_s)) - \frac{1}{n_t} \sum_{x_t \in \mathcal{D}_t} \log (\sum_{k=1}^K  p^{st}_{k+K} (x_t)) 
\end{equation}

where $\sum_{k=1}^K p^{st}_k (\cdot)$ and $\sum_{k=1}^K  p^{st}_{k+K} (\cdot))$ are effectively the probabilities of classifying a sample as source and target domain respectively.

Given a discriminator $C^{st}$, we aim to learn an invariant feature extractor $G$. Domain-adversarial training methods generally achieve this using domain confusion loss between the output predicted domain labels and a uniform distribution over domain labels \cite{tzeng2015simultaneous}. SymNet on the other hand adopts a two-level domain confusion training that is based on a domain-level confusion loss and a category-level confusion loss. The domain-level confusion loss uses only unlabeled target domain samples but obtains predictions from the domain-specific task classifiers.

\begin{equation}
    \mathcal{F}_{domain} (G,C^{st}) = -\frac{1}{2n_t} \sum_{x_t \in \mathcal{D}_t} \log(\sum_{k=1}^K p^{st}_k (x_t)) -\frac{1}{2n_t} \sum_{x_t \in \mathcal{D}_t} \log(\sum_{k=1}^K p^{st}_{k+K} (x_t))
\end{equation}

The category-level confusion loss on the other hand relies on labeled source samples, and uses a cross-entropy between discriminator predictions and a uniform distributions.

\begin{equation}
    \mathcal{F}_{category} (G,C^{st}) = -\frac{1}{2n_s} \sum_{x_s, y_s \in \mathcal{D}_s} \log (p^{st}_{y_s} (x_s))-\frac{1}{2n_t} \sum_{x_s, y_s \in \mathcal{D}_s} \log (p^{st}_{{y_s}+K} (x_s))
\end{equation}

Lastly, we have the entropy minimization loss term $\mathcal{M}(G,C^{st})$ which is also used to update the feature extractor. Variations of entropy minimization has been adopted as a standalone domain adaptation method in previous work \cite{zhang2018importance} as well as in this paper as described in section \ref{sec:entropy-min}. SymNets employ an entropy minimization objective that facilitates discrimination at the level of the label categories across domains. We achieve this by summing over the probabilities at each source-target pair of category-corresponding neurons in $C^{st}$:

\begin{equation}
\label{eq:symnet-em}
    \mathcal{M}(G,C^{st}) = -\frac{1}{n_t} \sum_{x_t \in \mathcal{D}_t} \sum_{k=1}^K q_k^{st} (x_t) \log (q^{st}_{k} (x_t))
\end{equation}

where $q_k^{st}(\cdot) = p_k^{st}(\cdot) + p_{k+K}^{st}(\cdot)$. In practice Eq \ref{eq:symnet-em} is only used to update the feature extractor $G$ and not $C^{st}$. This minimizes the chance that target samples get stuck early in training in wrong label predictions due to too large a domain shift \cite{zhang2018importance}. As observed in the full SymNet training objective (Eq \ref{eq:symnet-loss}), a hyperparameter $\lambda$ is applied to the domain confusion $\mathcal{F}_{domain} (G,C^{st})$ and the entropy minimization $\mathcal{M}(G,C^{st})$ loss terms to control them, since they can be quite noisy at the start of training due to being dependent on unlabeled target samples.


\subsection{Adaptive batch normalization}
Similar to other frameworks, adaptive batch normalization takes the core assumption that domain transfer is challenging due to differences between source and target domain distributions. It however goes further to posit that models primarily capture domain-related knowledge in the statistics of the batch normalization layer. (while task related is store din the wieght matrix). To adapt a model, one only needs to update these BN statistics with the target domain and bob is your friend. This framework is nice in that it does not require additional components and is parameter-free ie no parameters to tune

Batch normalization \cite{ioffe2015batch} has become a standard component in many neural network architectures \cite{he2016deep, szegedy2016rethinking}. It enables models to converge faster and also improves accuracy -- all by guaranteeing that input distributions of each layer remain unchanged across different mini-batches during training. Considering an input $\mX \in \mathbb{R}^{n \times p}$, where $n$ is the batch size and $p$ is the number of features, BN transforms a feature $j \in \{1\dots p\}$ in two steps: (1) computes a standardized version $\hat{x}_j$ of each feature in a mini-batch, then (2) compute a new neuron response $y_j$ based on a learned slope $\gamma_j$ and bias $\beta_j$, shared across all mini-batches. Formally:

\begin{equation}
\begin{aligned}
    \hat{x}_j = \frac{x_j - \mathbb{E}[\mX_{.j}]}{\sqrt{Var[\mX_{.j}]}}, \\
    y_j = \gamma_j\hat{x}_j + \beta_j,
\end{aligned}
\end{equation}

where $\mX_{.j}$ denotes the $j^{th}$ column of the input. During the testing, the global statistics of all training samples is used to normalize every minibatch of test data before computing the layer BN output $y_j$ using the learned slope and bias.  By stabilizing the input distribution, batch normalization facilitates faster model convergence with fewer iterations to more accurate, robust solutions.

Adaptive batch normalization (AdaBN) was motivated from the finding that different datasets have different batch norm statistics, even when passed through the same model with fixed weights. \cite{li2016revisiting} conducted an example with 2 image datasets -- Caltech-256 and Bing images search results, showing that a linear SVM could almost perfectly classify which BN statistics (i.e., means and variances) came from which datasets. Consequently, the AdaBN algorithm is quite simple. Given a pre-trained model on some source domain, we adapt it to a target domain by computing the global BN statistics using target domain data. Then during testing in the target domain, we normalize each sample by the saved statistics per neuron/feature, then compute the BN output using the already pre-learned slope and bias. Formally, we define AdaBN in algorithm \ref{alg:adabn} :

\begin{algorithm}
\caption{Adaptive Batch Normalization}\label{alg:adabn}
\begin{algorithmic}
\For{neuron $j$ in DNN}
    \State Concatenate neuron responses on all images of target domain t: \\ \qquad $\vx_j = [\dots, x_j(m),\dots]$
    \State Compute the mean and variance of the target domain: \\ \qquad $\mu_j^t = \mathbb{E}(\vx_j^t), \sigma_j^t=\sqrt{Var(\vx_j^t)}$
\EndFor
\For{neuron $j$ in DNN, testing image $m$ in target domain}
    \State Compute BN output $y_j(m)= \gamma_j \frac{(x_j(m) - \mu_j^t)}{\sigma_j^t} + \beta_j$
\EndFor
\end{algorithmic}
\end{algorithm}


\subsection{Target noise injection}
Sequel to the insights from section \ref{sec:pitch-distribution}, we propose a relatively simple approach for domain adaptation -- target noise injection (TNI). In this method, we segment, extract, and inject target domain noise into source samples during training. The intuition is that if the environmental noise in the target hospital is the primary source of domain shift, then training the source classifier to be robust to such noise could enable the classifier to learn effective, cross-domain representations. 

This approach has practical benefits. Data collection only requires recording noises in the target environment -- no need for labels or actual cry recordings -- which can be much cheaper and faster to accomplish. During model training, this method requires no modification to the loss function (standard cross-entropy is sufficient) unlike kernel regularization methods. It requires no change to the model architecture, and no complicated training paradigm such as in domain-adversarial training. 

Given an audio sample $x_i$ from the source domain, it is augmented as $x_i' = x_i + \alpha \eta_i$, where $\eta_i$ is the $i^{th}$ randomly sampled noise from the pool of target hospital noises, and $\alpha \in [0, 1]$ is a hyperparameter to trade-off the amount of target noise to consider. When applying the target noise injection the loss function we minimize is therefore:

\begin{equation}
    \min_{G, C} \mathcal{L}_{tni} = - \frac{1}{n_s} \sum_{i=1}^{n_s} y_i\log(p (x_i + \alpha \eta_i)),
\end{equation}

Target noise injection can be seen as a special case of noise pertubation in neural networks.

\section{Experiments}
To evaluate the proposed methods, we setup an unsupervised domain adaptation problem using the Ubenwa cry database described in chapter \ref{chap:ssl-infantcry}. We select two hospitals based on available sample sizes which we term Hospital A and Hospital B. The database consists of cry recordings taken in the respective hospitals with each recording annotated as either healthy or neurological injury based on clinical exams conducted by doctors. We study the domain transfer problem in both directions, that is using labeled data from Hospital A as source and unlabeled data from Hospital B as target (HA --> HB) and using labeled data from Hospital B as source and unlabeled data from Hospital A as target (HB --> HA). In the experiments described subsequently, we benchmark the 5 different methods -- discrepancy-based kernel regularization, entropy-based kernel regularization, domain-adversarial training, adaptive batch normalization and target noise injection -- on this two-way unsupervised domain transfer task.

\subsection{Experimental setup}
\subsubsection{Data description}
\label{sec:da-experiment-data}
Hospital A contained a total of 406 recordings which were split into train, validation and test sets of 186, 70 and 150 recordings, respectively. While Hospital B contained a total of  1507 recordings which were split into train, validation and test sets of 1335, 43 and 129 recordings, respectively. There were a total of 284 patients in the source hospital and 910 in the target hospital. When making the splits, we ensured that all recordings from an individual patient belonged to only one set. Recordings were between 30s to 3min long. The data is summarized in Table \ref{tab:da_data_split}.

\setlength\arrayrulewidth{1.0pt}
\begin{table}[h!]
    \centering
    \begin{tabular}{l ccc | ccc}
        \hline
        & \multicolumn{3}{c}{\textbf{Hospital A (Source)}} & \multicolumn{3}{c}{\textbf{Hospital B (Target)}} \\
         & Train & Val & Test & Train & Val & Test \\
        \hline
        \textbf{\# recordings} & 186 & 70  & 150 & 1335 & 43  & 129 \\
        \textbf{\# patients} & 128 & 50  & 106 & 792  & 28  & 90  \\
        \textbf{\# hours} & 1.75 & 0.71 & 1.52 & 12.07 & 0.32 & 1.13 \\
        \hline
    \end{tabular}
    \caption{Summary of data used to evaluate unsupervised domain adaptation models for infant cry sounds.}
    \label{tab:da_data_split}
\end{table}

Data was processed in 3 steps to prepare it for model training: cry activity detection, cry unit segmentation and spectrogram computation. First, given that all recordings were taken in real-world settings other sounds such as doctors talking, machine beeps etc were captured alongside infant crying. The goal of cry activity detection step was to select all regions of the audio sample that contained a continuous sounds of the baby crying. Next, cry unit segmentation was performed to isolate only sections of audio that contained clear expiratory cry units. Cry expirations contain the harmonic content of crying and are the main input for cry analysis algorithms. This step and the previous were first done manually for some samples then a model was trained perform this pre-processing steps automatically. 

Lastly, as is standard when applying neural network to audio data, we transformed each cry unit into an audio spectrogram. To compute the log Mel spectrograms, raw audio recordings are downsampled to 8kHz and the short-time Fourier transform were computed for overlapping frame sizes of 30 ms with a 10 ms shift, and across 40 Mel bands. For each frame, only frequency components between 20 and 4000 Hz are considered. For each recording, this results to a cry spectrogram that is of dimensions $T \times 80$, where $T$ is the frame length in the frequency domain.


\subsubsection{Implementation details}
Our model consists of a backbone encoder followed by a classifier. We adopt, as an encoder, a CNN14 \cite{kong2020panns} which has about 80 million parameters. CNN14 is a convolutional neural network with 14-layers, constructed where each layer includes a batch normalization, ReLU activation and pooling, in addition to the convolution. The encoder is schematically shown in Table~\ref{tab:cnn14}. 

The encoder is pre-trained on VGGSound database, a large generic audio dataset containing a curated collection of 550 hours of sound from YouTube ~\cite{chen2020vggsound}. Pre-training is crucial in this setting given the relatively small amount of audio available for training our problem. For the classifier, we add one feed-forward layer on top of the backbone encoder. 

\setlength{\tabcolsep}{1pt}
\begin{table}[h]
    \centering
    \begin{tabular}{ll|rcccl}
    \multicolumn{2}{c|}{Blocks} & \multicolumn{5}{c}{Output dimension} \\ \hline
     \multicolumn{2}{c|}{Log-Mel Filterbank} & 1 & x & T & x & 80 \\ 
     2x[Conv(64)+BN+ReLU] &$\rightarrow$Pool 2x2      & 64  & x & T/2 & x  & 40 \\
     2x[Conv(128)+BN+ReLU] &$\rightarrow$Pool 2x2     & 128 & x  & T/4  & x & 20 \\
     2x[Conv(256)+BN+ReLU] &$\rightarrow$Pool 2x2    & 256 & x & T/8 & x & 10 \\
     2x[Conv(512)+BN+ReLU] &$\rightarrow$Pool 2x2    & 512 & x & T/16 & x & 5 \\
     2x[Conv(1024)+BN+ReLU] &$\rightarrow$Pool 2x2   & 1024 & x & T/32 & x & 2  \\
     2x[Conv(2048)+BN+ReLU] &                           & 2048 & x & T/32 & x & 2  \\
     \multicolumn{2}{c|}{Global Average Pooling} & \multicolumn{5}{c}{2048} \\ 
    \end{tabular}
    \caption{CNN14 blocks and output dimensions when processing T audio frames. Conv refers to 3x3 convolutions. T is 400 at training and arbitrary length at inference}
    \label{tab:cnn14}
\end{table}

We report test AUC as the mean and standard error across 5 random seeds using a model trained on the best hyperparameters. All models were trained using the Adam optimizer with a batch size of 32. For each method, learning rates for the backbone encoder and classifier were tuned as separate hyperparameters using the validation sets. In the discrepancy-based kernel regularization experiments, the HAFN model's norm constraint R is set to 30, and SAFN's $\Delta r$ is set to 0.2 following guidance from \cite{xu2019larger}.

Hyperparameter tuning was done using the validation portion of the labeled source domain data. For all methods (except SymNet), we found an optimal learning rate (lr) of $1e-4$ for both the encoder and classifier. The SymNets model achieved the best validation score with a Backbone Encoder learning rate (lr) of $5e-3$ and a classifier lr of $5e-4$. 

\subsection{Results}
Results for unsupervised domain adaptation (UDA) on infant cry data for classifying neurological injury are summarized in Table \ref{tab:main-result}. 

We find that all domain-adapted models outperform the unadapted model in the target domain. This is expected and consistent with previous work in computer vision where these methods were found to to be effective in learning a classifier that is functional across domains. The best model based on the amount of improvement to target AUC was the domain-adversarial method (SymNet), achieving an AUC increase of 7.2\% . The smallest improvement was observed with entropy minimization (0.7\%).

\begin{table*}[htbp]
    \centering
    \scalebox{.9}{\begin{tabular}{ccccc}
\hline
Methods & Loss Function & \begin{tabular}[c]{@{}c@{}}Req. Arch. Mod./\\ Req. target data\end{tabular} & \begin{tabular}[c]{@{}c@{}}Target test \\ AUC improvement\end{tabular} & \begin{tabular}[c]{@{}c@{}}Source test \\ AUC improvement\end{tabular} \\ \hline
No DA & CE & - & - & - \\ \hline 
EM & CE + EM & no/yes& 0.7\% & -5.3\% \\
Unsupervised BN & CE & no/yes& 1.00\% & -3.55\% \\
SAFN & CE + DC & yes/yes& 4.89\% & \textbf{8.33\%} \\
HAFN & CE + DC & yes/yes& 2.30\% & 6.80\% \\
SymNets & CE + DC + DD & yes/yes& \textbf{7.20\%} & 4.42\% \\
TNI & CE & no/no& 0.87\% & 5.1\% \\ \hline
\end{tabular}}
    \caption{Performance of different unsupervised domain adaptation methods. CE=Cross Entropy, EM=Entropy Minimization, DC=Domain Confusion, DD=Domain Discrimination. Other abbreviations are as used in this paper.}
    \label{tab:main-result}
\end{table*}

We also compare the models along 2 other dimensions: impact on source domain and complexity of solution. When adapting models in the clinical setting, we are interested in not only improving model performance in the target hospital but also preserving source hospital performance, such that the model remains useful across the board. We find that only 2 methods negatively impact source AUC -- entropy minimization and unsupervised BN. We suspect that, in the case of BN, this is due to its post-hoc nature i.e., the adaptation step is applied after the model has been trained as opposed to simultaneous training and adaptation which other methods have. Entropy minimization on the other hand is consistent with previous work in that reducing entropy of target domain alone is not enough to optimally confuse the source and target distributions. SAFN and HAFN have the largest improvements to the source AUC indicating that regularization using discrepancy measures between source and target domain has benefits in both domain.

The complexity of the solution matters in order to reduce the effort and computational resources required to adapt models to each new hospital. We compare our models on the complexity of their loss function, whether or not the DA solution requires architectural modifications and whether or not the DA solution requires data from the target domain (see Table \ref{tab:main-result}). We find that the 3 least complex solutions: entropy minimization, unsupervised batch norm and target noise injection, were also the 3 worst performing models on the domain transfer task (see target test AUC improvement). This highlights the difficulty of the domain adaptation problem especially in the unsupervised setting where no labels are available in the target domain. It suggests that the complexity of discrepancy-based regularization techniques (SAFN and HAFN) or domain-adversarial training methods (SymNets) are actually warranted.

\begin{figure}[htbp]
\label{fig:da_result_target_test}
  \centering
  \centerline{\includegraphics[width=.7\linewidth]{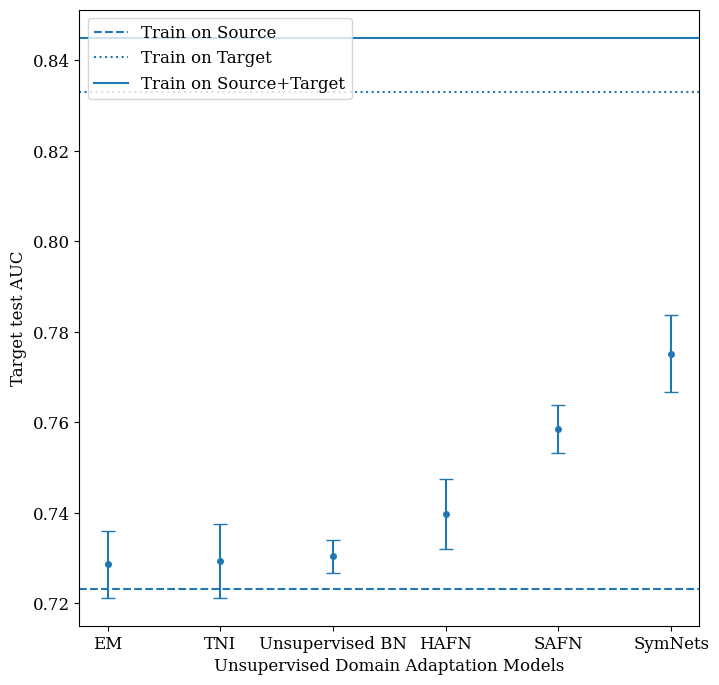}}
\caption{Performance of source $\rightarrow$ target adapted models on {\em target} test set.}
\end{figure}

\begin{figure}[htbp]
\label{fig:da_result_source_test}
  \centering
  \centerline{\includegraphics[width=.7\linewidth]{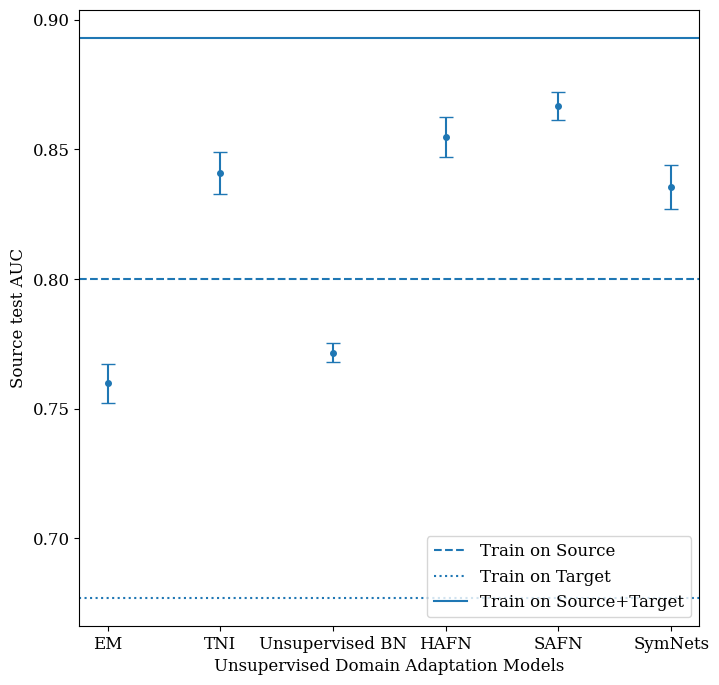}}
\caption{Impact of source $\rightarrow$ target adapted models on {\em source} test set.}
\end{figure}

We summarize our main results in more detail in Figs \ref{fig:da_result_target_test} and \ref{fig:da_result_source_test}. Here we compare our source-to-target unsupervised domain adaptation methods to 3 benchmarks: a model trained on source domain data (“trained on source”), a model trained on target (“trained on target”) and a model trained on data from both source and target domains (“trained on source + target”). In Fig \ref{fig:da_result_target_test}, we show the evaluation of all models on the target test set. All models perform better than the source only model at 72.5\% AUC showing that domain adaptation helps to different degrees. However none surpass the performance of the models that have access to labeled data in the target domain.

In Fig \ref{fig:da_result_source_test}, we examine in better detail, the impact of unsupervised domain adaptation when returning the models to the source domain. As already noted, only EM and unsupervised BN degrade the performance of the original source-only model, but again we find that the model with access to more labeled data even from a different domain outperforms all adapted models.

\subsection{Analysis}



\begin{figure}[htbp]
\label{fig:noise_sample_size}
  \centering
  \centerline{\includegraphics[width=.7\linewidth]{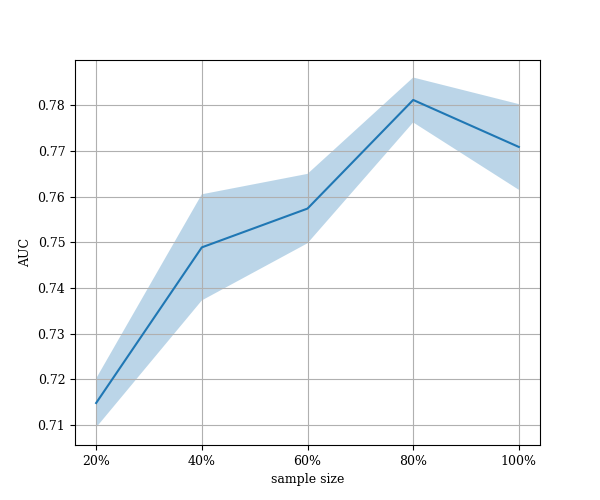}}
\caption{AUC in target domain as the amount of target domain noise is increased.}
\end{figure}
\begin{figure}[htbp]
\label{fig:tni-alpha}
  \centering
  \centerline{\includegraphics[width=.7\linewidth]{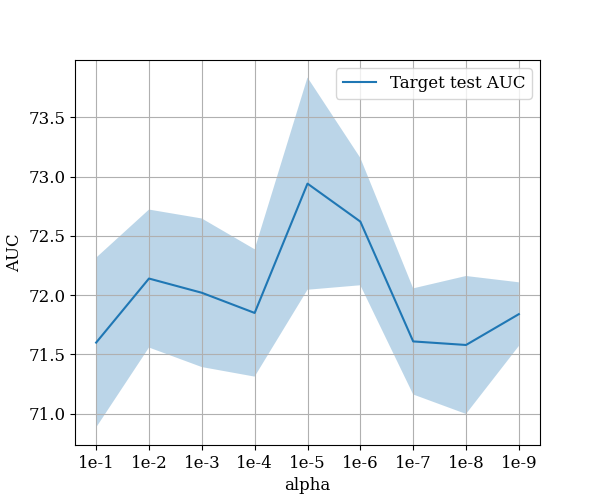}}
\caption{AUC in target domain for different values of $\alpha$, the weight on the target domain noise.}
\end{figure}

Though target noise injection (TNI) achieved a relatively small improvement of 0.87\% over the unadapted model on the target test set, sample size experiments (Fig \ref{fig:noise_sample_size}) indicate that the model is far from saturated and it improves as more target noise is collected. This is also supported by the fact that it achieves a much larger improvement of 5.1\% on the source test set. In Fig \ref{fig:tni-alpha}, we see as well that the value of $\alpha$ impacts the quality of the adaptation. Too small values would mean not enough signal to reap benefits, while too large values could degrade performance due to too much noise.

In terms of complexity, unsupervised batch norm and target noise injection offer the best deals. Both do not require a domain confusion loss and also require no modification to the original neural net architecture. Futhermore TNI, requires no training data in the target domain. It is worth noting that other more complex methods attain better improvements on the target test AUC. This leaves open the question of if those architectural updates are necessary in domain adaption or whether with access to more target domain noise, TNI might achieve comparable performance.


\section{Discussion}
Addressing domain shift and dataset bias is critical for the effective deployment of machine learning models in clinical applications, particularly for infant cry analysis. Variability arising from geographic, environmental, and recording conditions poses significant challenges to model generalization. This study demonstrates the efficacy of adapting domain adaptation methodologies originally developed for computer vision to the unique temporal and spectral characteristics of audio data. Techniques such as domain-adversarial training and adaptive normalization are systematically tailored to align with the intricacies of audio signals, thereby enhancing cross-domain generalization in medical audio datasets.

The introduction of Target Noise Injection (TNI) provides an innovative yet computationally efficient approach to mitigating domain discrepancies. By incorporating representative noise profiles from target domains during the training phase, TNI facilitates robust model performance in settings where labeled target data is unavailable. However, the method's reliance on the accurate representation of target noise as a proxy for domain-specific variability introduces potential limitations. TNI may fail to adequately address more complex or subtle domain shifts that extend beyond noise characteristics, particularly in scenarios involving significant divergences in acoustic properties or recording conditions.

An interesting observation from this work is the improvement in source domain performance (test AUC) when employing adaptation models. This counterintuitive result can be explained by the regularization effects of domain adaptation techniques, which encourage the model to focus on invariant and task-relevant features rather than overfitting to domain-specific idiosyncrasies. By aligning feature distributions across domains, these methods inadvertently refine the representation of the source domain as well, leading to improved performance.

While these methodologies achieve considerable improvements in domain-invariant representation learning, challenges remain in addressing extreme heterogeneity across datasets. The adaptation techniques assume a degree of similarity between source and target domains that may not always be present in highly diverse clinical data. Future research should investigate hybrid frameworks that integrate multiple domain adaptation strategies to account for broader variability, potentially leveraging advances in unsupervised and self-supervised learning to further enhance model robustness.



\chapter{CryCeleb: A Public Dataset for Cry-Based Infant Recognition}
\label{chap:cryceleb}
\section{Overview}
This paper describes the Ubenwa CryCeleb dataset - a labeled collection of infant cries - and the accompanying CryCeleb 2023 task, which is a public speaker verification challenge based on cry sounds. 
We released more than 6 hours of manually segmented cry sounds from 786 newborns for academic use, aiming to encourage research in infant cry analysis. 
The inaugural public competition attracted 435 submissions from 59 participants in 37 countries, 11 of whom improved the baseline performance. 
The top-performing system achieved a significant improvement scoring 25.8\% equal error rate, which is still far from the performance of state-of-the-art adult speaker verification systems. Therefore, we believe that there is room for further research on this dataset, which could extend beyond the verification task.

\section{Introduction}

Clinical research on the analysis of infant cries goes back to the 1960s~\cite{wasz1985twenty}. 
These days, machine learning techniques are demonstrating promising results in cry-based detection of cry reasons (hunger, pain, etc) and, more importantly, health pathologies, such as neurological injury~\cite{ji2021review,parga2020defining,gorin2023self}.

When deployed in hospitals or households with multiple babies, a cry analysis system should be able to identify the infant associated with the cry. Training such a model requires data with multiple recordings per infant. Given the complexity of data collection from newborns, such resources are extremely scarce. The most popular and diverse Chillanto database has 127 newborns but only one recording per infant~\cite{reyes2004system}.

In this work, we present the Ubenwa CryCeleb dataset, a first-of-its-kind collection of cries labeled with anonymized infant identities. Comprising 786 infants and 6.5 hours of cry expirations, with 348 infants recorded at least twice in different time frames (right after birth and pre-discharge from hospital), we aim to foster research in cry verification and, more broadly and importantly, to advance the field of infant cry analysis using verification as a proxy task. The dataset is available online\footnote{\url{huggingface.co/datasets/Ubenwa/CryCeleb2023}} under Creative Commons license.
In addition, we report on the first public baby verification competition and summarize the results.

\section{Dataset description}

CryCeleb2023 is a curated and anonymized subset of a large clinical study conducted by Ubenwa Health. In the next section, we describe the data collection and pre-processing steps. 

\subsection{Data preparation}
The original cry recordings were made either within an hour of birth or upon discharge from the hospital (typically within 24 hours of birth up to a few days). 
The cries were collected by medical personnel using the Ubenwa study application~\cite{onu2017ubenwa} and a Samsung A10 smartphone held at 10-15 cm from the newborn's mouth. 

Each recording was manually segmented by a human annotator into `expiration', `inspiration' or `no cry' segments. 
The CryCeleb dataset consists solely of the expiration segments, which we refer to as cry sounds. Inspirations (breath) are excluded as they are generally too short, hard to detect, and less likely to convey information about the vocal tract. Also, we manually removed any cry sounds containing personally identifiable information, such as human speech.

\subsection{Metadata and descriptive statistics}

This section summarizes the information about audio files included in the dataset and the associated anonymized metadata available for download.
Table~\ref{tab:dataset_stats} provides general statistics of the dataset.

\begin{table}[h]
\centering
\begin{tabular}{l|r}
\hline
Number of cry sounds (expirations)         & 26093 \\
Number of original recordings & 1372  \\
Number of infants             & 786   \\
Total cry time (minutes)      & 391  \\
\hline
\end{tabular}
\caption{Summary statistics of the CryCeleb dataset.}
\label{tab:dataset_stats}
\end{table}

The audio is accompanied by a metadata file with fields summarized in Table~\ref{tab:metadata_fields}. The 26093 rows of csv file provide complete information about the cry audio files.

\begin{table}[h]
\centering
\begin{tabular}{r|l}
\hline
\multicolumn{1}{c|}{\textbf{Field}} & \multicolumn{1}{c}{\textbf{Description}} \\
\hline
baby\_id            & Unique infant identifier. \\
period              & Recording time (birth or discharge). \\
duration     & Length of cry sound in seconds. \\
split               & Split for the challenge. \\
chrono\_idx & Chronological ordering of cries. \\
file\_name          & Path to cry sound. \\
file\_id            & Cry sound unique identifier. \\
\hline
\end{tabular}
\caption{Metadata fields available in the CryCeleb dataset.}
\label{tab:metadata_fields}
\end{table}

\begin{figure}[t]
\centering
\includegraphics[width=1.0\columnwidth]{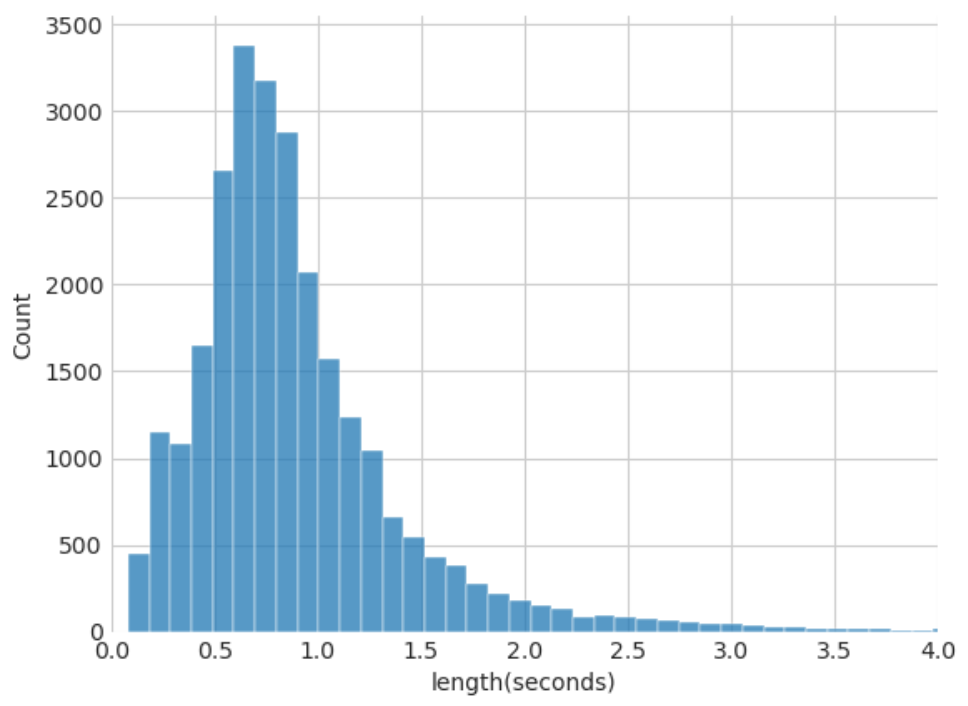}
\begin{minipage}[t]{0.9\columnwidth}
\caption{Histogram of cry sound durations.}
\label{fig:hist-len}
\end{minipage}
\end{figure}

\begin{figure}[t]
\centering
\includegraphics[width=1.0\columnwidth]{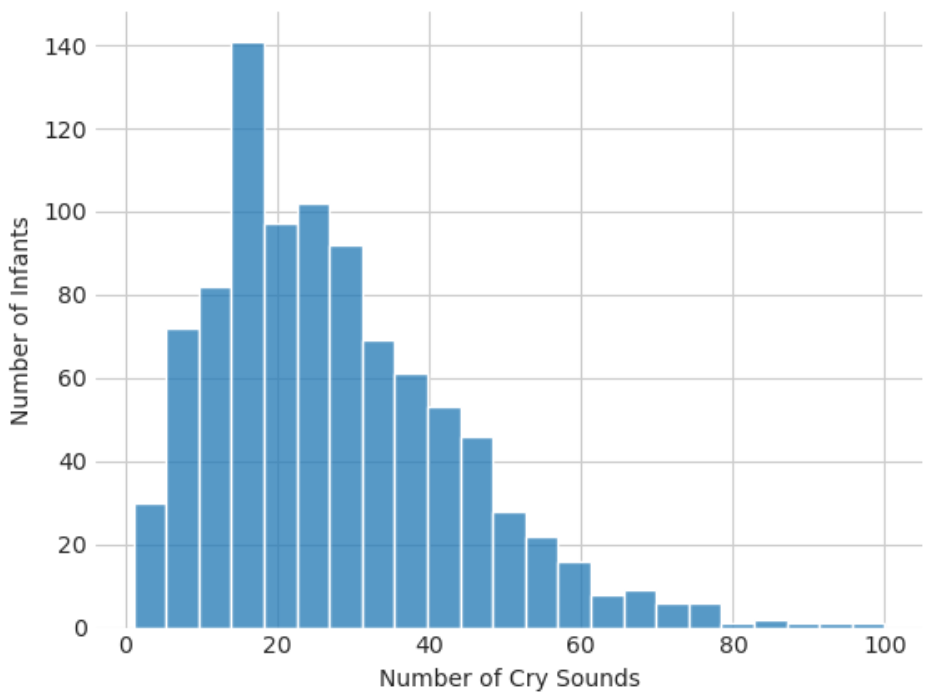}
\begin{minipage}[t]{0.9\columnwidth}
\caption{Number of infants per number of cry sounds.}
\label{fig:hist-numinf}
\end{minipage}
\end{figure}

Figures~\ref{fig:hist-len} and~\ref{fig:hist-numinf} show the distribution of cry sound duration and number of cries per infant. Most of the cry sounds are short (0.5 - 1.0 seconds) with only 0.3\% of expirations longer than 4 seconds. At the same time, there are multiple cry sounds corresponding to each infant. However, cry sounds (expirations) collected within one recording session, tend to have similar acoustic characteristics.

\section{CryCeleb 2023 Challenge}

\begin{figure}[h]
\centering
\includegraphics[width=0.9\columnwidth]{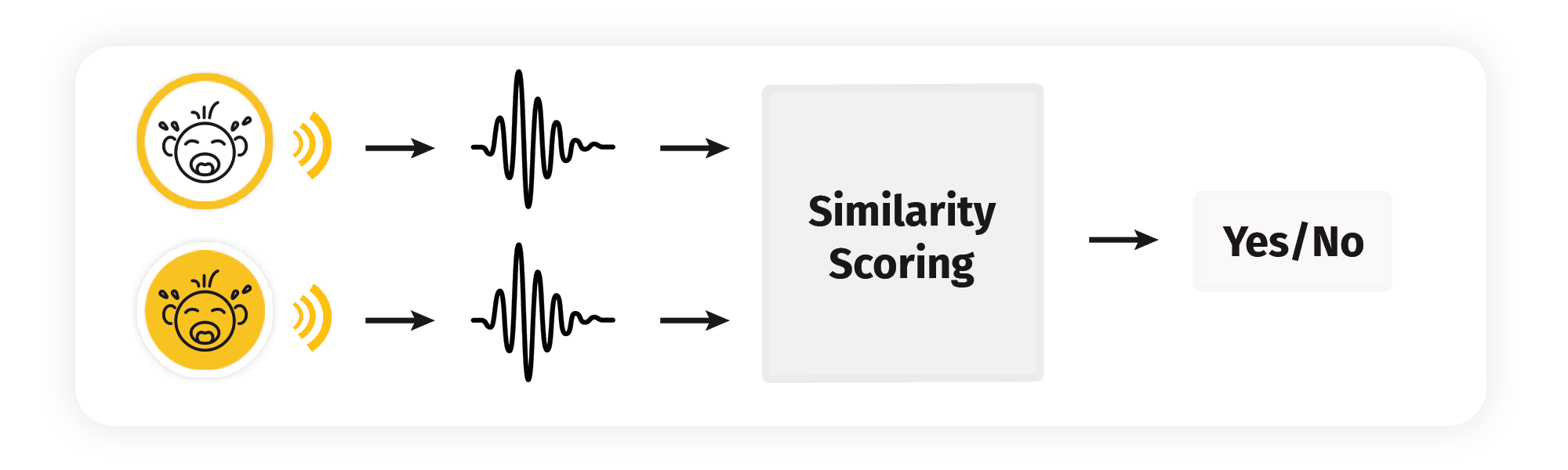}
\begin{minipage}[t]{0.9\columnwidth}
\caption{CryCeleb challenge verification task. Given two recordings, predict if they belong to the same infant}
\label{fig:verification}
\end{minipage}
\end{figure}

CryCeleb 2023 was a two-month machine learning competition\footnote{\url{huggingface.co/spaces/competitions/CryCeleb2023}} where contestants were asked to develop a system capable of determining whether two distinct cry recordings originated from the same infant (see Figure~\ref{fig:verification}). 

\subsection{Challenge design details}

The system should analyze any pair of cries and assign a similarity score to determine if the two sounds belong to the same baby, with an ideal system always assigning higher scores to positive pairs (two cry sounds from the same infant) than to negative pairs. 
For decision-making, a threshold can be applied to the assigned scores. If a score is greater than the threshold, it will indicate that the system accepts the two cries as belonging to the same infant. 

Submissions are ranked using the Equal Error Rate (EER)~\cite{brummer2006application}. The EER is the point on the ROC curve at which the false acceptance rate equals the false rejection rate. 
The lower the EER, the better.

For the CryCeleb2023 challenge, we have partitioned all infants into three sets: train (586 infants), dev (40 infants), and test (160 infants). All infants in the dev and test sets have recordings from both the birth (B) and discharge (D) periods. This is not true for all infants in the train set.

\begin{table}[h]
\begin{center}
\begin{tabular}{|c|c|c|c|}
\hline
 & \multicolumn{3}{c|}{\textbf{Split}} \\
\cline{2-4}
\textbf{Time(s) of Recording(s)} & \textbf{\textit{train}}& \textbf{\textit{dev}}& \textbf{\textit{test}} \\
\hline
Both birth and discharge & 348 & 40 & 160  \\
\hline
Only birth & 183 & 0 & 0  \\
\hline
Only discharge & 55	 & 0 & 0  \\
\hline
\multicolumn{1}{c|}{} & 586 & 40 & 160 \\
\cline{2-4}
\multicolumn{4}{l}{}
\end{tabular}
\caption{Number of infants by split and recording period.}
\label{tab:split}
\end{center}
\end{table}

To tune the algorithms, participants were provided with suggested development pairs - cross-product of the birth and discharge recordings (all possible combinations) of the dev infants. Similarly, test pairs consist of the $B\times D$ cross product for infants in the test set. Test pairs were provided without answers.

\begin{table}[h]
\centering
\begin{tabular}{l|c|c|c}
\hline
\textbf{Split} & \textbf{\# of +ive pairs} & \textbf{\# of -ive pairs} & \textbf{Total \# of pairs} \\
\hline
dev            & 40                            & 1540                           & 1580 \\
test           & 160                           & 25440                          & 25600 \\
\hline
\end{tabular}
\caption{Number of pairs in dev and test.}
\label{tab:positive_negative_total_pairs}
\end{table}

Each verification pair in both dev and test sets comprises one birth and one discharge recording. 
Pairing different recordings rather than cry sounds from the same recording is more representative of real-world applications for such a verification system, which may involve verifying an infant over multiple days. Additionally, we observed that verifying separate segments from the same recording is easier, possibly because an infant exhibits consistent traits within a single crying "episode" but not across different episodes. 

It's important to emphasize that the dev and test infants were not chosen randomly. Instead, they were randomly sampled from the top 200 infants with the highest cosine similarities between their birth and discharge embeddings, as calculated using the initial non-fine-tuned baseline model described in Section 3.2 and the first row of Table~\ref{tab:eer_baselines}. We opted for these relatively easier pairs due to the difficulty in recognizing an infant in an unseen recording within this dataset. By selecting easier-to-verify pairs, our aim was to add variance to the leaderboard and make the challenge more engaging.

For the competition, the test set was further subdivided into public (1,024 pairs) and private (24,576 pairs) parts at random. The participants could score their system on a public subset throughout the competition with a limit of 3 times per day. For the final evaluation, participants were asked to provide the top 5 performing systems for evaluation on the private subset.

\subsection{Baselines}
We consider two baselines based on ECAPA-TDNN speaker verification model~\cite{desplanques2020ecapa}. 
First, the ``naive'' baseline is pre-trained using a large adult speaker verification corpus - VoxCeleb~\cite{Chung18b} without any adaptation on cry data. We refer to the open-source SpeechBrain implementation~\cite{speechbrain} for further details with the model available in Hugging Face~\cite{voxcelebModel}. It yields 37.9\% and 38.1\% EER on the dev and test pairs respectively.

Second, the VoxCeleb model is fine-tuned on CryCeleb training data, specifically focusing on the 348 infants with both birth and discharge recordings (Table~\ref{tab:split}, top left). By limiting the dataset to this subset, we can train the model on all birth recordings while reserving the discharge recordings for validation. This approach enables us to assess the model's ability to generalize patterns learned from birth recordings to discharge recordings, in some sense simulating the verification setting. 
Alternatively, we could have fine-tuned the model on both birth and discharge recordings from the 348 infants or even expanded it to all recordings from the 586 train infants. The former option introduces more data but removes the ability to validate the classification performance. The latter allows for even more data, however, it also increases the number of classes, which could hinder the model's learning. 

The model is trained on 3-5 second random chunks from concatenated cry sounds at each iteration. The best 5 epochs, determined by validation accuracy, are saved, and these 5 checkpoints are then evaluated on the dev pairs using the EER (verification task). The checkpoint with the lowest EER is chosen as our final fine-tuned model. The fine-tuned model achieves an EER of 27.8\% on the dev set and 28.2\% on the test set. It is open-sourced\footnote{\url{huggingface.co/Ubenwa/ecapa-voxceleb-ft2-cryceleb}} along with fine-tuning code\footnote{\url{github.com/Ubenwa/cryceleb2023}}.

\subsection{Baseline performance}

Table~\ref{tab:eer_baselines} summarizes the performance of two baselines on dev and test sets and the following section provides more details about these systems.

\begin{table}[ht]
\centering
\begin{tabular}{l|c|c}
\hline
\textbf{ECAPA-TDNN baseline} & Dev & Test\\
\hline
No pre-training (CryCeleb only) & 32.0\% & 31.8\%  \\ \hline 
VoxCeleb pre-training (no fine-tuning) & 37.9\% & 38.1\%\\
 + CryCeleb fine-tuning & 27.8\% & 28.2\% \\
\hline
\end{tabular}
\caption{Performance (EER) of baseline models. First row indicates the performance obtained with the model pre-tained on VoxCeleb and the second row indicates the performance where the first model is fine-tuned on CryCeleb.}
\label{tab:eer_baselines}
\end{table}



\begin{figure*}[h]
\centering
\centerline{\includegraphics[width=\linewidth]{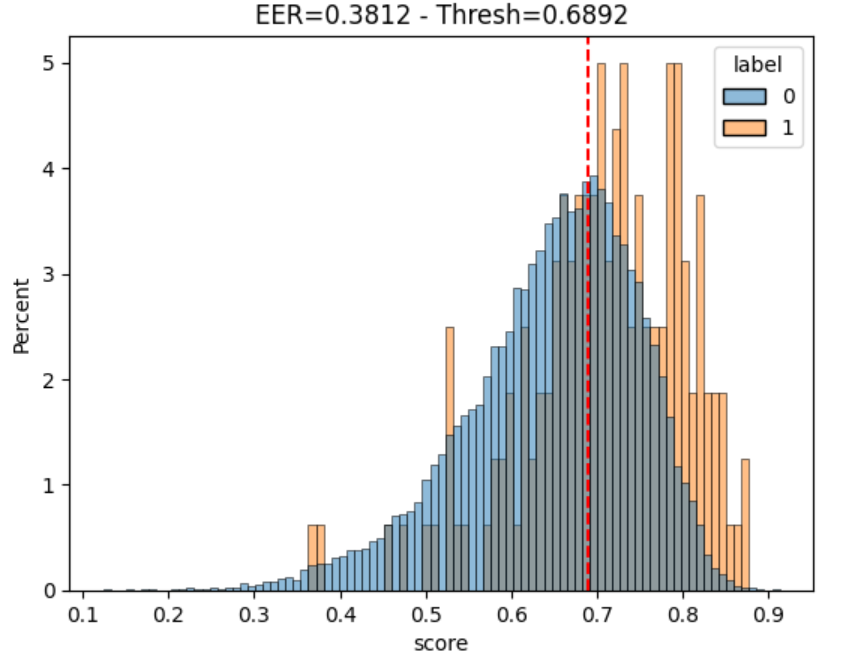}}
\caption{Verification scores for negative and positive pairs produced by the ECAPA-TDNN pre-trained on speech data VoxCeleb. Classification threshold (red dashed line) is selected to minimize EER. }
\label{fig:hist-perf-vox-only}
\end{figure*}

\begin{figure*}[h]
\centering
\centerline{\includegraphics[width=\linewidth]{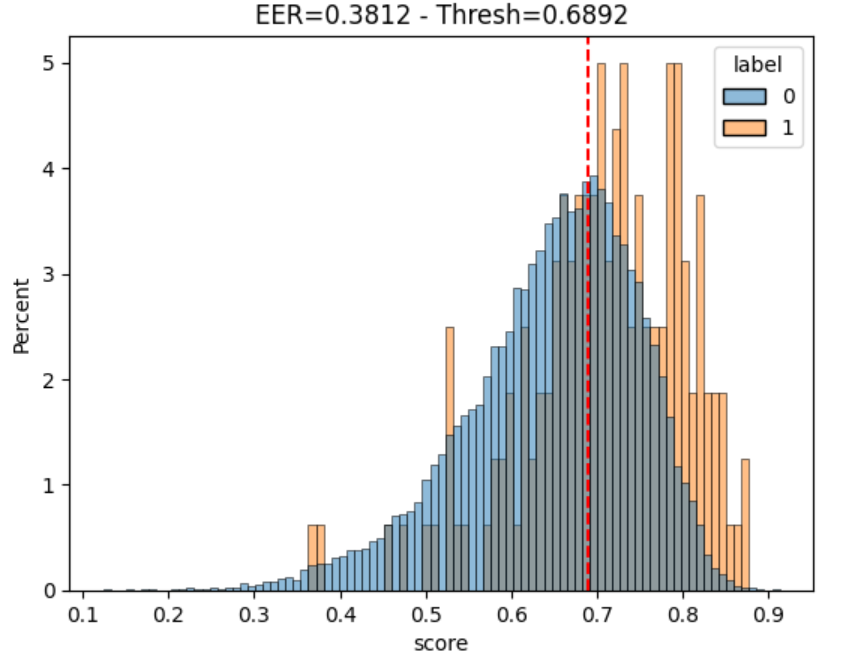}}
\caption{Verification scores for negative and positive pairs produced by the ECAPA-TDNN pre-trained on speech data VoxCeleb and fine-tuned on CryCeleb . Classification threshold (red dashed line) is selected to minimize EER. }
\label{fig:hist-perf-vox-plus-cry}
\end{figure*}

\begin{figure*}[h]
\centering
\includegraphics[width=1.0\columnwidth]{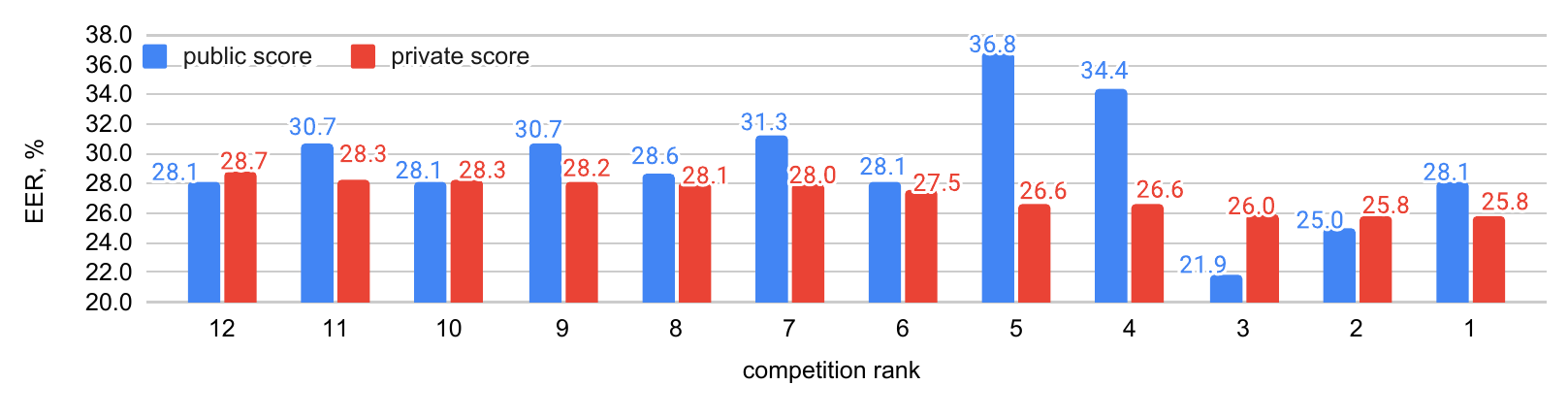}
\caption{\vspace{-2pt}Equal error rates of CryCelb 2023 competition submissions. The public subset of test set was available throughout the competition while private subset was used only after the competition end to evaluate 5 submissions selected by each participant.}
\label{fig:leaderboard}

\end{figure*}

Figures~\ref{fig:hist-perf-vox-only} and \ref{fig:hist-perf-vox-plus-cry} shows the histograms of the scores for positive pairs (orange) and negative pairs (blue), with the y axis normalized separately for each color. The red vertical line indicates the threshold where the EER is achieved.
First, we observe that fine-tuning the VoxCeleb model with cries leads to improved verification performance, as evidenced by the lower EER and more visually distinct distributions. Second, we notice that the scores for negative pairs in the tuned model form a bell-shaped distribution centered closer to zero. This is intuitively more reasonable compared to the naive VoxCeleb model, where the most common score for a negative pair is 0.7.

\subsection{Challenge results}

In total, 224 people from 37 countries registered for the challenge, and 59 participants made at least one submission. In total, we received 435 submissions. 

While many participants simply went through a starter notebook that demonstrated usage of the baseline, 11 experimented with the challenge a lot more and improved the baseline. Figure~\ref{fig:leaderboard} summarizes performance on public and private subsets of the challenge for these participants.

Even for the top-performing systems, the EER is considerably high, which shows that the problem is in fact quite challenging. We hypothesize that overall it is not easy to distinguish baby identity. In addition, over time between birth and discharge the physiological characteristics may change and impact the cry sounds.

The top three-performing systems were based on ECAPA-TDNN baseline. All three participants applied and reported and improvement from test-time data augmentation~\cite{kim2022improving}. The top-performing model\cite{finalists-1} did not use pre-training and took advantage of more parameters and label smoothing~\cite{szegedy2016rethinking}. 
The second system\cite{finalists-2} benefited from parameter tuning and training dataset enriched with development pairs. 
The third system\cite{finalists-3} achieved an overall significantly better EER 21.9\% on the public evaluation but was ranked third on a large private portion. It was built with a triplet loss\cite{li2017deep} instead of AAM-softmax\cite{deng2019arcface} of ECAPA-TDNN.

\section{Discussion}

The CryCeleb dataset offers the research community a robust foundation for exploring this nascent domain of cry-based infant recognition. The dataset’s extensive coverage, including multiple recordings per infant across different time periods, enables a detailed examination of cry variability and its implications for speaker verification tasks. The CryCeleb 2023 challenge, which attracted participants across 37 countries, has sparked diverse approaches to improving verification performance, revealing the complexities inherent in identifying infants through their cries. While fine-tuning on CryCeleb demonstrates considerable improvements over pre-trained baselines, the relatively high Equal Error Rates (EER) achieved by the top systems highlight the difficulty of this task and underline the need for further methodological investigation.

One of the key findings of this work is the gap between the current state of cry-based recognition and state-of-the-art performance in adult speaker verification. This discrepancy may be attributed to the inherent challenges of cry signals, including their high variability, short duration, and susceptibility to environmental noise. The challenge results underscore the importance of developing domain-specific algorithms tailored to cry data, rather than directly transferring methods from adult speech processing.

Looking forward, the CryCeleb dataset opens several avenues for future research. Enhancing the dataset by incorporating more diverse recording conditions and longitudinal data could provide deeper insights into cry dynamics and identity retention over time. Methodologically, exploring advanced machine learning techniques, such as self-supervised pre-training or domain adaptation tailored to cry signals, may further bridge the performance gap. By addressing these challenges, CryCeleb has the potential to serve as a catalyst for the development of robust cry analysis tools with applications in both clinical and household settings, ultimately advancing our understanding of infant vocalization and its utility in health diagnostics.

\chapter{Conclusion}
\label{chap:conclusion}
\section{Discussion}
This thesis explored three major challenges when applying deep learning to medical audio, specifically in analyzing infant cries for the early detection of health conditions: representation learning, model compression, and domain adaptation.

The study used neural transfer learning to address the issue of limited annotated infant cry data. By leveraging large adult speech datasets, we demonstrated that transfer learning can be effective for cry-based medical analysis. Our approach utilized pre-trained models from adult speech to learn rich representations for the classification of perinatal asphyxia. We found that this transfer learning approach significantly outperformed traditional SVM baselines and showed enhanced robustness in noisy and challenging environments, consistent with earlier findings in cross-domain transfer learning \cite{Pan2010, he2019rethinking}. These results support the growing body of evidence that transfer learning from large, related datasets can improve performance on medical applications \cite{shin2016deep}.

To ensure the models could be deployed in low-resource environments, we developed and implemented an end-to-end model compression technique using tensor decomposition for RNNs. This method achieved significant reduction in the number of model parameters while retaining high predictive performance, as evidenced by compression ratios up to 300x with only marginal performance degradation. This outcome highlights the potential of compressed deep learning models in resource-constrained healthcare settings, akin to findings from recent work on efficient model compression of deep neural networks. \cite{yang2017tensor,tjandra2017, yin2020}.

Addressing dataset bias and domain shift was another critical focus of the thesis. We applied domain adaptation techniques, including target noise injection, to ensure that the models generalize well across different clinical environments. These techniques enhanced the models' robustness, leading to improved generalization, especially when cry data were collected in diverse geographical settings. This contribution is aligned with the success of domain adaptation in more general purpose applications in speech and image recognition. \cite{ganin2016domain, tzeng2017adversarial}.

Through a large multi-center clinical study, the first of its kind focused on investigating the relationship between cry patterns and neurological conditions, we showed that some of the techniques developed in this work are applicable in the real world. Our system achieved 92.5\% AUC in detecting neurological injury from cries \cite{onu2023help}. In addition, by releasing a large portion of this database to the public, we made an important contribution to research into infant cries and physiological sounds at large. \cite{budaghyan2024crycelebspeakerverificationdataset}.

In summary, the work we conducted demonstrates the feasibility of using deep learning for medical audio analysis, particularly for infant cries. It provides evidence that transfer learning, model compression, and domain adaptation techniques can be effectively combined to create robust, deployable AI systems for healthcare.

\section{Limitations}
While this thesis made significant contributions to advancing the application of deep neural networks for medical audio analysis, several limitations must be acknowledged to provide a balanced perspective. One of the primary constraints of this work lies in the datasets used. The reliance on relatively small and specialized datasets, such as the Chillanto Infant Cry Database, limits the generalizability of the findings. The Ubenwa Cry Database moves the needle on this, as it is more than 10 times larger than Chillanto, yet it is still small in comparison to other audio datasets like VGG. The proposed methods demonstrated promising results, but their performance on larger, more diverse datasets remains to be explored.  This could be addressed by incorporating more extensive and varied datasets to enhance the robustness and applicability of the developed techniques.

Another limitation pertains to the domain-specific nature of the proposed methods. The approaches for transfer learning and domain adaptation were tailored specifically to infant cry data, which might restrict their applicability to other types of medical audio signals. While the physiological and acoustic principles underlying infant cries provide a unique opportunity for study, the adaptability of these methods to other domains of medical sound analysis remains an open question.

Additionally, the methods developed have not undergone extensive validation in real-world clinical settings. Although collaborations with clinical partners enabled meaningful insights, the models have yet to be tested comprehensively in diverse and unpredictable environments where medical audio is recorded. The challenges posed by noisy and variable recording conditions -as would be encountered in actual clinical scenarios- represent a critical area for further research. Such validation is essential to establish the practical reliability and robustness of these models.

Lastly, while the thesis addresses model compression for improved portability, further work is needed to ensure optimal deployment on low-resource hardware, such as mobile devices or embedded systems. Although significant strides were made in reducing model size and computational requirements, the transition from research prototypes to real-world applications demands more refinement. Testing under diverse hardware configurations and real-time scenarios would help bridge the gap between theoretical promise and practical implementation.

This thesis nevertheless provides a foundation for future research (discussed in the next section) to build upon its findings and address the challenges identified.

\section{Future work}
A promising direction for future work is the improvement of self-supervised learning (SSL) methods to generate even richer and more nuanced representations from infant cries. Early results here show that SSL could reduce the reliance on labeled datasets and leverage the large amounts of unlabeled cry data available. Furthermore, exploring multimodal learning by incorporating other data types, such as video or physiological measurements (e.g., heart rate), could improve the performance and robustness of cry-based health diagnostics.

To address the limitations of dataset size and diversity, more effort is needed in the development of larger, more representative datasets of infant cries across different geographies, ethnicities, and medical conditions. This is a cost-intensive but necessary step which can be augmented by synthetic data generation through methods like generative adversarial networks (GANs).

Another avenue of future research is to develop personalized models that adapt to individual infants. Using few-shot learning approaches, models could be fine-tuned with just a few cry samples from each infant, improving accuracy for individual cases \cite{finn2017model}. Such approaches are particularly promising for creating tailored diagnostics for infants with unique conditions.

Given the practical importance of deploying these models in low-resource environments, further work could focus on optimizing the models for deployment on edge devices, such as smartphones or neonatal monitoring units. Techniques like federated learning \cite{li2020review} could enable secure model training across distributed devices while preserving patient privacy.

To foster greater trust among clinicians and caregivers, future work should incorporate explainability frameworks into cry analysis models. Techniques like SHAP \cite{lundberg2017unified} and LIME \cite{ribeiro2016should} could help elucidate which aspects of the cry are indicative of particular conditions, ensuring that diagnostic decisions are transparent and understandable for clinicians.

Finally, the methodologies presented in this thesis can be applied to other types of medical audio, such as heart and lung sounds. Developing generalized audio models for a wide range of medical applications could help expand the utility of deep learning in affordable healthcare, making AI-driven diagnostics available to more people worldwide.

\printbibliography


\end{document}